\documentclass [11pt,letterpaper]{JHEP3}
\usepackage{amsmath}
\usepackage{epsfig}
\usepackage{psfrag}
\usepackage{cite}       
\usepackage{bm}         
\usepackage{verbatim}   
\advance\parskip 1.0pt plus 1.0pt minus 2.0pt   
\def\href#1#2{#2}	

\def\coeff#1#2{{\textstyle {\frac {#1}{#2}}}}
\def\half{\coeff 12}
\def\C{{\cal C}}
\def\N{{\cal N}}
\def\T{{\cal T}}
\def\P{{\cal P}}
\def\I{{\cal I}} 
\def\G{{\cal G}}
\def\None{\N\,{=}\,1}  

\def\R{{\mathbb R}}

\def\tr{{\rm tr}}

\def\Nc{N_{\rm c}}
\def\Nc{N}
\def\Nf{N_{\rm f}}
\def\Z{{\mathbb Z}}

\def\Dslash{{\rlap{\raise 1pt \hbox{$\>/$}}D}}
\def\O{{\cal O}}

\def\Im{\mathrm {Im}}

\preprint{SLAC-PUB-12376}

\title
    {%
    \boldmath 
Phases of  $\Nc= \infty$ QCD-like gauge theories 
on $S^3 \times S^1$  and nonperturbative 
orbifold-orientifold equivalences
    }%

\author
    {%
    Mithat \"Unsal$^1$\footnote{\email{unsal@slac.stanford.edu}}~
    \\${}^1$ SLAC, Stanford University, Menlo Park, CA 94025
    \\\; Physics Department, Stanford University, Stanford, CA, 94305
    }%

\abstract
    {%
We study the phase diagrams of $\Nc= \infty$  vector-like, asymptotically free 
gauge theories as a function of volume, on 
$S^3\times S^1$.   The theories of interest 
are  the ones with fermions in two index 
representations 
[adjoint, (anti)symmetric, and bifundamental abbreviated as 
QCD(adj),  QCD(AS/S) and QCD(BF)],
 and are interrelated  via orbifold or orientifold projections. 
The  phase diagrams  reveal interesting phenomena such as 
disentangled realizations of chiral 
and center symmetry,   confinement without chiral symmetry 
breaking,   
zero temperature chiral transitions,  and  in some cases, exotic phases 
 which  spontaneously  break 
the discrete symmetries  such as C, P, T as well as CPT.
In a regime where the theories are perturbative,  
the deconfinement temperature  in SYM, and QCD(AS/S/BF)  coincide.    
The thermal phase diagrams of thermal orbifold  QCD(BF), 
orientifold QCD(AS/S),  and $\N=1$ SYM    
coincide, provided charge conjugation symmetry for QCD(AS/S)  and $\Z_2$ 
interchange symmetry of the QCD(BF) are not broken in the phase continously 
connected to $\R^4$ limit. 
When the $S^1$ circle is endowed with periodic boundary conditions, the 
(nonthermal) phase diagrams of orbifold and 
orientifold QCD are still the same, however,  
both theories possess chirally symmetric  
phases which are absent   in $\None$ SYM. The match and  mismatch of the 
phase diagrams depending  on the spin structure of fermions along the $S^1$ 
circle is naturally explained  in terms of the necessary and sufficient 
 symmetry realization conditions which determine the validity of the 
  nonperturbative orbifold orientifold  equivalence.
        }%

\begin{document}

\section{Introduction and results}
\label{sec:intro}
Understanding the dynamics of the strongly coupled, vectorlike 
gauge theories is an outstanding problem of contemporary physics.  Various 
methods have been developed to come to  grips with QCD and QCD-like 
theories.  Among such  lattice gauge 
theory and effective field theory have been the most successful 
and reliable, and 
supersymmetry has been useful  in  addressing  some questions.
A more qualitative method is  the  $1/N$ expansion.

In this paper, we examine the dynamics and  phase diagrams of certain 
asymptotically free,  confining vectorlike 
gauge theories (without fundamental scalars) in the 
 infinite number of color limit by benefiting from the available techniques 
mentioned above. 
Since  $\Nc= \infty$  is a 
 thermodynamic  limit, it is not necessary  to take the volume to infinity 
to attain  phase transitions \cite{Gross:1980he,Narayanan:2005en, 
Narayanan:2003fc }.   In fact, dialing 
the volume (and temperature), which  acts as an external parameter,  
leads to rich phase diagrams.
We  work on the 
 compact space  $Y^3 \times S^1$, where $Y^3$  is some  
three manifold. Our primary  choice of the three manifold is $S^3$ for reasons 
explained below. We will discuss   the decompactification limits
$\R^3 \times S^1$, $\R \times S^3$ and  $\R^4$ 
where the hardest problems of the QCD-like theories lie. 
The study of the dynamics of  the QCD-like 
theories as a function of radii reveals new 
phenomena  such as  disentangled realization of chiral and center symmetry,  
confinement without chiral symmetry breaking,  
zero temperature chirally symmetric phases, as well as the 
appearance  of chirally asymmetric  phases at weak coupling.  
Most  of these  phenomena   are testable on  the lattice. 

The QCD-like theories that will be of interest are the ones with fermions 
in double index (or tensor) representations. 
The virtue of the double index representation 
is that  unlike the fundamental quarks, a finite number of the tensor flavors 
 are not kinematically 
suppressed in the  large $\Nc$ limit. 
In fact, too many tensor quarks may overwhelm the asymptotic freedom which 
our discussion assumes. Therefore, we   restrict the number 
of flavors to at most five, $n_f \leq 5$.      
In the determination of the vacuum structures, thermodynamics 
and other properties, the double index quarks are  as  
important as gluons and sometimes, as we will see, the  fermionic contributions
play the decisive role.  
There are three classes of double index representations for the complex 
$U(N)$ color gauge  group:  adjoint, antisymmetric, symmetric.  For the 
product gauge group  $U(N)\times U(N)$, 
there is also  two index bifundamental 
representation.    We   abbreviate these theories as
QCD(adj), QCD(AS/S) and QCD(BF). 
At first glance,  it may sound strange
to include a product gauge group. There are two good reasons for doing so: 
First, it is a vectorlike gauge theory, and second,  more 
interestingly, its phase  diagram turns out to be identical to QCD(AS/S).   
\footnote{As a matter of convention,  
by the  $n_f$ flavor vectorlike (or QCD-like) gauge theory,  
we refer to  $n_f$ Dirac fermions for 
QCD(AS/S/BF) since  (anti)symmetric and bifundamental representations are 
complex,   
and  $n_f$ Majorana (or Weyl) adjoint  fermion for QCD(adj) 
because  the adjoint representation is real.
Since  a gauge invariant mass term is allowed in any of these theories, 
they are all vectorlike by the standard  definition.  
The ratio of the number of fermionic  degrees of 
freedom to the bosonic degrees of freedom is just $n_f$ for 
 $n_f$ flavor QCD(adj) and QCD(BF), and is  $n_f(1\mp \frac{1}{N})$ 
for $n_f$ flavor QCD(AS/S).}

The QCD-like  theories formulated on $\R^4$ (or $\R^{3,1}$), 
unlike  conformal field theories such as ${\cal N}=4$ SYM,  
do not have a tunable coupling constant. 
However, formulating  this theory on $S^3 \times  S^1$, where the radius  
of either $S^3$ or $S^1$ is much smaller than  the strong confinement scale 
$\Lambda^{-1}$, 
i.e, $\min(R_{S^3}, R_{S^1}) \ll \Lambda^{-1}$,  we can benefit from asymptotic 
freedom. 
In this regime, the coupling constant is weak 
$g^2(1/\min(R_{S^3}, R_{S^1})) \ll 1$, and perturbative methods are reliable 
on the scale of the smaller radius. 
%

Formulating the theory on small $S^3 \times S^1 $,  as shown in 
\cite{Sundborg:1999ue, Aharony:2003sx, Aharony:2005bq} 
has the advantage of making  the center symmetry realizations  
accessible to perturbation theory.  (Also see 
\cite{Schnitzer:2004qt,Schnitzer:2006xz,Yamada:2006rx} for related work.)
Since  the three-sphere $S^3$ is simply-connected just like $\R^3$, 
its   spin structure is fixed.  On the other hand, 
depending on the choice of the boundary conditions for the  
fermions in the  $S^1$ direction (antiperiodic or  periodic), the partition 
function  will either be the usual thermal ensemble $Z= \tr e^{- \beta H}$ 
or   a twisted partition function ${\widetilde Z}= \tr e^{- \beta H} (-1)^F$. 
$F$  is the  fermion number  operator.  
Depending on the choice of the boundary conditions for fermions on  
$S^1$,  the center symmetry should be viewed  as temporal 
 $\G_t$ for antiperiodic 
or spatial center symmetry  $\G_s$ for periodic boundary conditions. 
Accordingly, there are  two distinct classes 
of center symmetry changing transitions, one 
associated  with   temporal Wilson lines 
(thermal Polyakov loop)  and the 
other associated with the spatial Wilson lines.
The temporal center symmetry realization forms a criterion for  the  
deconfinement, confinement transition, and  the change in symmetry realization 
is triggered by thermal fluctuations. The transition manifests itself as an 
abrupt change in the   thermodynamic free energy density ${\cal F}$. 
${\cal F}$ may be extracted 
from the minimum of the  effective 
potential for thermal Polyakov loops, hence from the partition function 
${\cal F}= -   \frac{1}{\beta V_{S^3}} \log  Z $ where $V_{S^3}$ and  
$\beta$ are respectively volume of the three sphere and inverse temperature. 
But the  spatial Wilson  lines 
do   not  constitute  a deconfinement criterion.  The change in the spatial 
symmetry realization is due to zero temperature 
quantum mechanical fluctuations (rather than thermal) and the phase transition 
is accompanied by an abrupt change in the vacuum energy density. Therefore it 
is more naturally  viewed  as a quantum phase transition.  
The vacuum energy density 
  may be extracted from the 
effective potential of the spatial Wilson line and the twisted partition 
function $
{\cal E}= - \frac{1}{ V_{S^3} \times V_{S^1 }} \log  \widetilde Z 
$ where $V_{S^3} \times V_{S^1}$ is the volume of the four manifold 
$S^3 \times S^1$. \footnote{Strictly speaking, on $S^3 \times S^1$, if one 
wants  to have an operator interpretation, the only way to define a transfer 
matrix (or Hamiltonian)  is to interpret $S^1$ as temporal 
(and there is no second choice.) \cite{LGY}. On the other hand, on $\R^3 \times S^1$ 
limit, one can define the transfer matrix  for either choices of 
$S^1$, temporal or spatial.
 The two choices of the $S^1$ have different physical interpretations. 
The case for temporal  $S^1$ (which is widely studied in literature) 
is  the finite temperature field theory on $\R^3$. In the case 
$S^3 \times ({\rm temporal} \; S^1)$, one can  define a 
transfer matrix. 
 If $S^1$ is a spatial 
circle, then this should be viewed (on $\R^3 \times S^1$) as a zero 
temperature field theory on a space with one compact direction, where zero 
temperature  axis corresponds to one of the noncompact $\R$. Or 
in Minkowski space, one can think 
of  an Hamiltonian formulation of a gauge theory on 
$\R^{2,1} \times {\rm spatial}\; S^1$.  Therefore,  a quantum mechanical 
interpretation of  the correlators in the functional integral is possible.  
The reader may be worried that the Euclidean  theory on 
$S^3 \times ({\rm spatial} \; S^1)$  does not have an operator interpretation 
(or even any  temperature related interpretation)
in the sense of the last statement at any finite value of $S^3$. Nonetheless, 
we can define an effective action for the  spatial Wilson lines  
by compactifying  the functional integral 
representation of $\tr e^{-\beta H}(-1)^F$
on $\R^3 \times S^1$ into $S^3 \times S^1$. The 
effective action defined in this way turns out to be a very useful tool as 
we will see in the course of the study of phases. 
Therefore,  we regard the effective action (for spatial Wilson lines) on 
$S^3 \times S^1$  
as an ``auxiliary'' device, and borrow  its terminology such as 
``ground state, correlation function, vacuum  energy ''  
from its decompactification limit, where a quantum mechanical 
interpretation makes sense.
 The twisted partition function, though lacking any operator 
interpretation on $S^3 \times S^1$,   
turns out to be beneficial  due to the smooth volume 
dependence conjecture for (spatial) center  symmetry realization that we 
will discuss momentarily. }
The reader  should  keep  in mind that the  realizations  
and physical interpretations  of the temporal and spatial center symmetries 
for a given theory  are different, and the existence of one do not imply the 
other, see for example
\cite{Aharony:2005ew, Unsal:2006pj,DeGrand:2006qb}

{\bf Phases of $N= \infty$ QCD(adj) and QCD(AS/S/BF) on $S^3 \times S^1$:}
For thermal $\None$ SYM and QCD(AS/S/BF), we analytically demonstrate 
the temporal 
center symmetry 
changing  confinement deconfinement transitions in a perturbative regime 
of the theory. These transitions are 
 associated with a jump in the 
free energy density from   $\O(N^2) $ in the deconfined plasma phase to being 
  $\O(1) $ in the confined phase.  In the regime where the radii of $S^3$ and 
$S^1$ are small compared to the strong confinement scale $\Lambda^{-1}$,  
$\max (R_{S^3}, \beta) \ll \Lambda^{-1}$ and a perturbative one loop 
analysis is  reliable,    we find that  the transition occurs
 at exactly the same  temperature: 
\begin{equation}
T_d^{\rm SYM}=  T_d^{\rm QCD(AS/S)}=T_d^{\rm QCD(BF)}
\label{Eq:temp}
\end{equation} 
At first glance, this result is rather surprising and its underlying reason, 
based on nonperturbative large $N$ orbifold/orientifold equivalence,  
will be explained below. 
We do not know how to calculate the deconfinement temperature in the strongly 
coupled, nonperturbative regime of these theories. Therefore, a priori there 
is no way to tell whether Eq.\ref{Eq:temp} will hold in the nonperturbative 
regime.  The implication of the large $\Nc$ orbifold/orientifold equivalence 
is that   Eq.\ref{Eq:temp} should be true in the $\Nc=\infty$ vectorlike 
gauge theories even in the nonperturbative regime.    The matching of the 
deconfinement temperatures in 
 Eq.\ref{Eq:temp} generalizes easily  to the multiflavor 
case as well where  $T_d^{\rm SYM}$ is replaced by $T_d^{\rm QCD(adj)}$.

If $S^1$  is endowed with periodic boundary conditions,  we do not 
observe any spatial 
center symmetry changing transition in SYM and its multiflavor 
generalization  QCD(adj). This is by itself surprising and just mimics the 
behaviour of these theories on $\R^3 \times S^1$ where spatial center symmetry 
realization is independent of the size of the $S^1$ circle 
\cite{Unsal:2006pj}, and is unbroken at any radius. 
On the other hand, we do observe a center symmetry changing transition 
in QCD(AS/S/BF) 
occurring at exactly the same radius of  $S^1$.
\begin{equation}
  R_{{S^1}, s}^{\rm QCD(AS/S)}=R_{S^1, s}^{\rm QCD(BF)}
\label{Eq:rad}
\end{equation} 
   This transition is 
associated with  the breaking of the spatial center symmetry $\G_s$ 
and is accompanied  with a  change in the vacuum energy density. In the 
$\G_s$ symmetry broken 
phase, the vacuum energy density is  $\O(N^2) $ while  in the unbroken phase, 
it is  $\O(1)$.    

The chiral properties of these vectorlike theories are 
 equally interesting. 
Upon  thermal compactification,  we find similar  behaviour for all the 
QCD-like  theories.  
At high temperatures or high curvatures (i.e, at small thermal 
 $S^1$ or small $S^3$), there is no condensate and 
the theory is chirally  symmetric.   The phase at small radius of $S^3$ 
but low temperature, shows confinement without chiral symmetry breaking.  
When 
$\min(R_{S^3}, \beta ) \gg \Lambda^{-1}$, the chiral symmetry is broken.  
The  $n_f=1$  QCD-like theories possess a discrete chiral symmetry  
$\Z_{2h}$ where $h= N$ for QCD(BF) and SYM and $h= N\mp2$ for QCD(AS/S). 
The chiral  symmetry  is spontaneously broken to $\Z_2$ leading to $h$ isolated 
vacua which are distinguished by the phase of the condensate.  
The  $n_f>1$ case is accompanied with both discrete and continuous chiral symmetry 
breaking, therefore leading to both domain walls and Goldstone bosons. There 
are  $h$ isolated vacuum manifolds distinguished by the phase of the 
determinant of the rank $n_f$ chiral condensate matrix.  The isolated vacua in 
the $n_f=1$ case  are 
replaced by suitable coset spaces.   The chiral properties 
are discussed in detail in subsections \ref{sec:SYM4},  \ref{sec:orb4}, 
and \ref{sec:orienti4}.

The case where  $S^1$  is endowed with the 
periodic boundary condition is drastically different between QCD(adj) 
and QCD(AS/S/BF).   The chiral properties 
of   QCD(AS/S/BF) are essentially the same as in the thermal case, 
i.e., a 
chirally symmetric phase at  high curvature and on a small 
$S^1$ circle $\min(R_{S^1}, R_{S^3}) \ll \Lambda^{-1}$,
 and a chirally asymmetric phase in the opposite limit. 
However, the underlying physical reasons behind  
the restoration  of chiral symmetry on small $S^1$  differs from the 
thermal case  (where it is  due to tree 
level antiperiodic boundary conditions for fermions),  
as it  is a  one loop quantum  effect. 
However for SYM (or QCD(adj)), we find the chiral symmetry realization is 
independent of the  size of the $S^1$ circle, and it only 
depends on the curvature of $S^3$. At large curvature, the theory is 
chirally symmetric while at small curvature, the chiral symmetry is broken. 
The physical reasons 
 why the chiral symmetry realizations are so different between 
QCD(adj) and QCD(BF/AS/S) and why the former (latter) has a chirally 
asymmetric (symmetric)  
phase at small $S^1$  is explained in \ref{sec:SYM4}, and \ref{sec:orb4}.

{\bf Large $N$ equivalences, small $n_f$  universalities:}
The results of the phase diagrams would be completely mysterious were it not 
for the nonperturbative large $N$ equivalence. The QCD-like theories described 
above, for each value of $n_f$, 
are related 
to each other via a chain of orbifold and orientifold projections 
(see refs.\cite{Kovtun:2003hr, Kovtun:2004bz, Tong:2002vp, Armoni:2004ub,
Armoni:2003yv, Armoni:2003fb,Armoni:2003gp} and references 
therein). 
The case where $n_f=1$ is particularly interesting, as it relates 
supersymmetric $\None$ SYM theory  to nonsupersymmetric theories such as 
QCD(BF/AS/S). 
Ref.\cite{Kovtun:2003hr, Kovtun:2004bz}  demonstrated that {\it as long as 
the discrete symmetries defining the neutral sectors are unbroken,  there is a 
nonperturbative large $N$ equivalence among the neutral sectors 
of  the theories related to each other via orbifold and orientifold 
projections.}   
\footnote{In 
\cite{Kovtun:2003hr}, the necessary  criteria for the validity of the 
equivalence is derived by comparing the generalized loop  equations (the 
Schwinger-Dyson equations for the correlators of the gauge invariant operators)
on lattice regularized parent and daughter theories.   If the symmetries 
defining the neutral sectors is unbroken (broken), then the loop equation do 
(not) coincide.  This demonstrates the necessity of the symmetry realizations.
However,  the loop equation are nonlinear and typically 
has multiple solutions. Therefore,  coinciding loop equations do not constitute 
a proof outside the strong coupling, large mass phase of the lattice gauge 
theory  where one can prove uniqueness of the solution. The more abstract 
coherent state approach plugs this hole and demonstrates that the unbroken 
symmetry is the sufficient condition for large $N$ equivalence
\cite{Kovtun:2004bz}.}
The large $N$ equivalence does not make any reference to properties such as 
supersymmetry, conformal symmetry, spacetime dimension, or the topology 
of the spacetime manifold or other details.  Its validity only relies on 
the realizations of  symmetries defining the respective neutral sectors of our 
vectorlike theories. 

If the symmetries defining neutral sectors are unbroken, then the  
$N=\infty$ neutral sector dynamics coincide.
This implies a well defined mapping between the 
expectation values of Wilson lines and Wilson loops, the 
expectation values of the  
chiral condensates, the  nonperturbative particle spectra such as 
 glueball and meson masses, the multibody decay amplitudes, as well as  
the thermodynamics. 
 In other words, the large $N$ equivalence provides a powerful  $N=\infty$ 
duality without solving neither  theory (which is hard) 
\cite{Kovtun:2003hr, Kovtun:2004bz}.   


In our case, the neutral sectors amount to  the bosonic 
subsectors of the $U(N)$ QCD(adj) theory (which is neutral under the fermion 
number 
modulo two $\Z_2= (-1)^F$) , the charge conjugation symmetry 
$\C$ even subsector of 
$U(N)$ QCD(AS/S)  and $\Z_2$ exchange even subsector  of $U(N)\times U(N)$ 
QCD(BF).  Therefore, 
the validity of orbifold and orientifold equivalences is crucially  tied to 
their respective unbroken $\Z_2$ symmetries. 
If the corresponding $\Z_2$ is unbroken,  a prediction of large $N$ equivalence 
is that the phase diagrams of $n_f$ flavor large $N$ QCD(adj)  and  
QCD(AS/S/BF) should be the same.  
 In the thermal case, where the properties of  QCD(AS/S/BF) and 
SYM coincide,  the $\Z_2$  charge conjugation symmetry for QCD(AS/S) and 
 the $\Z_2$  exchange symmetry  for  QCD(BF)   are unbroken in any phase of 
these theories. (More precisely, there are three phases and in two of them, 
we can analytically demonstrate that these  symmetries are unbroken. In the 
third phase where $\min(R_{S^3}, \beta) \gg \Lambda^{-1}$, the current knowledge 
is   consistent with unbroken  $\Z_2$.)
  In the nonthermal case, where the phase diagram of QCD(adj) 
is different from QCD(AS/S/BF),  
there exists a phase in which the  
$\Z_2$ symmetries for QCD(AS/S/BF) are spontaneously  broken.  
Outside that 
phase, the large $N$ equivalence is still valid. Therefore,  coinciding phase 
diagrams in the thermal case 
among our QCD-like gauge theories may  naturally be 
viewed as a  consequence of large $N$ orbifold/orientifold equivalence.  
The differences  between  the phase diagrams of  QCD(adj) and QCD(AS/S/BF) 
in the nonthermal case may be viewed 
as an eminent demonstration of the symmetry realization conditions.  

{\bf A conjecture on the spatial and temporal center symmetry realization:}
 The examination of  the center symmetry realizations  of QCD-like theories 
on  $S^3 \times S^1$ leads us to a smooth volume dependence conjecture for the 
asymptotically free, confining gauge theories.
\begin{quote}
 {\it In confining 
vectorlike, asymptotically free  large $\Nc$ gauge theories formulated 
 on $S^3 \times S^1$,   
 the (spatial and temporal) center symmetry realizations are independent of the 
size of the $S^3$  sphere in the following sense:
 If a  perturbative center symmetry changing transition 
 exists in small $S^3 \times S^1$, 
 it smoothly interpolates into a nonperturbative transition 
 in  $\R^3\times S^1$.  
 If there is no change in center symmetry realizations on small $S^3 $, the 
theory   on $\R^3\times S^1$ does not undergo a center symmetry changing 
transition as a function of $S^1$ volume  either. 
}
\end{quote}
In QCD(AS/S/BF), 
there is a change  in  the spatial and temporal center symmetry realization 
on $S^3 \times S^1$, and we expect this to continue to $\R^3\times S^1$.  
For thermal QCD(adj), there is a change in the 
temporal center symmetry realization 
on small $S^3 \times S^1$ which is expected to interpolate all the way to 
$\R^3 \times S^1 $. However, for QCD(adj) with periodic boundary conditions, 
the spatial center symmetry is unbroken on $S^3 \times S^1$,  and  this result 
also holds on $\R^3 \times S^1$.  
This conjecture is a natural generalization 
of the one  by Aharony et.al. \cite{Aharony:2005bq} 
in the case of pure Yang-Mills theory. 

\section{Generalities of vector-like gauge theories}
\label{sec:gen}

Let us first review some generalities of the asymptotically 
free, confining  vectorlike gauge theories,  and  set the tools for the 
analysis.  
Throughout this paper, we will examine both zero temperature and finite 
temperature phases of certain large $\Nc$ vectorlike theories on 
Euclidean spacetime manifolds, 
${ Y^3 \times S^1}$.  The three manifolds may be  chosen to be 
${Y^{3} = \{ T^3, \R^3,  S^3\}}$. Our primary choice is $S^3$ even 
though we will comment on the decompactification limit $\R^3$, and  the flat 
three space $T^3$.  

The reason for choosing $S^3  \times S^1$ is,  as illustrated  in 
\cite{Aharony:2005bq, Aharony:2003sx, Sundborg:1999ue}, 
the ability to observe a confinement 
deconfinement transition in the perturbative regime of the theory 
if $S^1$ is interpreted as a thermal circle.  As we will observe momentarily,  
$S^3 \times S^1$ also admits other center symmetry changing transitions 
which are not associated with abrupt change 
in the free energy density, but vacuum energy 
density.  The corresponding center symmetry should be viewed as spatial.  
 We will observe the existence of both  spatial and temporal 
types of center symmetry changing transitions for 
QCD-like theories. 

Strictly speaking, a  theory at  finite spatial volume and 
at finite $\Nc$,  with a finite number of  
local  fields cannot have a phase transition and 
 spontaneous symmetry breaking, but just rapid crossovers.
The thermodynamic limit in the theories of interest   is attained in the 
$\Nc \rightarrow  \infty $ limit, and phase transitions at  
$\Nc = \infty $ are  perfectly sensible \cite{Gross:1980he}.  In the 
$\Nc = \infty $ limit, the radii of the $S^3 \times S^1$ space take the roles 
of external tunable parameters.    
The size of 
the  $S^3 \times S^1$ also tells us  whether the theory is strongly or weakly 
coupled due to asymptotic freedom. In the regime where $\min(R_{S^3}, R_{S^1}) \geq  \Lambda^{-1}$, the 
theory    is strongly coupled and we lack  analytical tools to establish 
the details of the dynamics and phase transitions. 
However, by benefiting from the existing 
lattice result in the strongly coupled regime, we will be able to infer 
the qualitative structure of the   phase diagrams. 

\subsection{Thermal and twisted partition functions}  
\label{sec:gen1}
We wish to analyze the phase diagrams of   the QCD-like theories (the $n_f$ 
flavor  QCD(AS/S/BF) and   QCD(adj)) 
 on 
$S^3 \times S^1$ in the $\Nc= \infty$ limit.  To do so, we need to 
write the partition function of the theory on $S^3\times S^1$.
The spin structure on $S^3$ is fixed. 
Depending on the choice of the boundary conditions of 
fermions on $S^1$ circle (antiperiodic or  periodic), the partition 
function  will either be the usual thermal ensemble $Z= \tr e^{- \beta H}$ 
or   ``a twisted partition function'' 
${\widetilde Z}= \tr e^{- \beta H} (-1)^F$. The operator  $F$ is the 
fermion number operator. 
 In particular, 
the twisted partition function $\widetilde Z$ can be used to 
probe phase transitions, for 
example, the ones associated with spatial center symmetry, or chiral symmetry. 
Since it depends on the phase of the theory, it is not an index. 
\footnote{The twisted partition function is conceptually more profound than 
supersymmetric index.
It is a useful tool to study the phases of  all of the 
QCD-like theories, not only the supersymmetric ones. In fact, among all the  
QCD-like theories examined in this paper, there is only one whose underlying 
Lagrangian is supersymmetric. This 
is QCD(adj) with $n_f=1$  or   $\None$ SYM.  
 Even for 
the   supersymmetric theory,   
the  $\tilde Z = 
\tr \left[e^{- \beta H} (-1)^F\right]$ is not generally a supersymmetric 
 index on curved backgrounds.
 Formulating $\N=1$ SYM theory on  a curved manifold breaks the supersymmetry.  
(Because of the absence of the  covariantly constant spinors on curved spaces, 
 one can not  define global supersymmetry.) At a more mundane level, 
 we will explicitly see in sections 
 \ref{sec:SYM1}, and  \ref{sec:SYM4} that $\None$ SYM
on $S^3 \times ({\rm spatial} \; S^1)$  has at least two phases: 
one with unique vacuum and the other with $\Nc$ vacua, therefore $\tilde Z $ 
depends on the phase, which means it can not be an index. 
 If the three manifold $S^3$  
is replaced by a flat 
space such as $T^3$ or $\R^3$ where global supersymmetry is restored,  
then the 
 $\tr \left[e^{- \beta H} (-1)^F\right]$ 
 corresponds to a supersymmetric  Witten index \cite{Witten:1982df}. 
Therefore,  in the large radius 
limit i.e.,  $R_{S^3} \Lambda \gg 1$  where we can approximate $S^3$ with 
$\R^3$,  
the leading  large $\Nc$ behaviour of $\tr e^{- \beta H} (-1)^F$ 
may be viewed as an   index. The main point is  $\tilde Z$ reduces to 
a supersymmetric index only under extremely special circumstances.
}
It is useful to express the formulae 
for these partition functions to understand them better.  Let ${\cal B}$ and 
${\cal F}$  be the  bosonic and fermionic subsectors of the physical 
Hilbert space graded according 
to the fermion number operator.   Physical states  in  
${\cal B}$ and ${\cal F}$ will satisfy  
\begin{equation}
(-1)^F |B \rangle =  |B\rangle, \qquad (-1)^F |F\rangle =  - |F\rangle 
\end{equation}
grading.
Therefore, the explicit form of the thermal 
partition function and twisted partition 
function are  
\begin{eqnarray}
&&Z= \tr e^{- \beta H} = \sum_{\cal B} e^{-\beta E_n } +  
\sum_{\cal F} e^{-\beta E_n } =   \int dE \; [ \rho_B(E) + \rho_F(E) ] \;
e^{-\beta E} = Z_B + Z_F \cr
&&{\widetilde Z}= \tr e^{- \beta H} (-1)^F= 
\sum_{\cal B} e^{-\beta E_n }  - 
\sum_{\cal F} e^{-\beta E_n } =
\int dE \; [ \rho_B(E) - \rho_F(E) ] 
\; e^{-\beta E} =  Z_B - Z_F \qquad \qquad
\label{eq:twisted}
\end{eqnarray}
where $\rho_B(E)$ and $\rho_F(E)$ are bosonic and fermionic 
density of states. $Z_B$  and $Z_F$ 
are defined as the contribution to partition function from the bosonic 
${\cal B}$  and 
fermionic ${\cal F}$ subsectors of the physical 
 Hilbert space, respectively.  Referring to 
$\widetilde Z$ as twisted partition function 
is justified since the sum over all 
fermionic states contributes with a  minus sign as opposed to the 
usual partition function where the  sum is over all states without alternating, 
 hence the moniker  ``twisted''.  
Throughout this paper, we will benefit from both partition functions 
to establish phase diagrams of the QCD-like 
theories. In particular, the free energy 
can be derived from the thermal partition function 
\begin{equation} 
{\cal F}= -  \frac{1}{\beta V_{S^3}} \log  Z 
\end{equation}
where $\beta$ is  inverse temperature, the radius of the 
thermal $S^1$. Similarly,   
the  vacuum energy density can be extracted from the 
twisted partition function
\begin{equation} 
{\cal E}= - \frac{1}{ R_{S^1}V_{S^3  }} \log  \widetilde Z 
\end{equation}
where the $R_{S^1}$ is just the radius of a spatial $S^1$. 

\subsection{Temporal and spatial center symmetry }
\label{sec:gen2}
The manifold $S^3\times S^1$ has     
the fundamental group $\pi_1(S^3 \times S^1)=  \pi_1( S^1) \equiv \Z$ 
just like  
the $\pi_1(\R^3 \times S^1) \equiv \Z $.  
The   existence of compact directions for which 
the fundamental group is nontrivial generates  a global symmetry 
called center symmetry, $\G$.  
The center symmetry is the invariance of the gauge 
theory under gauge transformations which are only periodic up to an element of 
the center of the gauge group.  Since both $\R^3$  and $S^3$ are 
simply connected, the center symmetry realization 
 will be determined by the  Wilson line 
winding around the $S^1$ circle.   

Whether this center symmetry should be 
regarded as spatial or temporal depends on the spin structure on the $S^1$ 
circle. For the antiperiodic boundary condition for fermions 
around the $S^1$ circle, the 
center symmetry is naturally interpreted as the  temporal center symmetry 
$\G_t$.  The corresponding QCD-like  theory on $S^3 \times S^1$ may be 
regarded as a finite temperature field theory on $S^3$. In the 
decompactification limit of the $S^3$, we obtain the theory on 
$\R^3 \times S^1$, a thermal field theory on $\R^3$. 

If the fermions are endowed with periodic boundary conditions on the $S^1$ 
circle, then the theory on $S^3 \times S^1$ does not have any thermal 
interpretation at finite $S^3$.  
In the limit  where $S^3$  decompactifies to  $\R^3$,  
the theory becomes formulated on $\R^3 \times S^1$.  Here, 
$S^1$ is a compact spatial dimension unlike the thermal case where it is 
viewed as a temporal (or thermal) dimension.  The QCD-like theory on 
$\R^3 \times S^1$ may be viewed as a zero temperature field theory
where the decompactified thermal circle is viewed as one of the infinite 
dimensions in $\R^3$. Therefore, the 
 center symmetry along the $S^1$ circle is a spatial center symmetry, which we 
denote by $\G_s$. 
 
On $\R^3 \times S^1$ where $S^1$ is temporal, 
   the behaviour of thermal Wilson-Polyakov loops  forms a criteria
for  deconfinement  and  is  related  to  the  static  quark-antiquark
potential.  If the $S^1$ circle is spatial, the  spacelike Wilson  loops, 
even though  they can be  used as 
order  parameters   of  the  (spatial) center  symmetry,  do   not  constitute  
a deconfinement criteria.  They merely  measure the correlations  of the
spatial gauge fields  \cite{Gross:1980br}.

Lattice formulations of the vectorlike gauge theories usually 
deal with a discretized four torus $T^4$ which is not simply connected, 
and for which the fundamental group is $\pi_1( T^3 \times S^1) 
\equiv (\Z)^4$. This means on $T^4$,  
the topology 
permits noncontractible Wilson lines winding around each direction separately.
 In this case,  more elaborate 
breaking patterns probing  $(\G_s)^3 \times \G_t$ symmetry realizations 
should occur as in  pure Yang-Mills theory on $T^4$ 
\cite{Narayanan:2005en, Narayanan:2003fc }. 
\footnote{For the  pure gauge theory on $T^4$ or $S^3 \times S^1$, 
the distinction between  $\G_s$ and $\G_t$ disappears due to the absence of 
fermions.}
In this regard, 
$S^3  \times S^1$ topology seems to capture an important 
ingredient of infinite volume theory on   $\R^3 \times S^1$, i.e, the 
 only 
order parameter measuring the  center symmetry is  the one winding around the 
$S^1$ circle.  
On the other hand, if the size of the three manifolds $T^3$ and $S^3$ is much 
larger than the   strong confinement scale,  
we may approximate both manifolds as $\R^3$ and the physics of the 
theory should become the one  on $\R^3 \times S^1$. 
 
\subsection{Symmetry and effective action}
Our interest is in $U(N)$ color gauge theories with dynamical fermions in the 
two index representation. The center symmetry of a  gauge theory  is in general 
different from the center of  the gauge group 
(unless the matter is in the adjoint representation), and is determined 
by  the charges of its dynamical fermions under the 
center of the group.  The center for the group $U(N)$ is $U(1)$.    

A dynamical   fundamental representation fermion has charge $+1$ under the 
$U(1)$ center of the group. 
The symmetric and antisymmetric representation fermions, 
which  carry     two upper indices has charge $+2$, and the 
adjoint representation, 
which carries one upper and one lower index has charge $0$ under the center.
 Let $a \in \Z $ be the charge of an external, nondynamical test 
quark. It is 
easy to see that a fundamental fermion  can 
screen any external charge without any cost, 
the AS/S fermion can screen any even charge, and 
convert any odd charge to $+1$, while an adjoint fermion cannot screen any. 
 Therefore, we can define the  equivalence classes for static 
external charges  as 
\begin{equation}
a \cong a+ q_d   
\label{Eq:class1}
\end{equation}
where $q_d$ refers to the charge of the dynamical quark under the center.  
The quotient of the charge lattice  of the static  test quarks to  the 
conjugacy class of the dynamical quarks is essentially the center symmetry 
of  the theory.  The quotient groups are respectively, 
$\Z/(q_d \Z)\cong  \{ 1, \Z_2, \Z \} $ in the presence of the fundamental, 
(anti)symmetric and adjoint representation dynamical quarks.
Therefore, for QCD  with fundamental quarks, center symmetry is absent,
for  QCD(AS/S) it is $\Z_2$ and QCD(adj), it is $U(1)$.
\footnote{The center of the  group $SU(N)$ is $\Z_N$. If $N$ is odd, 
dynamical    quark with $N$-ality $2$ can screen any external test quark. 
Therefore,  no center symmetry remains just like when one introduces 
fundamental quarks.  If $N$ is even, then the odd 
test charges cannot be screened and the center symmetry of the QCD(AS/S) 
theory is just $\Z_2$ as above.}
   
The QCD(BF) is a vectorlike  theory with  gauge group 
$U(N) \times U(N)$ and with bifundamental fermions. The center of the  
group is therefore 
$U(1) \times U(1)$. However, 
the massless bifundamental fermions have charges under the center given by  
\begin{equation}
(\lambda_1)^{i}_{\;\;j}: (1,-1), \qquad (\lambda_2)_{i}^{\;\; j}: (-1,1)  \; .
\end{equation}
This just means the dynamical bifundamental 
 fermions can  screen any test charge whose charge is an  
integer multiple of $(1,-1)$.   Suppose an external test quark has 
doublet of charge $(a, b)$ on the 
center charge lattice $\Z \times \Z$. 
The equivalence  class of $(a, b)$  is defined by 
\begin{equation}
(a,b) \cong (a, b) + (1, -1) 
\label{Eq:class2}
\end{equation}
The charge lattice modulo this congruence represents the charges that cannot 
be screened in the QCD(BF) theory and is given by  
$(\Z \times \Z)/\Z\cong \Z$. 
Therefore, the center symmetry of  QCD(BF)  is $U(1)$, just like 
QCD(adj).   
\footnote{If one consider an $SU(N) \times SU(N)$ quiver gauge theory with 
bifundamental 
fermions,  then the charge lattice becomes  $\Z_N \times \Z_N$.  Since the 
dynamical bifundamental fermion can screen any integer multiple of $(1, -1)$, 
the center symmetry reduce to   $(\Z_N \times \Z_N)/\Z_N\cong \Z_N$.}

The form of the effective action is dictated by the symmetries of the 
fundamental  theory. Since center symmetry is a symmetry of the 
original QCD-like theories, 
it has to be a symmetry of our effective theories. 
In the subsequent section, we will explicitly 
construct the one loop effective potentials 
for both temporal  and spatial Wilson lines 
wrapping the $S^1$ circles.  Before doing so, 
we may express what we should expect on symmetry grounds.  
The  effective action  should be 
\begin{equation}
S[U] = \sum_{\cal R} \sum_{n=1}^{\infty}  
a_n( Y^3 \times S^1, {\cal R} ) \;  \tr_{\cal R}U^n
\label{Eq:master}
\end{equation}
The coefficients $a_n( Y^3 \times S^1, {\cal R} )$  
depend on the  detailed structure 
of  the underlying theory, such as the 
topology 
of the three-manifold $S^3$ or $\R^3$,    
the   matter content,  the representation ${\cal R}$  under the color 
gauge group, as well as the Lorentz symmetry group and  
the statistics of particles
(fermions or bosons), 
and  finally on the boundary conditions imposed on the $S^1$ circle,  
periodic or antiperiodic.  Whether the action Eq.\ref{Eq:master} should be 
interpreted  as the effective action for the spatial or temporal  Wilson line 
depends on the spin structure  of fermions on the $S^1$ circle. For the 
periodic (antiperiodic) choice of boundary conditions,  Eq.\ref{Eq:master} is 
the 
effective action  for spatial (temporal) Wilson line. 

The trace  in the effective action  Eq.\ref{Eq:master}  
corresponds to the specific representation of the fields and in fact, makes 
the center symmetries manifest.    
For example, for  one and two index representations 
of the color gauge group, we have 
 \begin{equation}
\tr_{\rm fund.} U = \tr U, \qquad \tr_{\rm adj}U = \tr U \; \tr U^{\dagger}, 
\qquad  
\tr_{\rm S/AS} U = \frac{1}{2}(\tr U \; \tr U \pm \tr U^2).     
\label{Eq:trace}
\end{equation}
As expected, the $U(1)$ and $\Z_2$  
center symmetry invariance, acting as 
$U \rightarrow e^{i \alpha} U$ and $U \rightarrow -  U$
 is manifest for the adjoint and AS/S   representations respectively.   
Introducing  
fundamental representation fermions in the original theory destroy center 
symmetry completely, and this is also manifest in the  effective action.
\footnote{The effective action (in the presence of fundamental fermions in the original theory) does 
not possess center symmetry just like Gross-Witten model 
\cite{Gross:1980he}. Schnitzer 
demonstrates that the theory undergoes a third order phase transition 
\cite{Schnitzer:2004qt, Schnitzer:2006xz}.  
Such large $N$ phase transitions are not  associated with spontaneous 
breaking of any symmetry. Also see \cite{Dumitru:2004gd,Wadia:1980cp}}
  Similarly, 
the trace over the bifundamental representation 
of  a  $U(N)_1\times U(N)_2$ gauge group is   
\begin{equation}
\tr_{\rm BF} U = \tr U_1 \; \tr U_2^{\dagger}
\end{equation}
where the invariance under the  $U(1)$ center symmetry 
$U_i \rightarrow e^{i \alpha} U_i$ is manifest.

The one loop effective potential for the (spatial or temporal) Wilson line 
is just $V_{\rm eff}[U]= \frac{S[U]}{V_{S^3 \times S^1}}$.   In the regime where 
both  $S^1$ and $S^3$ are small and of comparable size,  the short wavelength 
degrees of freedoms can be integrated out perturbatively.
Then effective 
potential   may be conveniently regarded as a $d=0+0$ dimensional unitary 
matrix model. Our analytical calculations in the subsequent section 
will demonstrate the presence (or absence) of   weak coupling 
transitions in the regime where $\beta \sim  R_{S^3}$ for temporal and 
 $R_{S^1} \sim   R_{S^3}$  spatial center symmetry. 

If the $R_{S^3} \gg \Lambda^{-1}$ while $R_{S^1} \; ({\rm or} \; \beta) \ll 
\Lambda^{-1}$, 
then one can perturbatively 
integrate out the heavy modes in the theory. These  are 
the Kaluza-Klein  modes 
along the $S^1$  circle. 
 If $S^1$ is thermal, then this theory  may be regarded 
as a canonical  textbook  example of thermal field theory formulated on 
$\R^3 \times S^1$.  The theory at the scale of the small radius (which 
corresponds to high temperature) is perturbative. On the other 
hand, the long distance physics (at scales much larger than compactification 
radius $\beta$) is intrinsically nonperturbative, and understanding its 
dynamics requires  relevant  effective field theories.  We will not  
pursue such  a goal here. On the other hand,  we will 
explicitly demonstrate that  the  
well-known  effective thermal one loop potentials  for Wilson 
lines  on $\R^3 \times S^1$ can be recovered  by taking the appropriate 
decompactification limit of $S^3$. The  
    effective one loop potential obtained in this 
way is   $d=3+0$ dimensional, and depends 
on $U({\bf y})$ where ${\bf y} \in \R^3$ is a  slowly varying function 
of coordinates, and just coincides with the thermal one loop potential in 
\cite{Gross:1980br}. The one loop effective potential in 
$\R^3 \times S^1 $ can only tell us the realization of the center symmetry, 
whether it is broken or not. But, unlike the perturbative case above, it can 
not be used to demonstrate the phase transition as the temperature is lowered. 
In such  cases, we rely on existing lattice results.  


\subsection{Chiral anomaly at large $N$}

In the limit where the size of the base space is sufficiently 
large, $\min(R_{S^3}, R_{S^1}) \gg \Lambda^{-1}$, we may regard the theory as 
it is on $\R^4$ to a good approximation.   It is important to 
 recall some basics of the axial chiral anomaly relation for two index 
fermions.  
One feature of  double index representation fermions which is different 
from fundamental fermions on $\R^4$, 
is that the anomaly does not vanish in the infinite number of color limit. 
The anomaly relation is \cite{'tHooft:1976up} (see \cite{Terning:2003th} 
for a review) 
\begin{equation} 
  \partial^{\mu} J^{5}_{\mu} = \frac { g^2 n_f}{16 \pi^2}\;  \tr_{\cal R} 
F_{\mu \nu}  \tilde   F^{\mu \nu} =   \frac { g^2 h  n_f}{16 \pi^2}\;  
\left\{ 
\begin{array} {ll}
\tr F_{\mu \nu}  \tilde   F^{\mu \nu} & \qquad {\rm QCD(adj/AS/S)} \\
\tr   F_{1}  \tilde   F_{1} + 
\tr   F_{2}  \tilde   F_{2} & \qquad   {\rm QCD(BF)}  
\end{array} 
\right.
\label{Eq:anomaly}
\end{equation}
where  the  $\tr_{\cal R}$ in the first trace denotes the 
representation of the  fermions forming  the chiral current  
$J^{5}_{\mu}$. For $U(N)$ gauge group,    
$\tr_{\cal R} F \widetilde F =   h \; \tr  F \widetilde F $ where in the latter 
the trace is over defining representation. Here, 
$$
 h= \{ N, N,  N-2, N+2 \}, \qquad  {\rm \;for\; QCD(adj),\; QCD(BF), \; 
QCD(AS/S)} 
$$
respectively, and the fact that it is proportional to $N$ is the reason why 
the anomaly is not suppressed in the large $N$ limit for two index 
representation fermions.  
For fundamental fermions,  $h=1$  and therefore, in  't Hooft's 
large $N$ limit where $g^2N$ = fixed,   the coefficient of the 
right  hand side is $\O(n_f/N)$,  and anomaly vanishes \cite{Witten:1979vv} 
unless $n_f$ scales with $N$. 
For two index fermions,  there is no suppression and the classical axial 
$U(1)_A$ symmetry at the quantum level  is $\Z_{2hn_f}$.  
Note that by taking  the number of fundamental representation 
flavors $n_f$ scale with $N$ by keeping
  $n_f/N $fixed as $N
\rightarrow \infty$, one can also keep the fundamental quarks ``alive'' in 
the $N= \infty$ limit \cite{Veneziano:1976wm}, as well as the anomaly.  
But we will not pursue it here.    

The discrete symmetry $\Z_{2hn_f}$ is  determined  by finding the 
number of   fermionic
zero modes in a one instanton background for the given gauge group and 
representation ${\cal R}$ of fermions. 
$2h$ is the number of fermionic zero modes for each flavor, 
in the representation $\cal R$.  Therefore, 
 the simplest
nonvanishing fermionic correlator  in a one instanton background  has to have 
$2hn_f$ fermionic insertions, so that the Grassmann integral over the zero 
modes will not trivially be zero. 
Let us call the  operator with $2hn_f$ fermionic insertions  $\O$. 
Under a  generic 
$U(1)_A$ transformation, $\O \rightarrow  e^{i\alpha 2N_cn_f} \O$.  The
nonvanishing of expectation value of $\O$ requires   
$e^{i\alpha 2N_cn_f}= e^{2 \pi i k}$, where $k$
  is an integer. Hence the phase $\alpha$ must take values in discrete group 
  $\mathbb Z_{2hn_f}$. This implies, for the quantum theory 
the discrete axial symmetry is  $\mathbb Z_{2hn_f}$.

\section{The $\None$ SYM  on $S^3 \times S^1$}
\label{sec:SYM}

\begin{FIGURE}[ht]
{
  \parbox[c]{\textwidth}
  {
  \vspace*{-30pt}
  \begin{center}
  \psfrag{inf}{$\infty$}
  \raisebox{3pt}{\includegraphics[width=3.0in]{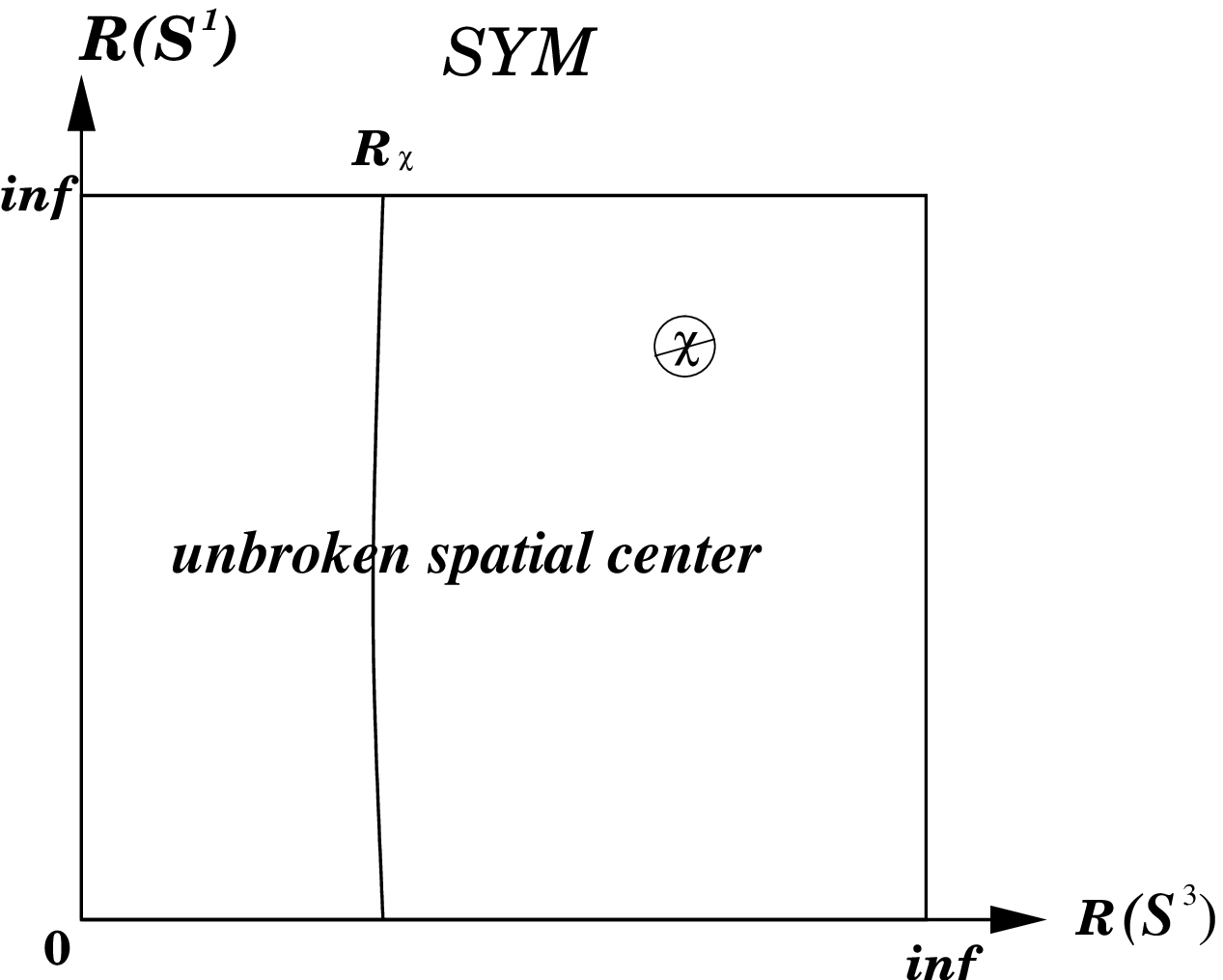}}
  \includegraphics[width=3.0in]{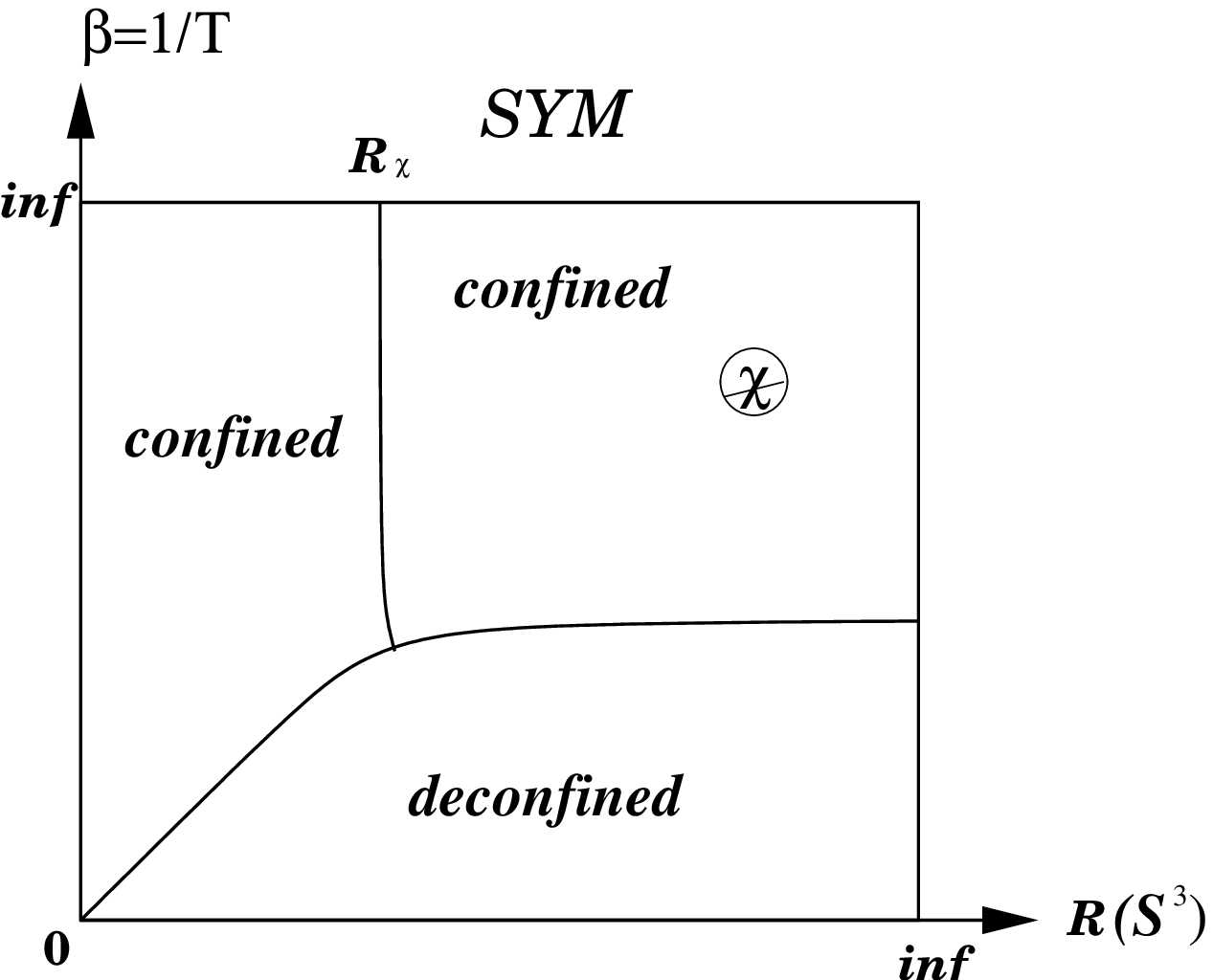}
  \vspace*{-15pt}
  \caption
    {%
    Schematic phase diagrams of $U(\Nc)$ $\None$ SYM 
     as a function of the radii $R_{S^1} (\beta) $ and $R_{S^3}$. 
    The left figure is the nonthermal case where  fermions obey
     periodic boundary conditions along $S^1$, and the 
     right one is the thermal case corresponding to 
     anti-periodic  boundary conditions for fermions.  
    Thermal 
    $\None$ super-Yang-Mills has two confining low temperature phases
    in which temporal center symmetry is unbroken. The 
    discrete chiral symmetry is spontaneously broken at sufficiently large 
    radius $R_{S^3}> R_{\chi}$ while  small radius  $R_{S^3}< R_{\chi}$ shows 
    confinement without chiral symmetry breaking.     
    The theory also  possesses a    
    deconfined high temperature phase with unbroken chiral symmetry.  
    The nonthermal compactification respects spatial center symmetry throughout 
    the whole phase diagram.  Regardless of   the size of the $S^1$ circle, 
    for   sufficiently small radius, $R_{S^3} < R_{\chi}$,  the theory is in 
    a chirally symmetric phase. 
    For sufficiently large radius, $R_{S^3} > R_\chi$, the 
    discrete chiral symmetry is believed to be spontaneously broken.
    These sketches depict the simplest scenario in which only
    a single phase transition separates various  phases;
    more complicated scenarios with distinct deconfinement,
    chiral restoration transitions are also possible.
    }
  \end{center}
  }
\label{fig:phase}
}
\end{FIGURE}

\subsection{Phases as a function of volume}
\label{sec:SYM1}
To study the theory as a function of the volume of $S^3 \times S^1$
(not temperature),  we use  the twisted partition function 
$\widetilde Z= \tr e^{-\beta H} (-1)^F$.   In the Euclidean functional integral,
this means employing   periodic boundary conditions for fermions on the $S^1$ 
circle. Notice that this boundary condition is    supersymmetry preserving 
in the decompactification limit $\R^3 \times S^1$. 
If either of  the radii 
$R_{S^3}$ or $R_{S^1}$   or both is much smaller than the strong confinement 
scale, the theory is amenable to a perturbative one loop analysis. 
In particular, we are searching a transition in the regime where 
$R_{S^1} \sim  R_{S^3}$, where perturbative analysis can demonstrate 
its presence or absence  \cite{Aharony:2005bq}. 
By working in the background of  a constant  gauge field configuration $U$,
the  group valued Wilson line  $U \equiv e^{i \int A_0}$, 
we evaluate the one loop effective potential 
following \cite{Sundborg:1999ue,  Aharony:2003sx}.      
The twisted partition function of the theory on $S^3 \times S^1$  and the one 
loop effective potential is given in 
terms of the spatial Wilson line as
\begin{equation}
\widetilde Z (x) =  \int \; dU \; \exp{-S[x, U]},  \qquad 
S[x, U]= \sum_{n=1}^{\infty}  \frac{1}{n} \left\{(- z_V(x^n) + z_f(x^n))   
\left( \tr(U^n) \tr (U^{\dagger n}) 
\right)    \right\}
\label{Eq:effaction}
\end{equation}
where $x\equiv e^{-R_{S^1} / R_{S^3}}$,  $dU$ is the Haar measure and 
$S[U]$ is the effective one loop action for the spatial Wilson line. 
The functional form of the action $\left( \tr(U^n) \tr (U^{\dagger n}) 
\right)$   is dictated by the symmetries of the original theory, whereas the 
coefficients $(- z_V(x^n) + z_f(x^n))$ are due to  matter content and 
the topology of  underlying  manifold. 
Recall that the $\None$ SYM on $S^3 \times  S^1$ has a global 
$U(1)$ center symmetry associated with the Wilson lines in the $S^1$ direction 
 and this is manifest in the Eq.\ref{Eq:effaction} as the 
invariance $U \rightarrow e^{i \alpha} U$.   The coefficients for vectors and 
spinors are given by (the so called ``single particle partition function'')
\begin{equation}
z_V(x)= \frac{6x^2 - 2x^3}{(1-x)^3},   \qquad  z_f(x) =  
\frac{4 x^{3/2}}{(1-x)^3} \; .   
\label{Eq:singlepartition}
\end{equation}
It is convenient to express the Haar measure in the eigenvalue basis and 
combine the Jacobian  with the one loop effective action. 
On symmetry grounds,  
the Jacobian has to be 
a linear combination of center symmetry invariant double-trace 
operators and should not depend on the volume 
of the manifold $S^3 \times S^1$, i.e 
 $\sum_n c_n \tr(U^n) \tr (U^{\dagger n})$ where $c_n$ is a pure number.  
The Haar measure is 
\begin{equation}
\int \; dU= \int \prod_{i} dv_i J[v_i]=  \int \prod_{i} dv_i \prod_{i<j} \sin^2 \left[\frac{v_i -v_j}{2}
\right] 
=  \int \prod_{i} dv_i \; \exp\{ \sum_{i \neq j}
\log | \sin \left[\frac{v_i -v_j}{2}  \right] | \}
\label{Eq:Jacobian1}
\end{equation}
Using $\log|2\sin \frac{x}{2}| = -\sum_{n=1}^{\infty} \frac{1}{n} \cos nx$, 
the Van der Monde determinant (or Jacobian) may be rewritten as 
\begin{equation}
\ln J[U]= -\sum_{n=1}^{\infty} \sum_{i,j}  \frac{1}{n} \cos[n(v_i -v_j)] =  
- \sum_{n=1}^{\infty} \frac{1}{n} |\tr U^{n}|^2
\label{Eq:Jacobian2}
\end{equation} 
The Jacobian, in either form Eq.\ref{Eq:Jacobian1} or Eq.\ref{Eq:Jacobian2}, can be regarded as a potential among eigenvalues and  
and in effect,  provides a repulsive interaction among them. 

At this stage, we may express the twisted partition function in the eigenvalue 
basis of the spatial Wilson line as 
\begin{equation}
\widetilde Z(x) = \int \prod_i dv_i \; \;  e^{\ln J[U] -S[x,U]}
\end{equation}
where $\prod_i dv_i$ is the flat measure over the compact eigenvalues $v_i$. 
The Jacobian of the measure can be absorbed into the definition of the
 action and    
we may define the ``effective action'' with a flat measure.  
This gives 
\begin{equation}
S_{{\rm eff}}[x,U] \equiv  S[x,U] - \ln J[U]=  
\sum_{n=1}^{\infty}  \frac{1}{n} \left\{(1- z_V(x^n) + z_f(x^n))  
 \left( |\tr(U^n)|^2  
\right)    \right\}
\label{Eq:potsusy2}
\end{equation}
$S_{{\rm eff}}[x, U]$ governs   the dynamics of  the eigenvalues,
hence center symmetry realizations.
For any value of $x\in [0, 1)$, we have the inequality  (or positivity 
in the second form) 
\begin{equation}
(1- z_V(x^n) + z_f(x^n)) \geq 1   \qquad  {\rm or}  \;  
-z_V(x^n) + z_f(x^n) \geq 0 \qquad  {\rm for}  \;  x \in [0,1) 
\label{Eq:inequality1}
\end{equation}
and 
 moreover, $(1- z_V(x^n) + z_f(x^n))$ is a
monotonically increasing function of $x$.  This means, the effective 
action $S_{\rm eff}$ is always larger  than the Jacobian contribution to the 
action:
\begin{equation}  
S_{\rm eff}[x, U] \geq  - \ln J[U], \qquad  {\rm for}  \;  x \in [0,1) 
\label{Eq:inequality2}
\end{equation}
The  net combined effect of the  action  $S_{\rm eff}[x, U]$ is to provide 
a repulsive interaction among eigenvalues, and 
to force them to be maximally apart from one another. The minima 
of the action is located at 
$ U= e^{i \alpha} {\rm Diag}( 1, e^{i2 \pi/N}, \ldots  e^{i2 \pi (N-1)/N} )$ 
where $\alpha$ is some phase, which arises due to the fact that gauge group is 
$U(N)$ rather than $SU(N)$. 
Consequently, the Wilson line eigenvalues distribute uniformly over the unit 
circle, the expectation values of Wilson line 
\begin{equation} 
\langle \tr \;U \rangle= 0 \; . 
\label{Eq:minimum}
\end{equation}
and the spatial center symmetry  $\G_s$ is unbroken.
Therefore, there is no phase transitions 
associated with spatial center symmetry realizations around 
$R_{S^1} \sim R_{S^3} \ll \Lambda^{-1} $ where our analysis is reliable.
This is 
unlike the thermal case which has a deconfinement transition  associated with 
the change in the temporal center symmetry realization $\G_t$ around  
$\beta \sim R_{S^3}$, and two distinct phases at weak coupling. 
The $N$ scaling of the ``vacuum energy density'', or the 
expectation value $\langle S_{\rm eff} \rangle$ 
obtained by using the functional integral  Eq.\ref{Eq:effaction}, 
 is given by 
\begin{equation}
 {\cal E}^{\rm SYM}  = N^2( 0 + \O(1/N^2))= \O(1)
\label{Eq:vacSYM}
\end{equation}
where the leading zero piece is the classical,  
and the subleading is due to the fluctuations. A relevant analysis 
in  the  decompactification limit  
 $\R^3 \times S^1$   
where $S^1$ is  small in units of the strong scale
  demonstrates that 
the spatial center symmetry remains unbroken.   We also expect, on  
$({\rm large}\; S^1) \times ({\rm small} \; S^3)$  unbroken center symmetry.

The inequality Eq.\ref{Eq:inequality2} is a consequence of the 
matter content of 
$\None$ SYM and  the choice of periodic boundary conditions for fermions. 
In particular, for pure YM theory, there is no such constraint. In fact,  
for pure YM, $1-z_V(x)$ is monotonically decreasing function of $x$ and 
it becomes negative at some critical $x_c$ leading to a change in the symmetry 
realization, and confinement deconfinement  transition.
In the case at hand, 
the fermionic contribution overwhelms the vector boson contribution 
and leads  the the  positivity  Eq.\ref{Eq:inequality2}.  
As  we will see, there
is no analogous inequality for the thermal compactification of $\None$ SYM,  
or for the other thermal ensembles  that we will examine such as QCD(AS/S/BF). 
In fact, all such 
thermal ensembles  show characteristics similar to pure YM and possess 
a deconfinement  transition. 

However, even for the QCD-like theories  in which the fermions satisfy 
 periodic boundary conditions on $S^1$ 
(for example the nonsupersymmetric 
theories  related to $\None$ SYM via orbifold/orientifold 
projections) such as QCD(AS/S/BF), the inequality fails to hold and there are 
phase transitions associated with spatial center symmetry breaking. We will see 
such examples.  The last statement seems to favor   supersymmetric 
theories. This is not so. In particular, adding  multiflavor adjoint Majorana 
fermions into SYM (hence a nonsupersymmetric QCD with  $n_f$ 
adjoint representation fermions)   also satisfies the inequality (which now 
reads $(1- z_V(x^n) + n_f z_f(x^n)) \geq 1$)   
and does  not undergo a spatial center symmetry changing transition. 

Of course, the most interesting (and hardest) question at this stage 
is what happens 
as we make the volume of $S^3$ in $\R \times S^3$ and   $S^1$ in 
$\R^3  \times 
S^1$ larger?   With either of these two maneuvers, because of 
asymptotic freedom, the theory goes from a  weakly coupled regime 
 to  a strongly coupled regime and there is no  known 
controlled  approximation.
Below, we will argue that, in the sense of spatial center symmetry 
realizations, nothing interesting happens. The theory is always in the spatial 
center symmetric phase.  However, there are changes in chiral symmetry 
realization 
dependent on the size of  $S^3$, but not $S^1$. 
We will come back to this point in the next section.  

The theory in the  decompactification limit 
$\R^3 \times S^1$  is relatively simpler because of the 
restoration  of supersymmetry on flat space (zero curvature)  limit.  
(Recall that our boundary 
condition for fermion is periodic, hence supersymmetry preserving.)
In the supersymmetric compactification of the $\None$ SYM on 
$\R^3 \times S^1$, the one loop effective potential vanishes identically . 
This statement is true to all orders in perturbation theory due to 
supersymmetry.   However, an instanton induced 
 nonperturbative superpotential do get generated and 
dictates the vacuum structure of the theory. This regime 
is analysed in the literature in depth \cite{Davies:1999uw, Davies:2000nw}.  
It is  shown   that on $\R^3 \times S^1$, 
 the configuration 
Eq.\ref{Eq:minimum} is the minimum of the nonperturbative 
effective potential for $\None$ SYM.  
The most important consequence of the nonperturbative superpotential 
 is unbroken spatial center symmetry  $\G_s$  on  small $S^1$ circle in the 
$\R^3 \times S^1$ limit.  It is strongly believed  
that the spatial center symmetry $\G_s$ is unbroken at  large radius of 
the  $S^1$ circle as well. 
 The absence of a spatial center symmetry changing transition 
on $S^3 \times S^1$,  as well as in 
$\R^3 \times S^1$ is in favor of the conjecture made in the introduction.
\footnote{
Unbroken spatial center symmetry on $\R^3 \times S^1$ has important 
implications which may help us to gain greater  
understanding of  the large $N$ limit of 
vectorlike gauge theories.  The volume independence 
in the large $N$ gauge theories (generalization
 of the old Eguchi-Kawai reduction) 
can be formulated as an orbifold equivalence. This equivalence is valid for 
QCD(adj) on $\R^3 \times S^1$ so 
long as the spatial center symmetry is unbroken. 
Therefore, in large $N$ limit, small volume QCD(adj) is equivalent to large 
volume theory.  More precisely, the volume expansion and reduction may be 
formulated as orbifold projections for spatial dimensions for which the 
first homotopy group is nontrivial, and its validity relies 
on unbroken center symmetry.
This 
implies the theory on $S^3 \times S^1$ should be 
 independent of $S^1$ radius, but not 
$S^3$.   For a fuller discussion, see  \cite{Kovtun:2007py}.} 

\subsection{Finite temperature phases }
\label{sec:SYM2}
Choosing antiperiodic boundary conditions for fermions in the Euclidean 
functional integral  is equivalent to a thermal $\None$ SYM theory on $S^3$.  
The partition function is the usual thermal ensemble $Z= \tr e^{-\beta H}$. 
At finite temperature,  the only change in Eq.\ref{Eq:potsusy2} is the 
coefficient of the fermionic  contribution in  the expansion.  
The action for the temporal Wilson line is 
\begin{equation}
S_{\rm eff}[x, U]= \sum_{n=1}^{\infty} \frac{1}{n} a_n(x) \; |\tr U^n|^2, \qquad   
a_{n}(x) \equiv 1- z_V(x^n) + (-1)^n z_f(x^n) 
\label{Eq:thermal}
\end{equation}
where the alternating sign  is due to the 
change in the boundary conditions  of fermions. The variable 
$x= e^{-\beta/R_{S^3}}= e^{-1/(TR_{S^3})} $ has  a temperature dependence, 
 unlike the case  of spatial $S^1$ where $x$ is unrelated to temperature.  

In the large volume limit where $\min(R_{S^3},  \beta ) \gg \Lambda^{-1} $,
  the dynamics of 
the theory  should be independent of the  choice of the 
boundary  conditions.  However,  
in small volume, the choice of the boundary condition 
alters the dynamics drastically. 
With this  modification of the boundary conditions for fermions,
$a_{2n+1}(x)$ become a monotonically decreasing function   of $x$ which may 
become negative,  whereas 
$a_{2n}(x)$ is still monotonically increasing due to alternating $(-1)^n$ 
factor in Eq.\ref{Eq:thermal}. The function  
$a_1(x)$  is the decisive element in the 
phase  transition. In some loose sense,  $a_1(x)$ may be interpreted as
the mass of the simplest temporal Wilson line $\tr U$.   
(Recall that 
$x \rightarrow 0$ is low  temperature  $\beta \rightarrow \infty$, 
 and $x \rightarrow 1$ is the high temperature $\beta \rightarrow 0$ limit.) 

As  the temperature is increased,   $a_1(x)$ becomes zero at  some critical 
$x_c$ and there is no action (energy)  cost to have a nonvanishing expectation 
value of the temporal Wilson line (or thermal Polyakov loop) $\tr U$. 
In the regime   where $x >x_c$ 
$a_1(x)$ is negative and  the action is minimized with at  $\langle 
\tr U \rangle \neq 0$. 
Therefore, the $U(1)$  center symmetry  is spontaneously broken and the theory 
is  in a deconfined plasma phase with a characteristic  free energy density of 
$\O(N^2)$. 
 At low temperature $(x< x_c)$,  
all the  coefficient $a_{n}(x)$ are positive. The action provides a repulsion among the eigenvalues of the Wilson line,  and hence, eigenvalues are  uniformly 
distributed. Therefore,  $\langle \tr U \rangle = 0$.  
This implies an unbroken 
center symmetry and,   the theory is in the confined phase. 
The leading $\Nc^2$  order term in free energy vanishes identically, 
reflecting the $N$ independence of the number of physical gauge invariant 
states of the theory. The resulting free energy is  just $\O(1)$, and it 
is due to fluctuations. (In other words, the spectral density of the color 
singlets remains $\O(1)$ in confined phase. 
This means,  as a function of temperature, the 
 theory undergoes a   confinement deconfinement transition at some temperature 
of order $T \sim 1/R_{S^3}$. The vicinity of the $x\sim x_c$ in the case of pure 
YM theory is  examined in detail in   \cite{Aharony:2003sx}.  

It is also instructive to see how the effective action on $S^3 \times S^1$ 
reproduce the well-known perturbative thermal field theory result in 
$\R^3 \times S^1$ limit \cite{Gross:1980br}.
Since temperature enters into our partition function in 
combination  $\frac{\beta}{R_{S^3}}$, naively  
the high temperature is same as large 
$S^3$. This is true for some   conformal theories such as ${\cal N}=4$ SYM. 
It is, however, a false statement for confining  gauge theories. 
There is in fact 
another scale in our problem, the strong confinement scale $\Lambda$.
In order to have perturbative access to the theory on large $S^3$,  
the temperature must be much  larger 
than the strong confinement  scale, i.e., $T \gg \Lambda$. 
More precisely, $\beta \ll \min (R_{S^3}, \Lambda^{-1})$ has to hold in order 
to derive  thermal one loop potential in   $\R^3 \times S^1$
\cite{Gross:1980br}.  

Let  $\epsilon\equiv  \frac{\beta}{R_{S^3}} \ll 1$. 
The leading results for single particle partition functions
 are: $z_V(x^n)= \frac {4}{\epsilon^3n^3} +  \frac {18}{\epsilon^2n^2}$ and 
  $z_f(x^n)= \frac {4}{\epsilon^3n^3} -  \frac {6}{\epsilon^2n^2}$, where the 
subleading terms vanish in the evaluation  of potential in the 
$\R^3$ limit.  
The one loop effective potential for thermal SYM  
$V^{\rm SYM}_{\rm eff}[U] =  
\frac{S_{\rm eff}}{\beta V_{S^3}}$ can easily be extracted from 
Eq.\ref{Eq:thermal} 
as
\begin{equation}
V^{\rm SYM}_{\rm eff}[U] =  \sum_{n=1}^{\infty} 
 \Big[ \frac{1}{n \beta V_{S^3}} -  \frac{2 T^4 }{\pi^2}\frac{1}{n^4} (1-(-1)^n) \Big]
|\tr U^n|^2 
\label{Eq:thermalR3}
\end{equation} 
The second term is the standard   
thermal one loop potential for SYM theory on  $\R^3$.
The first term, due to Jacobian, 
provides an  eigenvalue repulsion, but is identically zero 
 in the infinite volume $\R^3$ limit, and  is kept for later convenience. 
 The potential  is minimized when all the eigenvalues coincide, 
 i.e, $U= e^{iv_0}1$ which corresponds to a Dirac-delta 
 distribution of eigenvalues, $\rho(v) = \delta(v-v_0)$. (The eigenvalue 
distribution $\rho(v)$ is normalized as $\int_{0}^{2 \pi} dv \rho(v)=1$.) 
This configuration  
spontaneously breaks 
the $U(1)$ temporal center symmetry $\G_t$. 
The theory is in the deconfined phase and the 
leading order free energy density is 
\begin{equation}
{\cal F}^{\rm SYM} = 
 -\frac{2N^2 T^4 }{\pi^2} \sum_{n=1}^{\infty} \frac{1}{n^4} (1-(-1)^n) ) =  
-  \frac{2N^2 T^4 }{\pi^2} \frac{\pi^4}{90}(1+ \frac{7}{8})= 
- \frac{\pi^2}{24} N^2 T^4 
\label{Eq:SB}
\end{equation}
reflecting the characteristic of deconfined phase. This is, as expected, 
the Stefan-Boltzmann result for a free gas of $\Nc^2$ bosonic and fermionic 
degrees of freedom. 

\subsection{Multiwinding loops and  perturbative widening  of 
eigenvalue distributions}
\label{sec:SYM3}
Even though the measure can be ignored in the infinite volume limit 
of $S^3$, it has a non-trivial role at any finite volume.   
Let 
$R_{S^3} \gg \Lambda^{-1}$  and $\beta \ll \Lambda^{-1}$.  The question is, 
how does the eigenvalue distribution (of Wilson line) 
behaves  as  the temperature is lowered? 
This question is interesting   for two reasons. Primarily,  the dynamics of 
eigenvalues  directly determines the center symmetry realizations. Also, 
  lattice gauge theory (by necessity), is formulated on a finite space.
The goal of the following discussion is to point out that the widening 
of the eigenvalue distributions on finite spaces 
has two sources,  perturbative and  
nonperturbative. On $\R^3 \times S^1$, this widening is intrinsically 
nonperturbative. 
 \footnote{Lattice gauge 
theory is traditionally formulated on $T^3 \times S^1$. Even though the 
spaces  $T^3$ and $S^3$  are topologically distinct, 
the local dynamics of strongly coupled gauge theory should be independent of the global topology in the large volume limit. Therefore, as long as the 
characteristic sizes of  $T^3$ and $S^3$ is much larger than the strong 
confinement scale $\Lambda^{-1}$,  one is free to replace the volume of 
three sphere $V_{S^3}$  
with the volume of three torus $V_{T^3}$ in the discussion 
below. In fact,  sufficiently large sphere or torus 
is essentially  on the same footing with $\R^3$ for dynamical considerations.}

In the limit  where  $R_{S^3} ( R_{T^3}) \gg \Lambda^{-1}$  
and $\beta \ll \Lambda^{-1}$
the separation between eigenvalues is much smaller than one,
the problem simplifies further and reduces to a well-know Hermitian Gaussian 
matrix  model which is exactly solvable. 
In this limit,  the effective action becomes
\begin{equation}
S_{\rm eff} (v_i) = - \sum_{i\neq j=1}^{N} \log |v_i - v_j|  + \frac{1}{2} 
\left(\frac{2 \pi^2 N }{\epsilon^3} \right)\sum_{i=1}^{N} v_i^2 
\label{Eq:HMM}
\end{equation}
 where the first term is the  usual repulsive eigenvalue-eigenvalue 
interaction  and the second term provides a 
trapping potential for eigenvalues.

 The trapping potential has  anomalously large coefficient compared to 
the repulsive potential.   
Therefore, Wilson line eigenvalues are spread on a very 
narrow support of width $w$, and their  distribution can be 
solved exactly, 
yielding the Wigner semi-circle distribution. 
\begin{equation}
\rho(v) =  \left\{ \begin{array}{ll} 
\frac{2}{ \pi w^2} \sqrt{ w^2 - v^2} \;\; & {\rm for} \; 
v \in  [-w, w]   \cr
0 \; \; & {\rm for } \; v \in (w, 2 \pi -w)
\end{array} \right.
\label{Eq:distribution}
\end{equation}
The width, in terms of the radii, is given by
\begin{equation}
 w= 
\frac{1}{\pi} \epsilon^{\frac{3}{2}} = \frac{1}{\pi} 
\left(\frac{\beta}{R_{S^3}}\right) ^{\frac{3}{2}} 
\label{Eq:width}
\end{equation}
Notice that the eigenvalue distribution Eq.\ref{Eq:distribution},
  in the limit $\epsilon \rightarrow  0$ is the  
Dirac-delta function $\rho(v) \rightarrow \delta(v)$, as expected.  The 
width of the distribution $w$ starts to widen as we make $\epsilon$ larger with 
an $\epsilon^{\frac{3}{2}}$ scaling.  This widening is perturbative 
and is unrelated to nonperturbative physics of deconfinement on 
$\R^3 \times S^1$.

Let us investigate the temperature  (or width) 
dependence  of the Polyakov loop in this regime. 
The expectation values of a  Wilson/Polyakov loop  with winding number $n$ is 
\begin{equation}
\langle \frac{1}{N} \tr U^n \rangle = \int_{0}^{2 \pi} \; dv  \rho(v) e^{i n v} = 
\frac{2}{w n}  J_1 (w n) = 1 - \frac{n^2  w^2}{8}  +  O(w^4)
\label{Eq:moments}
\end{equation}
In the $\epsilon \rightarrow  0$  limit, 
since the eigenvalues coincide [$\rho(v)  
\rightarrow \delta(v)$], we obtain the expected  result,  
$\langle \frac{1}{N} \tr U^n \rangle =1$ for all $n$.  
The formula Eq.\ref{Eq:moments}  shows some 
interesting features as well depending on the winding number.  
Let $w \ll 1$ be the  fixed width of the eigenvalue distribution. It is then 
natural to split the Wilson/Polyakov loops according to their winding numbers.  
The loops with low winding number $(n \ll 1/w)$ 
 show behaviour similar to single winding loop.  In particular, their
 expectation value if of order one. However, the expectation 
value of loops with high winding number  $n \gg 1/w $  are suppressed,
 and rapidly fluctuates around zero with an amplitude
 bounded by  $\frac{1}{(nw)^{3/2}}$.  In particular, in the scaling limit 
where    $\frac{n}{N}$ is fixed as we take $N \rightarrow \infty$,  
we have $\langle \frac{1}{N} \tr U^n \rangle \rightarrow  0$. 
This is not surprising. As long as there are fluctuations in the eigenvalues, 
the higher order Fourier coefficient will fall off.  Even 
though the eigenvalues of the Wilson/Polyakov line are localized on a very 
narrow support $v_i \in [-w, +w]$, because of the large winding number of the 
corresponding Wilson line (when the number of winding times  the width becomes 
of order one, i.e, $v_in \in [-wn, +wn] \sim [-\pi, \pi]$), 
the  expectation value turns into a sum over 
random phases in the $[0, 2\pi)$ interval. Explicitly, 
$\langle \frac{1}{N} \tr \; U^n \rangle= \langle 
\frac{1}{N} \sum_{i=1}^{N} e^{iv_i n} \rangle$.
By the random phase approximation, the averaged sum of random 
phases on the $[0, 2 \pi]$ interval is  suppressed, 
explaining the smallness of the multiwinding loops. 

Lowering the temperature is equivalent to increasing the width.  On the other 
hand,  inverse width $(1/w)$ draws a line between the Wilson/Polyakov loops 
which  (almost)  vanish and the ones which do not. In particular, for a given  
width $w$,  there are roughly $1/w$ many   loop  classes 
whose expectation values are still of   $\O(1)$.   
When the temperature approaches  the strong scale of the theory (from 
the high temperature side), the region of validity of the  perturbative, and 
narrow support  
approximations break down,  and we cannot  say anything about the regime 
where temperature is in the vicinity of the strong scale. (See, however 
ref.\cite{Pisarski:2006hz} for the recent attempts in this window.) 
The numerical 
lattice  results on the other hand   show that the trend continues in the 
nonperturbative regime, and expectation values of all the 
Polyakov loops vanish in the confined phase restoring the temporal 
center symmetry. 

\subsection{Disentangled chiral  and center symmetry realizations on 
$S^3 \times S^1$} 
\label{sec:SYM4}
There are multiple interesting lessons arising from the consideration of these 
theories on $S^3 \times S^1$.   One  outcome is the 
disentanglement of chiral and center symmetry realization. 
We have shown that 
for SYM theory endowed with 
periodic boundary conditions for fermions 
 there is no change in center 
symmetry realization when one extrapolates from large to small radius of $S^3$. 
The spatial center symmetry $\G_s$ is unbroken throughout 
this volume change. However, 
the situation for chiral symmetry is rather different.  At small (large) 
$R_{S^3}$, 
regardless of  the size of  $S^1$, chiral symmetry is unbroken (broken). 

In order to see the absence of the chiral condensate on sufficiently small 
$S^3$, it suffices to recall that 
the eigenvalue 
spectrum of the free Dirac operator  on a three sphere 
has a gap of order $1/R_{S^3}$ and there are no 
fermionic zero modes. \footnote{This is also the reason why the supersymmetry 
 is 
explicitly broken on curved spaces. Because of the absence of
 covariantly constant spinors, we cannot define global supersymmetries. 
In certain supersymmetric theories admitting twisting, hence spin zero 
fermions,  one can have 
globally defined supercharges. The $\None$ SYM does not 
admit a nontrivial twisting in $d=4$,
 and once carried to curved spaces (or discrete spaces such as lattice), 
does not  preserve supersymmetry.  On the other hand, the $\N=4$  SYM  has 
nontrivial twists with nilpotent scalar supercharge, hence can be formulated 
in curved spacetimes (and discretized spacetimes, i.e, lattice)
 by preserving a subset of supersymmetries 
exactly. The magic in both case is the coordinate independence of the 
scalar supercharges, $Q^2 \cdot =0$. 
The observation about curved spacetimes leads to the topological 
twisting of supersymmetric theories, and 
the latter leads to the supersymmetric lattice  formulation.  
But the essence of the idea, i.e, coordinate independence of scalar 
supersymmetry, is same in both cases.  
For progress in supersymmetric lattices,  see 
 \cite{Kaplan:2005ta,Catterall:2005fd}.}
 The  eigenfrequencies of a spinor on $S^3$  are given by 
\begin{equation}
\omega_n^2= ( n+ \half)^2 \frac{1}{(R_{S^3})^2} , \qquad  n=0, 1, 2, \ldots 
\label{Eq:spectrum}
\end{equation} 
with degeneracies $n(n+1)$ at the level $n$.    There are two  
simple regime determined by the ratio of the amplitude of smallest 
eigenfrequency to the strong confinement scale. 
 Let the radius of $S^3$ be much smaller than $\Lambda^{-1}$.
Since the fermionic modes are heavy, i.e., $\frac{|\omega_0 |} {\Lambda} \gg 1$, in the 
description of the  long distance 
dynamics of the theory  (at length scales much larger than the size of $S^3$), 
they can be  integrated out  perturbatively. 
Therefore, a fermionic condensate cannot form and the theory is in a 
chirally symmetric phase. 

What happens at large radius (or at small curvature), 
$R_{S^3} \gg \Lambda^{-1}$?   In this case,  
since $\frac{|\omega_0 |} {\Lambda} \ll 1$, there are a large number of modes 
below the strong scale, therefore  the theory on $S^3$ (at zeroth order) 
may be regarded as the theory on $\R^3$.   Therefore,   global 
supersymmetry  gets restored, and we can use techniques 
from supersymmetric theory  to infer the chiral properties.  
 At sufficiently small $S^1$, it is possible to calculate 
the chiral condensate reliably,  and  the result can be extended to arbitrary 
size of $S^1$ via holomorphy.  Therefore, the theory on $S^3 \times S^1$ should 
undergo a curvature induced chiral phase transition 
at some radius of $R_{\chi, S^3}$ around the 
strong scale $\Lambda^{-1}$  of the theory.  We have 
    \begin{eqnarray} 
    &&  \langle \tr \lambda \lambda \rangle = 0, \qquad  \langle \tr U \rangle =0 
 \;\;\; {\rm  small}\;\; S^3
  \cr
 && \langle \tr \lambda \lambda \rangle \neq  0,  \qquad  \langle \tr U \rangle =0 
 \;\;\; {\rm  large}\;\; S^3
\label{Eq:phases1}    
\end{eqnarray}
The simplest phase diagram consistent with this knowledge is 
illustrated in fig.\ref{fig:phase}. 
 The theory has two phases. A chirally symmetric phase in which 
 the discrete chiral $\Z_{2N}$ symmetry is unbroken,  and  a chirally asymmetric phase in which $\Z_{2N}$ is spontaneously broken down to $\Z_2$ of $(-1)^F$ 
by the formation of the fermion bilinear condensate.  
We expect the condensate to have a smooth evolution  above a critical 
radius  and 
extrapolate to the condensate of the $\None$ SYM theory in the limit of zero 
curvature. 
The interesting point of this transition is that it is completely disentangled 
from (spatial) center symmetry realizations.  
The spatial center symmetry [with periodic 
boundary conditions for fermions] is unbroken on $S^3 \times S^1$ 
with arbitrary radii.  
 
Notice that the absence of chiral condensate at small $S^3$ is a consequence 
of the gap in the spectrum of the free Dirac operator 
(which is due to the topology of 
$S^3$) combined with the largeness of 
this    gap compared to the strong scale.  In contrast, on a flat 
space  such as $T^3 \times S^1$ or $T^3 \times \R$
(with the periodic boundary conditions  for fermions 
in each direction),  $\None$ SYM  
undergoes  neither a  chiral transition, nor a  center symmetry changing 
transition. Therefore, at even small radii of the three torus, there is 
a chiral condensate.  In fact, in the flat space, the twisted partition 
function $\widetilde Z= \tr [e^{-\beta H} (-1)^F]$ becomes the supersymmetric 
Witten index 
counting the number of supersymmetric vacua \cite{Witten:1982df}. 
In this particular example, the index is  $\widetilde Z= N$.  

It should be noted that the original 
calculation of the index was performed  on a small $T^3$, by performing 
a Born-Oppenheimer approximation. On $\R^4$, however, the theory is strongly 
coupled. The common wisdom of strongly coupled gauge theories tells us 
that a condensate will form in the most attractive channel (MAC) 
\cite{Cornwall:1974hz, Raby:1979my}. 
In this 
case, 
the condensate $\langle \tr \lambda \lambda \rangle$ breaking the discrete chiral 
symmetry $\Z_{2N} \rightarrow \Z_2$, possess a nonzero  phase 
equal to an integer multiple of $ 2\pi/N$. 
  
There is one  more interesting feature of the  chiral transition on 
$S^3 \times S^1$.  
Let
$R_{S^1} \ll \Lambda^{-1}$  and let us  dial the size of $S^3$ from small to 
large.   It is known that  the chiral 
symmetry is broken  on large $S^3$ (which can be approximated as  
$\R^3 $)    where $S^1$ is small \cite{Davies:1999uw}, 
and as demonstrated above, it is unbroken 
on small $S^3$.  
Therefore, it is likely that  the theory undergoes 
 a weak coupling chiral transition.   
The  study of  this regime in more detail should be useful and is left 
for future work. 
The simplest phase  diagram of the nonthermal $\N=1$ 
theory consistent with our current knowledge 
is shown in Fig.\ref{fig:phase}.

At finite temperature, the phase diagram of the $\None$ SYM theory on $S^3$ is 
richer and is shown in Fig.\ref{fig:phase}.   In particular, the use of 
antiperiodic boundary conditions around the thermal $S^1$  circle (irrespective 
of the size of $S^3$) contributes  a tree level thermal mass to  fermions 
(recall that the fermions already have a classical mass gap due to the 
curvature of three-sphere).  Let us consider the case where 
$R_{S^3} \gg \Lambda^{-1}$ 
and  $\beta  \ll \Lambda^{-1}$, hence the curvature effects are negligible.
   In this regime, the thermal tree level masses are given by 
$\omega_n= 2 \pi( n + \half)T$ where  $n = 0, \pm 1, \pm2, \ldots $, and $T$ is 
temperature.   Hence,  
the frequencies are bounded,  in magnitude,  from below by $\pi T$.   
Therefore,   analysing the dynamics of the theory at distances large compared 
to $\beta= T^{-1}$, the fermions can be viewed as a heavy Kaluza-Klein tower of particles 
and can be integrated out 
perturbatively with no formation of any nonperturbative condensate, and 
consequently with no breaking of chiral symmetry. (For a rigorous proof of this 
argument in  
lattice formulations of vectorlike gauge theories, see 
\cite{TomboulisYaffe1,TomboulisYaffe2})
\footnote{
The reader will realize that the argument on the absence of chiral 
condensate at high temperature  is essentially same as the 
one we have  presented above in the case of high curvature space.  It therefore 
seems tempting to draw an analogy between temperature and curvature as they 
induce the same effect on chiral symmetry.   However, this interpretation does 
not go far  because of the center symmetry realizations, which clearly 
distinguishes  curvature and temperature.}  Therefore, in this phase, 
the temporal center symmetry $\G_t$ is broken and the chiral symmetry is 
unbroken. 

In the limit where $R_{S^3} \ll \Lambda^{-1}$, irrespective of the value of the 
temperature,  there cannot be any chiral condensate as discussed above. 
However,   as we discussed in the previous section, there is a change in the 
temporal center symmetry realizations 
when $TR_{S^3} \sim 1$, from a high temperature 
deconfined plasma phase to a low temperature confined phase. 
 
When  $ \min ( R_{S^3}, \beta ) \gg \Lambda^{-1}$, we do not expect any 
dependence on the   boundary conditions.  In this phase, chiral symmetry is 
expected to be spontaneously broken, but not the temporal center symmetry. 
(This conclusion for thermal $\N=1$  SYM is also reached in 
ref.\cite{Aharony:2003sx}.)
 Therefore, there are at least three  phases of the theory, 
   \begin{eqnarray} 
    &&  \langle \tr \lambda \lambda \rangle \neq 0, \qquad  
\langle \tr U \rangle =0 
 \;\;\;  \min(R_{S^3}, \beta) > \Lambda^{-1}  \cr
 && \langle \tr \lambda \lambda \rangle =  0,  \qquad  \langle \tr U \rangle =0 
 \;\;\;  R_{S^3}< \Lambda^{-1} \; {\rm and} \;  \beta > R_{S^3}   \cr 
 && \langle \tr \lambda \lambda
 \rangle =  0,  \qquad  \langle \tr U \rangle \neq 0 
 \;\;\;  \beta < \min ( R_{S^3} ,  \Lambda^{-1}) 
\label{Eq:phases2}    
\end{eqnarray}
The simplest possibility  for the phase diagram consistent with our current 
 knowledge
is shown in  Fig.\ref{fig:phase}.  
Clearly, the  strangest phase is the one in which  neither chiral 
symmetry, nor  the center symmetry is spontaneously broken which implies 
confinement without chiral symmetry breaking.  
 Unfortunately, 
this phase is not visible in lattice simulations which uses a discretized 
torus as base space. On flat torus, the deconfinement/confinement transition 
is usually entangled to the chiral transition \cite{Kogut:1982rt}, at least 
they cannot be parametrically different. 
This is what we have seen so far in lattice simulations. 
The curved background intelligently disentangles the 
two symmetry realizations.   

\section{Phases of orbifold QCD(BF)  theory on $ S^3 \times S^1$ }
\label{sec:orb}
\subsection{Twisted partition function and the phases as a function of 
volume}
\label{sec:orb1}
In this section,  we wish to analyse the phase diagram of the orbifold theory 
on $S^3 \times S^1$. In order to compare with the $\None$ SYM, we analyze the 
orbifold theory  under the same conditions.  First, we consider the case
where the fermions on the $S^1$ circle have periodic boundary conditions.  
As we discussed before, the periodic boundary conditions on the Euclidean 
functional integral correspond to the twisted 
 partition function  $\widetilde Z= \tr e^{-\beta H} (-1)^F$.    The study 
of the twisted 
partition function  will reveal a rich phase structure for the orbifold theory. 

The $n_f=1$ orbifold QCD(BF)  is a  gauge theory with a  product gauge group 
 $U(N) \times U(N)$  with gauge bosons in adjoint representation of each 
group and bifundamental fermions transforming as fundamental under one and 
antifundamental under the other.  
The  theory  has equal number of bosonic and fermionic 
degrees of freedom, but it is nonsupersymmetric 
due to differing   color quantum numbers of its elementary constituents.   
The mismatch 
 of the color representation of bosons and fermions has a significant effect on 
the eigenvalue dynamics and spatial center symmetry realization of 
the orbifold theory. 

The twisted partition function of the theory and 
the effective action for the spatial Wilson lines is given by 
\begin{equation}
\widetilde Z(x)^{\rm QCD(BF)}= \int \; dU_1 dU_2 \; \exp -S[x, U_1, U_2] 
\end{equation}
where $x= e^{-R_{S^1}/ R_{S^3}}$.
The Haar measure is 
\begin{equation}
\int \; dU_1 dU_2 =   \int \; \prod_{i} dv_i^1 \; dv_i^2  
\prod_{i<j} \sin^2 \left[\frac{v^1_i -v^1_j}{2} \right]  
\sin^2 \left[\frac{v^2_i -v^2_j}{2}
\right],  
\label{Eq:Jacobianorb}
\end{equation}
the decoupled product of the Haar measures of each gauge 
group.  Therefore, it does not provide a mutual repulsion between 
 the first cluster and second cluster of eigenvalues. However, as before, 
within each cluster, it provides a logarithmic repulsive potential 
among eigenvalues as seen in Eq.\ref{Eq:Jacobian1}. 

Therefore, we may express the effective action as in Eq.\ref{Eq:potsusy2}, in the 
form  
\begin{eqnarray}
&&S_{\rm eff}[x, U_1, U_2]= S[x, U_1, U_2] - \ln J[U_1]- \ln J[U_2] \cr 
&&= \sum_{n=1}^{\infty}  \frac{1}{n} \left\{(1- z_V(x^n))  
\Big( |\tr(U_1^n)|^2   + |\tr(U_2^n)|^2  
\Big)  +   z_f(x^n) \Big(\tr(U_1^n) \tr (U_2^{\dagger n})  + h.c.\Big) 
  \right\}
\label{Eq:potorb2}  
\end{eqnarray}

The first half of the equation  is  due to the Jacobian, 
the gauge bosons (and ghosts) and originates from the two 
   gauge group factors. It  has a manifest 
$U(1) \times U(1)$ center symmetry, associated with global rotations  
$U_i \rightarrow e^{i \alpha_i}U_i,\; i=1,2$. 
   The second term is 
due to  bifundamental Dirac fermion. 
Recall that introducing bifundamental fermions in the original theory  
reduce this  center symmetry to  a $U(1)$ diagonal.  This is also manifest 
in our effective action, which is only invariant under the restricted rotation 
with $\alpha_1= \alpha_2=\alpha$. The original theory is also invariant under 
the  $\Z_2$  symmetry of the orbifold  which interchanges the two gauge 
group  factors. We will denote this $\Z_2$ symmetry with $\I$.
Therefore, the symmetries of the effective action  
Eq.\ref{Eq:potorb2} are given by 
\begin{eqnarray}
&&{\G_s}: U_1 \rightarrow e^{i \alpha} U_1, \qquad {\G_s}: U_2 \rightarrow 
e^{i \alpha} U_2 \cr 
&& \I: U_1\leftrightarrow U_{2} 
\label{Eq:symmetries}
\end{eqnarray} 
The  low energy effective potential has  to possess all
the  symmetries  of  the   original  theory  (if  the  symmetries  are
non-anomalous)    and   this    is   manifest    in    our   effective
action. 

In order to grasp the spatial center and orbifold shift symmetry realizations 
easily,  it would be more helpful to express the action 
Eq.\ref{Eq:potorb2} in terms of $\I \times { \G_s} $ eigenstates.  
[This furthermore eases the comparison with the orientifold QCD(AS/S) theory, 
which we will discuss  later.] 
Let us define the basis which simultaneously diagonalize 
the $\I \times {\G_s} $ operations. 
The eigenfunctions and eigenvalues of spatial center  and orbifold 
shift symmetry are given by 
\begin{eqnarray}
&\tr  \Omega_{\pm}^k = \tr U_1^k \pm  \tr U_2^k , \;\;\;\;\;  k=1, \ldots \infty 
\qquad, \cr&  \cr
& {\G_s} \; \tr \Omega^k_{\pm}= e^{ i \alpha k}  \; \tr \Omega^k_{\pm}, 
\qquad \qquad
 \I \; \tr \Omega^k_{\pm}= \pm \; \tr \Omega^k_{\pm}
\label{Eq:orbeigen}
\end{eqnarray}
The operators $\tr  \Omega_{\pm}^k$ are even-odd linear combination of Wilson 
lines  of the two gauge groups with winding number $k$.   

The reason for keeping track of the $\I=\Z_2$  interchange  
symmetry of the orbifold  theory is tied to  the nonperturbative large   $\Nc$ 
equivalence. The necessary  (and sufficient) condition for the  the validity of
nonperturbative 
large  $\Nc$ equivalence is   unbroken $\I=\Z_2$   
symmetry of the orbifold  theory  \cite{Kovtun:2003hr,  Kovtun:2004bz}.
If the symmetry $\I$ is not spontaneously broken,  then  the dynamics 
of the large $N$ QCD(BF)  coincides  with the dynamics of SYM in their 
respective neutral  sectors. Therefore, while keeping track of the spatial 
center 
symmetry realization in the classification of phases, we should also understand 
the realizations of the orbifold interchange symmetry.

Expressing the effective action of the  orbifold theory  Eq.\ref{Eq:potorb2} 
in terms  of the ${\G_s} \times \I$ symmetry eigenstates  results in  
\begin{eqnarray}
&& S_{{\rm eff}}^{\rm QCD(BF)} [ \Omega_{+} , \Omega_{-} ] = \half 
\sum_{n=1}^{\infty} \frac{1}{n}
 \Big\{ a^{+}_n (x) 
\;   |\tr(\Omega_{+}^n)|^2   
  +  a^{-}_n (x)  \;  |\tr (\Omega_{-}^n)|^2     \Big\} 
\label{Eq:orb2}  
\end{eqnarray}
where the coefficients are given by 
\begin{eqnarray}
&a^{+}_n (x)= ( 1- z_V(x^n) +   z_f(x^n))  \cr
 &   a^{-}_n (x) = ( 1- z_V(x^n) -  z_f(x^n))
\label{Eq:universal}
\end{eqnarray}

The effective potential $S_{\rm eff}$ Eq.\ref{Eq:potorb2} determines the 
symmetry realizations for  the spatial center symmetry  $\G_s$ and 
the $\I=\Z_2$ interchange 
symmetry of the orbifold QCD(BF) theory and is fairly easy to analyse.   
In the interval $x \in [0, 1)$ (while keeping 
either $S^3$ or $S^1$  much smaller than strong scale), we observe the 
 analog of positivity Eq.\ref{Eq:inequality1} for the coefficient of 
even modes: $a^{+}_n (x) \geq 1$ for all $x\in [0, 1)$.   This means all the 
even modes have positive ``mass''  $a^{+}_n (x)$. 
On the other hand, 
the coefficients of the odd  $\tr \Omega_{-}^n$ modes,   
$a^{-}_n (x) = ( 1- z_V(x^n) -  z_f(x^n)) $ are monotonically decreasing 
functions  of $x$,  and become negative  at sufficiently large $x$.
Let $x_c$ be the locus of $a^{-}_1 (x)$.  Therefore,  
for $x>x_c$, $a^{-}_1(x) $ becomes  negative and leads to spontaneous 
breaking  of the spatial center symmetry $\G_s= U(1)$.  
Since this symmetry breaking is 
driven by an $\I$-odd mode,  it spontaneously breaks the $\I$ symmetry of the 
orbifold theory.

The visualization  of this  symmetry realization on $\R^3 \times S^1$ in the 
eigenvalue basis    was given by D. Tong in \cite{Tong:2002vp}, 
here we adopt his arguments. 
 We already argued that there are two clusters of 
eigenvalues, associated with each gauge group.  These two clusters, 
(each of which has $N$ eigenvalues in them), 
in the symmetry broken phase, wishes  to be maximally apart 
from each other. If the center of mass of one cluster is located at 
$e^{i v_0}$, the other cluster is located at the antipodal point,  
$- e^{i v_0 }$.  Therefore,  the vacuum expectation values 
of the  spatial Wilson lines in the two gauge groups  are 
anti-parallel  to each other. (see fig.\ref{fig:wilson}, small spatial).
We will in the next 
section see that in the thermal case, the positions of the 
two clusters coincide, making the temporal Wilson lines parallel and  
restoring the $\I$ symmetry and breaking just 
the temporal  center symmetry $\G_t$. (see fig.\ref{fig:wilson}, 
small thermal)

\begin{FIGURE}[ht]
{
  \parbox[c]{\textwidth}
  {
  \vspace*{-30pt}
  \begin{center}
  \raisebox{3pt}{\includegraphics[width=5.0in]{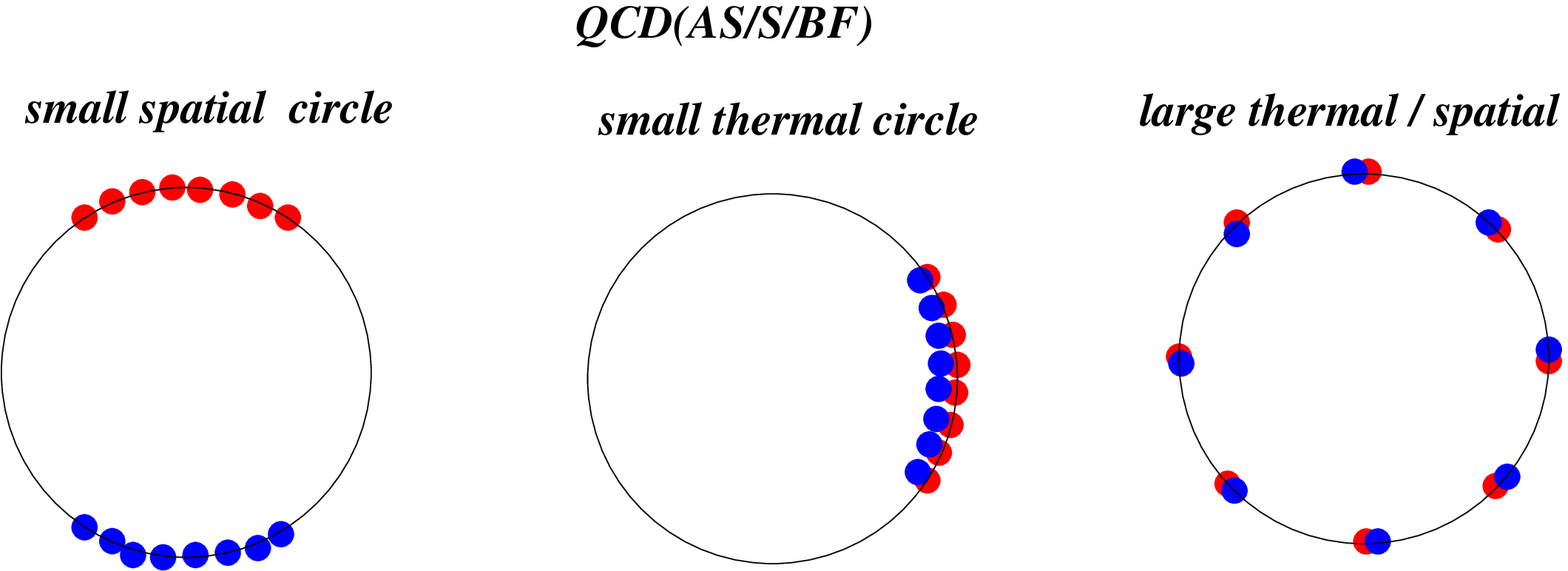}}
  \vspace*{-15pt}
  \caption
    {%
     The distribution of the eigenvalues of the spatial (and temporal) 
     Wilson lines for 
     $U(N)_1 \times U(N)_2$ QCD(BF) theory. 
     The red (blue) points label eigenvalues of the Wilson line $U_1 (U_2)$. 
     At small spatial $S^1$  circle, both spatial center symmetry 
      $\G_s$ and $\I$ interchange symmetry of orbifold are broken. (Wilson 
     lines  $U_1$ and $U_2$ are antiparallel.)     
     At small temporal $S^1$, the temporal center symmetry  $\G_t$ 
     is broken, but not 
     $\I$.  (Wilson lines are parallel.) 
     At large  radius of $S^1$ (either thermal or spatial), the center 
     symmetry is restored, as well as $\I$ in the case of spatial $S^1$. 
          Exactly analogously, one may regard the same picture for orientifold 
      $U(N)$ QCD(AS/S). In this case, the role of the $\I$ symmetry of orbifold 
     is taken by charge conjugation symmetry $\C$, and the role of the 
     Wilson line $U_2$ is played by the  mirror image, $U^*$.  Therefore, 
     the  red (blue) points label eigenvalues of the Wilson line $U$ ( 
     its complex conjugate $U^*$).  At small spatial $S^1$  circle,
     both       $\G_s$ and $\C$ of  QCD(AS/S) are broken. (antiparallel $U$ 
     and $U^*$)
     At small temporal $S^1$, $\G_t$  is broken, but not $\C$. (parallel 
      $U$ and $U^*$.)  
     At large  radius of $S^1$, all symmetries mentioned above 
    are restored. 
     (The blue and red points are split to guide the eye.)
     The orientifold 
     QCD(AS/S) will be discussed in detail in next section. 
     We should also point that for high temperature SYM, the eigenvalues clump 
     just like QCD(AS/S/BF). At low temperature, they delocalize and distribute      uniformly.   On the other hand, for spatial $S^1$ circle, the 
     eigenvalues 
     are uniformly distributed regardless of the size of the $S^1$.
    }
  \end{center}
  }
\label{fig:wilson} 
}
\end{FIGURE}

In the limit where  $x \rightarrow 1$, (small spatial circle), the width of 
the two  clusters of eigenvalues shrinks to zero in the sense described 
in section  \ref{sec:SYM3}, and the distributions of the eigenvalues 
become the Dirac-delta functions  given as 
\begin{equation}
\rho_1(v_1) = \delta (v_1 - v_0) , \qquad \rho_2(v_2) = 
\delta (v_2 - v_0 -\pi ) 
\end{equation}
where $v_0$ is some phase. Therefore, the analysis of the 
effective potential Eq.\ref{Eq:orb2}
reveals 
\begin{equation}  
\langle \frac{1}{N}\tr \Omega_- \rangle =  2  e^{iv_0}, \qquad
 \langle \frac{1}{N}\tr \Omega_+ \rangle = 0 
\end{equation} 
breaking the $\I  \times {\G_s}$ symmetry. (More detailed account of this
 symmetry   breaking and its consequences is discussed in section 
\ref{sec:orb3}. )
The vacuum energy density can be deduced from the twisted partition 
function in the limit $x= 1- \epsilon = 1- \frac{R_{S^1}}{ R_{S^3}} 
\rightarrow 1$ easily. 
\begin{eqnarray}
{\cal E}^{\rm QCD(BF)}(x \rightarrow 1) &&= -  \frac{1}{R_{S^1} V_{S^3}} 
\log \widetilde Z (x \rightarrow 1)
\cr 
&&=
 \frac{1}{R_{S^1} V_{S^3}}  \half \sum_{n=1}^{\infty} 
( \frac{-8}{ 
n^4 \epsilon^3}) |N (1- (-1)^n)|^2 = 
- \frac{\pi^2}{24} ({2N^2}) \frac {1}{(R_{S^1})^4}
\label{Eq:vacuumenergyBF} 
\end{eqnarray}
where the subleading terms in $\epsilon$ and $N$  are neglected. 
 The $2N^2$ is a 
kinematical factor reflecting that there are $2N^2$ gauge fields and 
bifundamental fermions in QCD(BF).

On the other hand,    for $x< x_c$ while keeping  
$\min(R_{S^3}, R_{S^1}) \ll \Lambda^{-1}$ for the validity of the 
perturbative  analysis,   the coefficients 
$a_n^{\pm}(x)$ are all positive.  This implies a net repulsive force among 
eigenvalues of Wilson line, and a uniform distribution. 
This means,  the  $U(1)$ spatial center symmetry and 
$\Z_2$  shift symmetry  of the orbifold  
QCD(BF) theory are restored 
within  this phase.  
The minimum of the effective potential is located at
\begin{equation} 
\langle \tr U_1 \rangle = \langle \tr U_2 \rangle =0, 
\qquad \langle U_i \rangle = e^{i \alpha_i} {\rm Diag}\;(1,
e^{2 \pi/N}, \ldots, e^{2 \pi(N-1)/N} )
\end{equation}
 The leading 
$\O(N^2)$ contributions to vacuum 
energy density vanishes identically within the symmetry 
restored phase, and hence the first nontrivial contribution  to the  $N$ 
scaling of the vacuum energy come from the fluctuations:
\begin{equation}  
{\cal E}^{\rm QCD(BF)} =  N^2(0 + \O(1/N^2))= \O(1) 
\end{equation}
where the  $\O(1)$ contribution is due to the quantum fluctuations.  
Therefore, the theory undergoes a phase transition which alters its ground 
state energy from $\O(N^2)$ in the symmetry broken phase to being $\O(1)$ in 
the   unbroken phase.

The   perturbative one loop analysis of the center symmetry realizations 
can be extended to cover the region $ \min(R_{S^1}, R_{S^3}) \ll \Lambda^{-1}$ 
where the  coupling at the scale of 
the smaller radius is small $g^2 \ll1$, and one can  construct the 
 relevant effective theories. 
In particular, our analysis  shows that the theory on $\R^3\times S^1$ 
(or large $S^3 \times S^1$) on the limit of small  spatial $S^1$ has a phase 
in which $\I= \Z_2$ interchange  symmetry of orbifold is 
spontaneously broken. This  
is in agreement with  D. Tong's result on $\R^3 \times {\rm small \;} S^1$ 
\cite{Tong:2002vp}. In this regime,  higher 
order corrections to the effective potential are small perturbations which can 
not alter the  conclusion  on symmetry realizations. On large spatial 
$S^1$, since the 
theory is strongly coupled, we can not directly check the realization of 
$\G_s \times \I$ symmetry.  
Nevertheless,  as we will discuss momentarily,   
for thermal compactification, the $\I$ symmetry of orbifold theory 
is unbroken at small  $S^1$.  Therefore, the $\I$   
symmetry realizations, probed by the $\I$ 
odd-combination of the Wilson line,   is {\it dependent} of the choice 
of the  boundary conditions for fermions, and should 
indeed disappear  above some critical radius 
of the order of the strong confinement scale, and in the large volume 
limit. 

It is very plausible that 
in the sense of spatial center symmetry and the $\I$ interchange    symmetry 
realization,  
these  are the only two phases of the theory. In particular, 
we expect the spatial center symmetry and shift symmetry  restoring transition 
in the perturbative regime $R_{S^1} \sim  R_{S^3} \ll \Lambda^{-1}$ on 
$S^3 \times S^1$  to extrapolate  all the way to a 
 strongly coupled transition on $\R^3 \times S^1$.   There is simply 
no evidence  indicating otherwise  \cite{Kovtun:2005kh}, although there are 
claims \cite{Armoni:2005wt}. Given this, 
there should be a strong coupling $\G_s \times \I$ restoring phase transition 
around the confinement scale of the QCD(BF) theory on $\R^3 \times S^1$.   
Since QCD(BF) is a vectorlike gauge theory, this should be testable on the 
lattice. 
 
\subsection{Finite temperature phases}
\label{sec:orb2}
In this section, we wish to examine the dynamics of the finite temperature 
orbifold field theory on  $S^3 \times S^1$. This corresponds to  choosing 
anti-periodic boundary conditions for fermions on the $S^1$ circle, while 
keeping the periodic boundary conditions for bosons. 
The resulting Euclidean functional integral corresponds to the  partition 
function $ Z= \tr e^{- \beta H}$  of a  thermal ensemble on  $S^3$ space 
with inverse temperature $\beta$.  

The net effect of the change in boundary conditions for fermions  
is to make the coefficient of fermionic term in Eq.\ref{Eq:potorb2} an 
alternating one: 
\begin{equation}
z_f(x^n) \rightarrow  (-1)^n z_f(x^n) 
\end{equation}
This, in turn, alters the prefactors in the action Eq.\ref{Eq:orb2} into
\begin{eqnarray}
&a^{+}_n (x)= ( 1- z_V(x^n) + (-1)^n   z_f(x^n))  \cr
 &   a^{-}_n (x) = ( 1- z_V(x^n) - (-1)^n z_f(x^n))
\label{Eq:universal2}
\end{eqnarray}
and the action should be interpreted as the effective action for the 
temporal Wilson line (or thermal Polyakov loop). 

As stated earlier, 
there is  a pleasant twist to the story in the thermal case. 
In particular, 
$a_1^{-} (x)$ which leads to the spontaneous breaking of $\Z_2$ symmetry of the 
orbifold theory is now positive definite.  On the 
other hand, $a_1^{+}(x)$ which is the mass of the 
$\I$ even mode $\tr\Omega^{+}$ is 
now  a monotonically decreasing function of $x$. 
The   mass $a_1^{+}(x)$ of the $\I$ 
even mode  passes through zero at the locus $x=x_c$, and is 
negative for $x>x_c$ leading to an instability. 
 
Therefore,  at sufficiently high temperature, 
($x \rightarrow 1$), 
the  $\G_t=U(1)$ temporal center symmetry breaks down spontaneously, but not 
the  $\Z_2=\I$ symmetry of the orbifold theory. In the limit $x \rightarrow 1$, 
the order parameters are  given by 
\begin{equation}
\langle  \frac{1}{N} \tr U_1 \rangle  = \langle \frac{1}{N} 
\tr U_2 \rangle  = e^{i v_0},  \qquad {\rm or} \;\;  
\langle \frac{1}{N} \tr \Omega_- \rangle =0, \;\; \langle 
\frac{1}{N} \tr \Omega_{+} \rangle =2 e^{i v_0}, 
\label{Eq:deconorb}
\end{equation}
In other words, the minima of the effective potential is the  parallel  
 nonzero thermal 
Wilson  line for the two gauge groups.  Needless to say, 
the eigenvalue distributions are given by 
$\rho_1(v_1)= \delta(v_1 - v_0)$ and $\rho_2(v_2)= \delta(v_2 - v_0)$. 
This means the two cluster of the eigenvalues, in the thermal case, 
are literally on top of each other, implying unbroken 
$\I$ symmetry of the orbifold theory.
Therefore,  in the high temperature phase, 
the breaking pattern is $U(1) \times \I \rightarrow  \I$.   The free energy 
density within the deconfined phase is order $N^2$. It simplifies in the  
$x \rightarrow 1$ limit to the usual Stefan-Boltzman result: 
\begin{eqnarray}
{\cal F}^{{\rm QCD(BF)}}(x \rightarrow 1)  && = -   \frac{1}{\beta V_{S^3}} 
\log Z  (x \rightarrow 1)
= - \frac{\pi^2}{24} ({2N^2}) T^4
\label{Eq:freeenergyBF} 
\end{eqnarray}
reflecting the fact that at high temperature there are $2N^2$ gauge bosons and 
fermions contributing to the free energy.  

At sufficiently low temperature ($x< x_c$), the discussion of the 
temporal center symmetry restoration 
is analogous to the spatial center symmetry discussion \ref{sec:orb1}.  
At low temperature, 
the order parameters are 
\begin{equation}
\langle  \frac{1}{N} \tr U_1 \rangle  = \langle \frac{1}{N} 
\tr U_2 \rangle  = 0, \;  {\rm or} \qquad 
\langle \tr \Omega_- \rangle =0, \;\;  \langle \tr \Omega_+ \rangle =0, 
\label{Eq:confinedorb}
\end{equation}
 In the confined phase, both temporal center symmetry and $\Z_2$ symmetry of orbifold theory are unbroken. 
The free energy density is  $\O(1)$ 
reflecting the $\Nc$ independence of the number of color singlet single trace 
operators.  

One can give a  conceptual derivation   demonstrating that 
 $\Z_2$ symmetry in  thermal  orbifold QCD(BF) theory 
(at sufficiently high temperature or small thermal circle) cannot be 
spontaneously broken. Let us recall some basic lore on thermal theories:
The thermal Polyakov loop, as usual, 
 measures the excess of the 
free energy that an external test charge generates in the  thermal medium.  
If the medium, via its thermal fluctuations, 
 is  unable to screen such a charge, the excess free energy is 
infinite, therefore, the expectation value of the Polyakov loop is 
zero (unbroken temporal center symmetry).  This is the confined phase.  
On the other hand, if the thermal fluctuations of color 
are large enough so that 
the medium  can  screen the test charge, then the excess free energy is 
finite and  the 
Polyakov loop acquires a vacuum expectation value breaking the temporal 
center symmetry.  This is the deconfined phase. 

Since the QCD(BF) gauge theory is a product gauge group $U(N) \times U(N)$, 
an  external test quark  has charges under both centers 
$(a, b) \in  \Z \times \Z$ as discussed in section \ref{sec:gen2}.  
The   $\langle  \frac{1}{N} \tr U_1 \rangle$ corresponds  to the excess free 
energy of an external quark whose charge is  $(1,0)$   and 
 $\langle  \frac{1}{N} \tr U_2 \rangle$ to the one whose charge is $(0,1)$.  
In the confined phase,  Eq.\ref{Eq:confinedorb} implies neither $(1,0)$ nor  
$(0,1)$ can be screened. In the deconfined phase,   Eq.\ref{Eq:deconorb} 
implies that the excess of free energy to have a test charge $(1,0)$  in 
thermal medium 
is equal to the one of having $(0,1)$.  On $\R^3 \times S^1$, the fact that 
$\Delta {\cal F}_{(1,0)} =  \Delta {\cal F}_{(0,1)}$  should be expected.  
The difference of these two charges is $(1,-1)$ and the dynamical 
 fermions (which have charges $\pm (1,-1)$) can easily 
convert one into the other. Since free energy is a class function, and 
$(1,0)$ and $(0,1)$ are in the same conjugacy class by Eq.{\ref{Eq:class2}, 
we should 
indeed expect that $\Delta {\cal F}_{(1,0)} - \Delta {\cal F}_{(0,1)}=0 $, 
implying $\langle \tr \Omega_{-} \rangle =0 $.
This   retrospectively   justifies
 that the $\Z_2$ symmetry in  thermal  orbifold QCD(BF) theory  cannot be 
spontaneously broken. 
  
\subsection{Digression: Topological classification of  order  parameters}
\label{sec:orb3}
As we discussed, 
the  orbifold  theory has $U(1)$ (spatial or temporal) center symmetry,  
and  $\Z_2$ shift symmetry 
exchanging the  two gauge group factors. These symmetries act on the 
Wilson lines as given in  Eq.\ref{Eq:orbeigen}.  

In order to define the breaking patterns unambiguously, we wish to analyze 
symmetries  in some more detail.  First note that the action of the 
$\I$ on the order parameter $\tr \Omega_-$ negates it, and so does a $U(1)$ 
center symmetry action by $e^{i \alpha}= e^{i \pi}$.  Therefore, the combined 
action of the two does not change the order parameter $\tr \Omega_-$. 
Therefore,  the vacuum expectation value of the  $\tr \Omega_-$ cannot break 
the combined $\Z_2$ gauge and $\Z_2$ shift symmetry. In order to isolate the 
precise symmetry breaking pattern, 
let us (artificially) split the center symmetry into a $\Z_2$ (global gauge rotations generated by $e^{i \pi}$),   
and the  quotient of the $U(1)$ by $\Z_2$. 
This  can be done by declaring equivalences among phases    
$e^{iv} \in  U(1)$  as $v \equiv (v+ \pi)$, i.e, identifying antipodal
points on the $S^1$ circle (the $U(1)$ group manifold).  
We will refer to this  coset space as $\widetilde{U(1)}$. Therefore,  
we may write $U(1) \equiv  \widetilde{U(1)} \times \Z_2  $.   
Up to this point $U(1)$ center symmetry could have been either spatial or 
temporal. Let us first analyze the periodic compactification along the $S^1$ 
circle. In order to see the difference between the breaking patterns in 
nonthermal and thermal case,  we  introduce the notation 
$\G_s=U(1) \equiv  \widetilde{U(1)} \times \widetilde{\G_s}   $ where 
 $\widetilde{\G_s}$ is the $\Z_2$ subgroup of spatial center symmetry $\G_s$. 

Notice that the $\widetilde{\G_s}$  subgroup of spatial center symmetry 
is not the same as the $\I$  interchange symmetry of the orbifold 
theory. Even
 though  their action on  $\tr \Omega_-$ is the same, 
the  $\tr \Omega_+$ 
is singlet  under interchange symmetry and negates under spatial  center 
subgroup $\widetilde{\G_s}$. Alternatively, one can check their action on 
Wilson lines  
$\tr U_1$ and  $\tr U_2$:
\begin{equation} 
\I: U_1 \leftrightarrow U_2, \qquad 
\widetilde {\G_s}: U_i \rightarrow -U_i, \qquad i=1,2 
\end{equation}
The symmetry of the theory may conveniently be written as 
\begin{equation} 
\widetilde{U(1)} \times    (\Z_2)^2 =  \widetilde{U(1)} \times 
\{1, \I, \widetilde \G_s, \I \widetilde \G_s\}
\label{Eq:symm}
\end{equation}  
and 
the pattern of spontaneous symmetry breaking  discussed in section 
\ref{sec:orb1} corresponds to 
\begin{equation} 
\widetilde{U(1)} \times    (\Z_2)^2      
\rightarrow  \Z_{2, D}= \{1, \I \widetilde \G_s\} 
\label{Eq:pattern}
\end{equation}  
The  vacuum of the  theory is not invariant  under the discrete operations
of $\{\I, \widetilde \G_s \}$,  
but  is invariant  under  the diagonal subgroup 
$\Z_{2, D}= \{1,  \I \widetilde \G_s \}$.  

There is one important corollary  we want to address in this pattern. 
The fact 
that the diagonal $\Z_{2,D}=  \{1,  \I \widetilde \G_s \}$ is unbroken 
has simple, 
yet non-trivial implications for local $\I$-odd order parameters.

{\bf  Corollary:}  Let  $\O(x)$ be  a local  order parameter  probing the
$\I$  interchange  symmetry  of the  orbifold QCD(BF) theory.   
The phase  of the
theory with  the breaking  pattern Eq.\ref{Eq:pattern} does  not admit 
local order parameters to acquire a vacuum expectation value.

Since  $\O(x)$ is  local, it  is a singlet  under center symmetry. 
Therefore,  we  have 
$$(\I  \widetilde \G_s):  \O(x) = \I   \O(x) =  -\O(x).$$
Since  the
vacuum  of  the  theory  is  invariant  under  $\I \widetilde \G_s$,  
the  vacuum
expectation value of the operator  $\O(x)$ must satisfy
$$
 \langle \O(x)  \rangle =  \langle (\I \widetilde \G_s)\O(x))\rangle =  - \langle\O(x)
\rangle=0
$$
Therefore, even though the theory is in the $\I$ broken phase, the {\it local} 
order parameters must have vanishing expectation values.
 
This curious fact arises due to intertwining of the spatial center symmetry 
 with the $\I$ symmetry as in Eq.\ref{Eq:pattern}. 
A genuine distinction arises  between the topologically trivial  and 
nontrivial  operators. 
The topologically trivial operators (which do not wind 
around the $S^1$ circle) are all singlet under the center symmetry. Since 
$\Z_{2, D}$ 
is unbroken in neither phase of the theory, and   
since the local  $ \I$ odd order parameters necessarily transform nontrivially 
under the  action of $\I \widetilde \G_s$, 
 the expectation 
value of all such operators must be  equal to zero. 
In particular,   $\O(x)= \frac{\tr}{N} F_1^2 - \frac{\tr}{N} F_2^2$ is such 
an operator, trivial under center symmetry and transforming under $\I$.   
Our result implies that 
\begin{equation}
\langle \tr F_1^2 - \tr F_2^2 \rangle =0, \qquad   {\rm  (either \;\;  phase)}. 
\end{equation}

This unambiguously demonstrates that $\langle \tr F_1^2 - \tr F_2^2 \rangle =0$ 
also in  $\R^3\times S^1$ where $S^1$ is small.
Our current knowledge on the QCD(BF) theory  is consistent with the assertion 
that this is true at arbitrary size $S^1$, in particular on  $\R^4$ limit.
On $\R^3 \times S^1$ where spatial  $S^1$ is small, we have the breaking 
pattern  Eq.\ref{Eq:pattern}. We believe, the theory  will 
undergo a center symmetry restoring  transition  around the strong scale, 
and restore its  spatial center symmetry along with $\I$ symmetry of the 
 orbifold QCD(BF).
Our  results are  in accord with \cite{Kovtun:2005kh,Tong:2002vp} and 
contradicts the claim  in ref.\cite{Armoni:2005wt}. 
On $\R^4$,  Ref.\cite{Armoni:2005wt} 
presents   
$\langle \tr F_1^2 - \tr F_2^2 \rangle \neq 0 $ as  a 
prediction of string theory, see the reference therein. 

The thermal case is simpler. With the analogous notation, 
the breaking pattern is 
\begin{equation} 
\widetilde{U(1)} \times    (\Z_2)^2 =   
\widetilde{U(1)} \times
\{1, \I, \widetilde \G_t, \I \widetilde \G_t\}    
\rightarrow  \Z_2= \{1, \I\}
\label{Eq:thermalpattern}
\end{equation}  
in the deconfined phase.  (Therefore, it is not even necessary to  
quotient the temporal center symmetry into subgroups, 
since it is fully broken in this case.)   
The $\I$ symmetry of orbifold theory is unbroken in the high 
temperature phase. 
In confined  phase, the $\widetilde{U(1)} \times    (\Z_2)^2$   symmetries 
 are  unbroken. 
In neither phase of the theory, since  the  $\Z_2$ symmetry of the orbifold 
 is  unbroken,  no 
 $\I $ odd operator can acquire an expectation value. \footnote{The fact that 
the expectation value of the topologically nontrivial order parameter 
$\tr \Omega_{-}$ is sensitive to the choice of the boundary conditions 
(non-zero with the use of periodic boundary conditions, and  zero 
when antiperiodic boundary conditions)   demonstrates that this breaking
is a finite volume effect, which should disappear in the large radius limit. 
Nevertheless, the scale at which this transition should occur is a 
nonperturbative  scale of the underlying theory and is physical the same way 
the deconfinement temperature is.  }

\begin{FIGURE}[ht]
{
  \parbox[c]{\textwidth}
  {
  \vspace*{-30pt}
  \begin{center}
  \psfrag{inf}{$\infty$}
  \raisebox{3pt}{\includegraphics[width=3.0in]{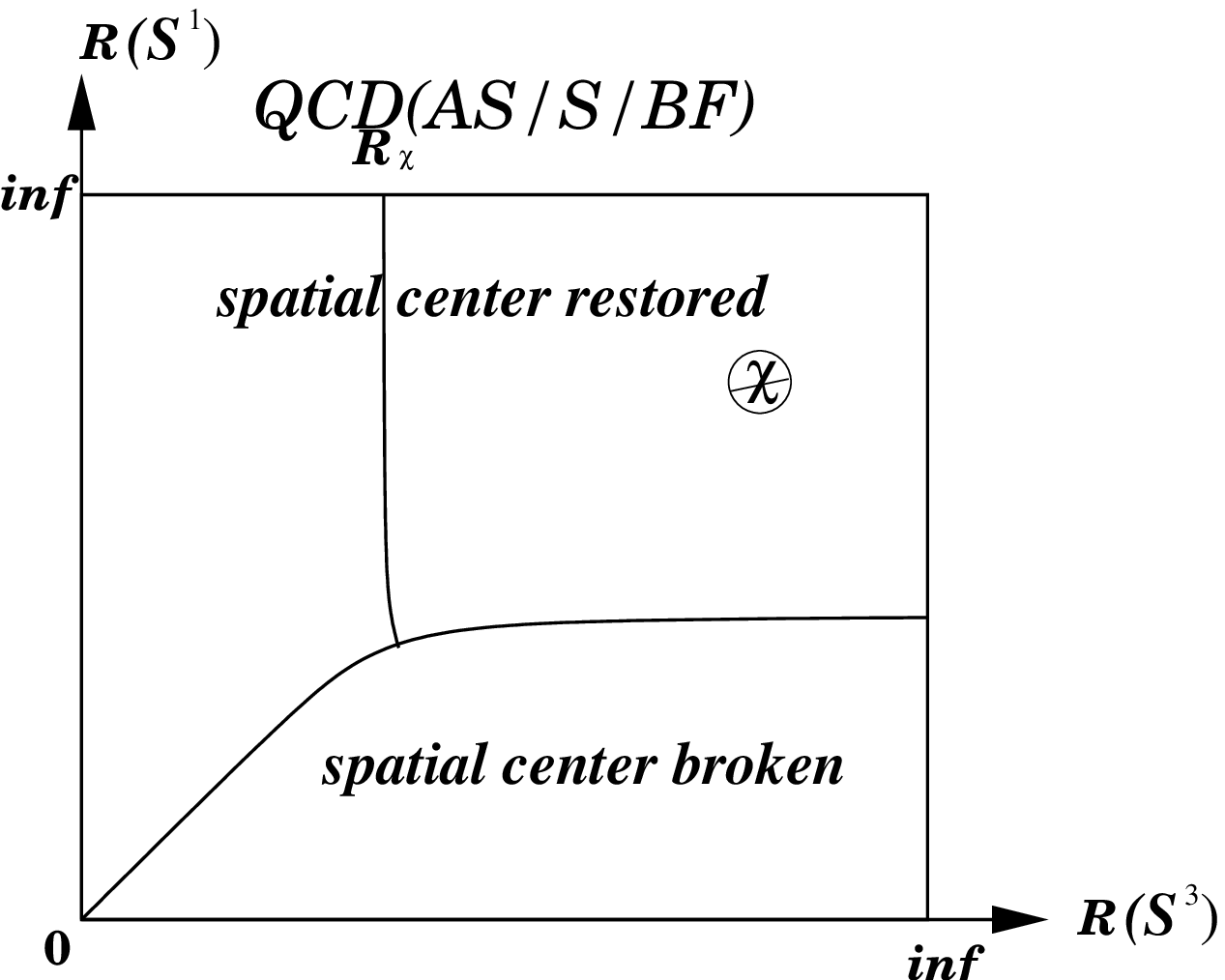}}
  \includegraphics[width=3.0in]{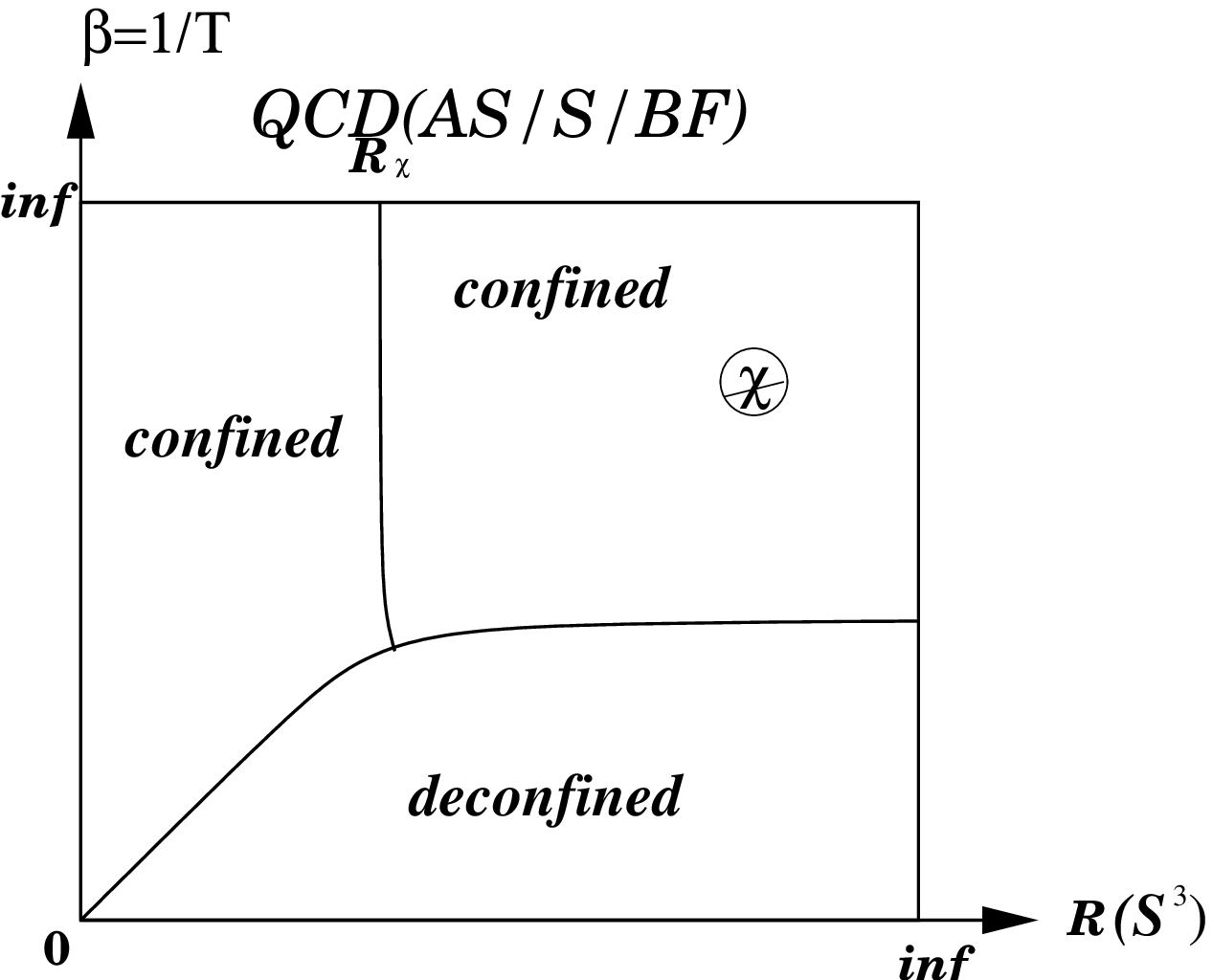}
  \vspace*{-15pt}
  \caption
    {%
    Schematic phase diagram of $U(\Nc)$ QCD(AS/S/BF) with periodic boundary 
    conditions for fermions (left)
    and with thermal (anti-periodic) boundary conditions  (right)
    as a function of radii $R_{S^1} (\beta)$ and $R_{S^3}$. Thermal 
    QCD(AS/S/BF)  has 
     two confining low temperature phases
    in which temporal center symmetry $\G_t$ is unbroken. These two phases are 
    distinguished by  the discrete chiral symmetry realization. 
    The  theory is chirally  symmetric at sufficiently small 
    radius $R_{S^3}< R_{\chi}$  and  asymmetric at large radius  
    $R_{S^3}> R_{\chi}$.   These theories 
    also possess a    
    deconfined high temperature phase with broken  $\G_t$   and 
     unbroken chiral symmetry.  
     For  the  nonthermal compactification where $S^1$ is a spatial circle, 
    the symmetry realizations are essentially same modulo the replacement 
   of temporal spatial symmetry with the spatial one $\G_s$. 
    As explained in the 
    text, the underlying 
    reason for unbroken chiral symmetry in the spatial symmetry 
    broken phase is a one loop quantum effect, unlike the thermal 
    compactification where it is due do  tree level mass of fermions. 
    These sketches depict the simplest consistent scenarios. 
    In the 
   thermal  case, the $\I$ and $\C$  symmetries  necessary for the validity 
   of large $N$ equivalence are unbroken, explaining the matching of phase 
   diagram to $\N=1$ SYM. In the nonthermal case, the phase which breaks 
   $\G_s$ also breaks $\I$ and $\C$, explaining the mismatch to  $\N=1$ SYM 
    within that phase.
    }
  \end{center}
  }
\label{fig:phase2} 
}
\end{FIGURE}

\subsection{Chiral symmetries }
\label{sec:orb4}
We first examine the chiral symmetry realizations of thermal orbifold theory  
on $S^3 $,  where fermions are endowed with antiperiodic boundary 
conditions on the thermal $S^1$ circle. In this case, there are no surprises and
the phase diagram is essentially same as  thermal $\None$ SYM. The  more  
interesting  case is the one where fermions obey periodic boundary condition, 
which  we will discuss afterwords. 

As discussed in the $\None$ SYM case, high curvature or high temperature 
does not allow a fermion condensate to form because of the same reason. In 
either case, the lowest fermionic Kaluza-Klein mode is much larger than 
the strong  
scale of the QCD-like theory and the fermions can be  integrated out 
perturbatively without any  formation of chiral condensate. 
In the limit where $\min (R_{S^3}, \beta )\gg  \Lambda^{-1}$,  the orbifold 
theory is strongly coupled. The theory has a vectorlike baryon number
 and axial chiral symmetry,  $U(1)_B \times \Z_{2N}$. 
 In vector-like gauge theories, the vector-like symmetries do not 
break down spontaneously while it is believed that  the axial 
symmetries do \cite{Vafa:1983tf}.
The discrete 
(non-anomalous) axial  symmetry    $\Z_{2N}$ acts on the bifundamental 
fermions as $\Psi \rightarrow e^{i \gamma_5 \alpha} \Psi$, where 
$\alpha$ is an integer multiple of $2 \pi/N$.   
The conventional wisdom tells us that a condensate  will form in the most 
attractive channel (MAC). The MAC in this example is 
$\langle \bar \Psi \Psi \rangle$, breaking the discrete chiral symmetry as 
$\Z_{2N} \rightarrow \Z_2$.  Therefore, we expect $N$ vacua distinguished 
by the phases, $e^{i 2 \pi k /N}$ where $k=0, \ldots N-1$  exactly as in SYM 
on $\R^4$.  

Therefore, as in the case of SYM, we expect QCD(BF) to possess at least 
three different phases. 
The simplest  phase diagram of the thermal theory consistent with 
all the knowledge that 
we have gathered so far is given in fig.\ref{fig:phase2}.

The case where fermions are endowed with periodic boundary conditions is more 
interesting. The clear distinction compared to the thermal case  (where 
fermions acquire a tree level mass term because of anti-periodic boundary 
 condition along the thermal circle)    is that the lowest fermionic mode  
 on the $S^1$ circle are  not 
classically  gapped. This is purely  a manifestation  of 
the periodic boundary condition. 
Let us first consider the theory on $S^3 \times S^1$ where 
$R_{S^3} \gg \Lambda^{-1} $ is fixed, and $S^1$ is dialed as desired.  Since the 
$S^3$  is large, we can approximate it as $\R^3$.  At tree level, the fermions
have a spectrum $\omega_n^{(0)}= \frac{2 \pi}{ R_{S^1}} n$ where $n=0, \pm 1, \pm 2, \ldots$. 
Let $R_{S^1} \ll \Lambda^{-1}$  so that the theory is amenable to a perturbative 
treatment. It is easy to take the $\R^3 \times S^1$  limit of the action 
 Eq.\ref{Eq:potorb2} and obtain the one-loop effective potential for the Wilson 
line in the orbifold  theory.   
The limit obtained in this way is identical to the result of 
\cite{Tong:2002vp}. 
The one-loop Coleman-Weinberg potential breaks the $\I$ interchange symmetry 
which is signalled by the   $\I$ odd combination of the spatial 
Wilson line acquiring a 
vacuum expectation value. This quantum effect provides a mass term for 
fermions, by shifting all the eigenfrequencies by a half integer.
It is easy to see that fermions becomes gapped due to this quantum effect.
 The quantum corrected eigenfrequencies are given by  
$\omega_n^{(1)}= \frac{2 \pi}{R_{S^1}} (n + \half) $ where  
$n=0, \pm 1, \pm 2, \ldots$.  \footnote{The four dimensional mass term is 
prohibited by the 
chiral symmetry itself.  The mass term generated in this case is a three 
dimensional mass, for the Kaluza-Klein tower of all the fermionic modes from 
the three dimensional point of view. This 
mass is allowed by the chiral symmetry}     Notice that these eigenmodes are 
identical to the eigenmodes of the thermal compactification if we were to 
identify  $R_{S^1}$ as inverse temperature $\beta$. However, it should be kept 
in mind that the latter is a tree level result, and in our case, the gap is a  
consequence of the one loop quantum correction.  
The mass is set in units of inverse compactification radius, $R_{S^1}$. 
Similar to the 
high temperature considerations, at length scales much larger then $R_{S^1}$, 
the  fermions can be regarded as a heavy Kaluza-Klein tower and 
therefore, can be integrated out perturbatively without the  formation of 
a condensate.  
Hence chiral symmetry is restored in this phase at sufficiently small $S^1$.
At large $S^1$, the theory is strongly coupled.  
At large radius, the 
dynamics of the theory should be independent of the choice of the boundary 
conditions.  Therefore we expect, based on MAC argument similar to the 
thermal case discussed above, a chiral condensate and a chirally asymmetric 
phase where $\Z_{2N}$ symmetry is spontaneously broken to $\Z_2=(-1)^F$. 

This transition is quite interesting in its own right. The reason is, the 
theory on $\R^3 \times S^1$ where fermions have periodic boundary conditions 
along $S^1$, may be regarded as a zero temperature field theory. Here, we are 
viewing   one of the noncompact directions $\R$ as a temperature circle, 
albeit decompactified,  corresponding to $T=0$. Therefore, what we see 
is a zero temperature chiral phase transition.  We expect this transition 
to take place around the strong scale of the orbifold theory and be entangled 
with  spatial center  symmetry realizations. \footnote{ It should be possible 
to see this transition on lattice simulations.  The lattice should be an 
asymmetric torus, and with say, three long, and one short direction. The 
thermal boundary conditions for fermions should be used in one of the 
long directions, so that the theory may be regarded as a low temperature field 
theory. Then, by dialing the size of the short direction (or equivalently, 
the bare coupling of the lattice theory), this transition should be seen  
around strong scale.}

Therefore, as long as $\min (R_{S^1}, R_{S^3}) \ll \Lambda^{-1}$,
 there should not 
be   any chiral condensate,  either due to curvature or  due to 
quantum effects rendering fermions massive.  Within this region of the 
phase diagram however,  there is a phase transition associated with 
spatial center symmetry $\G_s$ and $\I$ interchange  
symmetry of the orbifold theory 
at around $ R_{S^1} \sim R_{S^3}$. When  $\min (R_{S^1}, R_{S^3}) \gg \Lambda^{-1}$,
the chiral symmetry is broken, but spatial center symmetry is 
unbroken. The simplest 
phase diagram consistent with our current  knowledge is 
shown in fig.\ref{fig:phase2}.

We also note that the chiral properties of orbifold QCD(BF) theory are clearly 
different from those  of the $\None$ SYM  theory on $\R^3 \times S^1$. The 
 $\None$ SYM have a chiral  condensate even at small radius, 
which by holomorphy  is equal to the condensate at large radius 
(in units of strong scale).  
An implication  of holomorphy is radius independence  of 
the chiral condensate as demonstrated by Davies et.al. \cite{Davies:1999uw}.
  On the other hand, 
the orbifold  QCD(BF) theory  does not possess a chiral condensate in small 
radius.  
Ref.\cite{Dijkgraaf:2002wr} provides a recipe relating 
 the holomorphic  quantities in $\None$ SYM to their image in the 
nonsupersymmetric orbifold theory. As discussed above, we expect a  
chiral phase transition in the orbifold theory around the strong scale.
This in particular implies radius independence does not hold for  the 
chiral condensate in QCD(BF) in the  $\I$  broken phase. 
It is, on the other hand, perfectly sensible that the 
condensate proposed in ref.\cite{Dijkgraaf:2002wr}, 
will  coincide at  large radius, the  $\I$ 
unbroken phase.  Therefore, as long as the radius is larger than a critical 
size,  we expect  radius independence from QCD(BF) condensate, 
and an equivalence to $\N=1$ SYM.   

\section{Phases of orientifold  QCD(AS/S) theory on $S^3 \times S^1$}
\label{sec:orienti}
\subsection{The phases as a function of  volume}
\label{sec:orienti1}
We continue the analysis of the 
phases of the vector-like gauge theories with  QCD(AS/S), the orientifold 
partner of   $\None$ SYM. \footnote{While this paper was in the 
completion process, a preprint by  Hollowood and Naqvi 
ref.\cite{Hollowood:2006cq} appeared. The material of 
sections \ref{sec:orienti1} and \ref{sec:orienti2} has identical results with 
  \cite{Hollowood:2006cq}. Tim Hollowood also told me that he  
 obtained the  $\I$ restoring phase transitions in the case of 
orbifold QCD(BF) \cite{Hollowood}. 
I thank them  for communications related to their work.}
  To be able to compare with the other vector-like 
gauge theories examined, we work on   $S^3 \times S^1$, and benefit from the 
spin structure of $S^1$. We first study the case where fermions obey periodic 
boundary conditions. The fundamental quantity of interest is again 
  the twisted   partition function  $\widetilde Z= \tr e^{-\beta H} (-1)^F$. 

As in the orbifold QCD(BF) theory examined in the previous section,
the orientifold QCD(AS/S) is non-supersymmetric   irrespective of the  
background  space. On the other hand, in leading order in  $N$, it has 
identical numbers of bosonic and fermionic degrees of freedom. The fact that 
 the color 
quantum number of fermions (which is antisymmetric or symmetric representation) 
is different from the one of gauge boson (adjoint) affect the eigenvalue 
dynamics of the Wilson line, as well as the center symmetry realizations. 

The twisted partition function of the QCD(AS/S) theory is given by 
\begin{equation}
\widetilde Z(x)^{\rm QCD(AS/S)}= \int \; dU  \; \exp (-S^{\rm QCD(AS/S)}[x,U]) 
\end{equation}
where the effective action for the spatial Wilson line 
may be expressed as in Eq.\ref{Eq:potsusy2},
\begin{eqnarray}
&&S_{\rm eff}^{\rm QCD(AS/S)} [x,U]= S[x, U] - \ln J[U] \cr 
&&= \sum_{n=1}^{\infty}  \frac{1}{n} \left\{(1- z_V(x^n))  |\tr U^n |^2   
  +   z_f(x^n) \Big( \frac{(\tr U^n)^2 \pm \tr U^{2n}}{2}  + {\rm h.c.} \Big) 
  \right\}
\label{Eq:potorienti2}  
\end{eqnarray}

The first half of the equation  is  due to the Jacobian, and 
gauge bosons (and ghosts).  It  has a manifest 
$U(1)$ center symmetry, associated with global rotations  
$U \rightarrow e^{i \alpha}U$. 
   The second term is 
due to  the antisymmetric/symmetric  representation  Dirac fermion. 
Recall that introducing two-index representation 
 fermions in the original theory   reduced this  center symmetry 
to  a  $\Z_2$.  
This is manifest 
in our effective action, which is only invariant under the restricted rotation 
of $\alpha= \pi$. 
The QCD(AS/S)  theory is also invariant under 
the  $\Z_2$ charge conjugation 
  symmetry which we  denote by $\C$.
The theory also has other discrete spacetime symmetries, but 
the only   symmetry necessary for the validity of the nonperturbative 
large $N$ equivalence  between QCD(AS/S) and SYM is 
$\C$. Therefore, we classify phases according to 
charge conjugation symmetry and spatial center symmetry 
$\C \times \widetilde \G_s$.(The use of $\widetilde \G_s$ for spatial center 
symmetry rather than   $ \G_s$ will be seen below. But, 
as the reader may easily guess, we will identify it with the corresponding 
symmetry in orbifold QCD.)
The symmetries of the effective action  
Eq.\ref{Eq:potorienti2} are given by 
\begin{eqnarray}
&&{\widetilde \G_s}: U \rightarrow e^{i \pi} U = -U,  \cr
&& \C: U \rightarrow U^{*} 
\label{Eq:symQCDAS}
\end{eqnarray} 
consistent with the symmetries of the original QCD(AS/S) theory.

Notice that the action has 
both double  and single trace operators.  The leading large $N$ 
dynamics of the theory is dictated by the double trace operators with natural 
$\O(N^2)$ scaling,  and the effect of the single trace operator 
$\tr U^{2n}$ is $\O(N)$, thus  subleading by $\frac{1}{N}$.  
Therefore, in the analysis of the 
dynamics in the $N= \infty$ limit,  
we neglect the $\pm \tr U^{2n}$ term.  This removes any distinction between 
QCD(AS) and QCD(S). The subleading corrections to physical 
correlators in QCD(AS/S) are 
$\O(1/N)$, as  opposed to $\O(1/N^2)$ which is typical in orbifold QCD(BF). 

This purely kinematical large $N$ effect brings in 
a beautiful set of simplifications. In fact, exactly in the case of 
orbifold QCD(BF), we will be able to express the action as a sum of perfect 
modulus  squares. Naturally enough (analogous to the orbifold example),  
the completion to the 
sum of perfect squares occurs with the use of eigenfunctions which 
simultaneously  diagonalize the spatial center symmetry $\widetilde \G_s$ and 
charge conjugation symmetry $\C$. 

To do so,  let us define the linear combination of the 
Wilson lines in orientifold QCD(AS/S) theory in the eigenbasis of 
the spatial center symmetry and charge conjugation symmetry  
$\widetilde \G_s \times \C$.  
 The orientifold 
eigenvalue problem  can be expressed as
\begin{eqnarray}
&\tr  \Omega_{\pm}^k = \tr U^k \pm  \tr (U^*)^k , \;\;\;\;\;  k=1, \ldots \infty 
\qquad, \cr&  \cr
& {\widetilde \G_s} \; \tr \Omega^k_{\pm}= e^{ i \pi k}  \; \tr \Omega^k_{\pm}, 
\qquad \qquad
 \C \; \tr \Omega^k_{\pm}= \pm \; \tr \Omega^k_{\pm}
\end{eqnarray}
where we constructed $\C$ even-odd combinations of Wilson lines with winding 
number $k$ for a given  orientation and its conjugate with opposite 
orientation.  Notice that the action of $\widetilde \G_s$ and   
$\C$ on eigenfunctions 
$ \tr  \Omega_{\pm}^k $  is exactly identical to the action of  
$\widetilde \G_s$ and $\I$
in the case of orbifolds, hence the notation. 
[Recall that in  QCD(BF), the full symmetry is 
$\widetilde{U(1)} \times \widetilde \G_s \times \I $.   
The   $ \widetilde{U(1)}$ is not 
present in QCD(AS/S), the  center symmetry that is present is just
$\widetilde\G_s$.]
\footnote{\label{footnote:centerAS} For SYM and QCD(BF),  the 
$ U(1)$ center symmetry, in the unbroken center symmetry  phase, 
guarantees  that any topologically nontrivial Wilson line 
with arbitrary number of winding  will have a zero vacuum expectation value.
This is a consequence of unbroken symmetry. On the other hand, for QCD(AS/S), 
the center symmetry is  just $\Z_2$.  This implies only the Wilson line with 
odd winding number will have a zero vacuum expectation value due to symmetry. 
What about the Wilson lines with even winding number? Do they acquire 
 a (vacuum or thermal)  expectation value, and  if so, 
does this invalidate orientifold equivalence 
even in the large volume phase?  
The answer to the first question is,   yes. If an operator is not protected by 
a symmetry, it will acquire a vacuum expectation value. However, it is possible 
to show that, by employing the loop equations, 
 the vacuum expectation value of even winding number Wilson lines
is suppressed in the large $N$,  $\langle \frac{1}{N} 
\tr U^{2n} \rangle = \O(1/N)$.  Therefore, neither symmetry considerations, 
nor large $N$ equivalence is compromised.  We expect this type of behaviour, 
i.e., {\it  vanishing of certain correlators without symmetry reasons},
 will give us 
important hints about the dynamics of large $\Nc$ limits,  and should admit a 
greater understanding of large $\Nc$ dynamics.  
This prediction of large $\Nc$ orientifold equivalence 
may be tested on lattice simulations.
}    

Expressing the orientifold QCD(AS/S)  effective 
actions    Eq.\ref{Eq:potorienti2}  in 
terms  of the ${\widetilde \G_s} \times  \C$   symmetry eigenstates  gives 
\begin{eqnarray}
&& S_{{\rm eff}}^{\rm QCD(AS/S)} [ x, \Omega_{+} , \Omega_{-} ] = 
\frac{1}{4} \sum_{n=1}^{\infty} \frac{1}{n}
 \Big\{ a^{+}_n (x) 
\;   |\tr(\Omega_{+}^n)|^2   
  +  a^{-}_n (x)  \;  |\tr (\Omega_{-}^n)|^2     \Big\} 
\label{Eq:oo}  
\end{eqnarray}
where $a^{\pm}_n (x)$ are exactly the same as in the orbifold theory given in 
Eq.\ref{Eq:universal}. The range of the modulus 
$|\tr \Omega_{\pm}|$  coincides precisely in 
orbifold QCD(BF) and orientifold QCD(AS/S), hence the extremization problems 
are identical. (The extra $\widetilde {U(1)}$ symmetry present in  QCD(BF) only 
changes the phase of  $\tr \Omega_{\pm}$ in QCD(BF) without altering its 
modulus.  Since the action only depends on the modulus,  
the presence of $\widetilde {U(1)}$ symmetry  
does not interfere with the extremization of action. Nevertheless, it is 
significant in the discussion of quantum fluctuations, see footnote
 \ref{footnote:centerAS}. 
The difference between the 
overall actions by a factor of $\half$ is a purely  kinematic factor (which 
should indeed be there) 
reflecting that the number of colors in $U(N) \times U(N)$
 QCD(BF) is $2N^2$ and number of colors $U(N)$ in QCD(AS/S) is $N^2$, and 
does not influence  the dynamics.

The exact equivalence of the symmetry realization of $N=\infty$ QCD(AS/S) 
and QCD(BF) on   $S^3 \times S^1$ is one of the main result of this paper.  
If $\Z_2$ symmetry of the orbifold QCD(BF) and $\C$ symmetry of the orientifold 
QCD(AS/S) are unbroken in the $\min (R_{S^3}, R_{S^1}) \gg \Lambda^{-1}$ 
regime as well, there is an exact  $N=\infty$ 
equivalence  between them  at arbitrary radii.  Currently, there is no 
evidence (and 
neither a proof) that 
$\I$ and $\C$ are broken on large radii or on $\R^4$.  
But both are strongly unlikely.    
Below, we will see that all
 the discussion of QCD(BF) can be carried verbatim to 
QCD(AS/S) by just replacing wherever one sees the $\Z_2=\I $ symmetry of the
 orbifold theory by $\C$ charge conjugation symmetry of the orientifold 
QCD(AS/S) theory.  This, in particular means, if there is a phase in which 
$\Z_2$ symmetry of orbifold theory is spontaneously broken, so is  $\C$ in 
orientifold theory. If for 
some reason  $\Z_2$  gets  restored or remains unbroken in QCD(BF), 
so does $\C$ in QCD(AS/S). 
\footnote{The simplest way to understand this companionship between $\C$ in 
QCD(AS/S) and $\I$ in QCD(BF) is to go to basics, and see how these theories 
are obtained by genuine projections.  Let us consider a $U(2N)$ and $SO(2N)$ 
parent $\None$  SYM theories and  apply identical   projections
by global $\Z_2$ gauge symmetry times fermion number modulo two $(-1)^F$, 
i.e,  $\Z_2(-1)^F$. 
The projections  are indeed identical, but have different names: 
orbifold projection (if one starts with unitary group) 
and orientifold (if one starts with orthogonal group). 
There are two possible orbifold projections by $\Z_2(-1)^F$ depending on the 
fermion number assignment $(\pm)$: 
If one use $(+)$ fermion number assignment, 
the daughter theory is a decoupled 
$U(N)_1 \times U(N)_2$ $\None$ SYM theory.  If one uses $(-)$, then 
fermions  comes in bifundamental representation, i.e, 
fundamental under $U(N)_1$, and  antifundamental under $U(N)_2$. 
Analogously, 
starting with $SO(2N)$, and doing projections by using $(+)$ assignment, 
 one obtains again a decoupled 
``product'' gauge group, $U(N) \times U(N)^*$ $\N=1$ SYM. The $U(N)^*$
is the analog of  $U(N)_2$ of the orbifold theory,  and is just an auxiliary 
 mirror image of the $U(N)$  group.
 Incorporating fermion number 
into the projection  yields two index representation 
fermions transforming as ``bifundamentals'' of 
$U(N)\times U(N)^*$, fundamental under $U(N)$ and antifundamental under 
$U(N)^*$. Since the antifundamental of $U(N)^*$ is  fundamental of $U(N)$,  
the fermions are in two index antisymmetric representation.  
Therefore, 
the $\I$ symmetry which interchange the $U(N)_1$ and $U(N)_2$, is replaced 
by a $\C$ symmetry which interchange $U(N)$ with its mirror image   $U(N)^*$. 
If $\I$ and $\C$ are not spontaneously broken, there are valid parent-daughter 
equivalences, which in turn implies daughter-daughter equivalence. 
There are infinitely many order parameters which can probe the breaking of 
either symmetry on $\R^4$.
Therefore,   the orbifold and 
orientifold equivalences of 
QCD-like theories are precisely on the same footing, and all their 
differences are cosmetic.}

Nevertheless, let us spell out  shortly  the symmetry realizations 
in QCD(AS/S). 
The effective potential $S_{\rm eff}$ in Eq.\ref{Eq:potorienti2} determines the 
symmetry realizations for  the spatial center symmetry $\widetilde \G_s$  and 
charge conjugation symmetry $\C$.   Just above $x_c$, the mass of the 
lightest $\C$
 odd mode   $\tr \Omega_{-}$ becomes negative and leads to a $\C$ and 
$\widetilde \G_s$  
breaking  instability. The even combination $\tr \Omega_{+}$ remains massive. 
In the limit 
$x \rightarrow 1$, the potential has a two-fold degenerate minima located at 
$\tr U = \pm i $, or  
\begin{equation}
\langle \frac{1}{N}\tr \Omega_{-} \rangle  = \pm 2 e^{i \pi/2}, \qquad 
 \langle \frac{1}{N}\tr \Omega_{+} \rangle  = 0
\end{equation}
spontaneously breaking the spatial center symmetry 
and charge conjugation symmetry to 
its diagonal subgroup.  
The meaning of this formula, in the basis of eigenvalues of the Wilson line 
is analogous  
to the orbifold example we examined. The two  clusters of 
eigenvalues  of orbifold theory is now replaced by 
 one cluster and its mirror image. The mirror image    is the effect of 
 complex conjugation.   The  cluster, therefore, analogous to 
orbifold example,  wishes to be maximally away  
from its image. (see fig.\ref{fig:wilson}.) 
The eigenvalue distribution may be   either 
$\rho(v)= \delta(v - \frac{\pi}{2})$ or  $\rho(v)= \delta(v + \frac{\pi}{2})$ 
depending on the choice of vacua. In vectorlike theories with unbroken 
$\C$ symmetry,   the eigenvalue distributions must satisfy
$\rho(v) = \C \rho(v) = \rho(-v)$.  For example, this is the case in  
$\None$ SYM, and  as well as other  QCD(adj).  However, QCD(AS/S) possess a 
phase in which 
 $\rho(v) \neq  \rho(-v)$, and $\C$ is spontaneously broken. 
The vacuum energy density in leading 
order in $N$  is particularly simple in the  $x\rightarrow 1$ limit, and 
 is given by 
\begin{eqnarray}
{\cal E}^{\rm QCD(AS/S)}(x \rightarrow 1) &&= 
-  \frac{1}{R_{S^1} V_{S^3}} \log \widetilde Z(x \rightarrow 1) =  
- \frac{\pi^2}{24 (R_{S^1})^4} N(N \mp \frac {7}{15})
\label{Eq:vacuumenergyAS} 
\end{eqnarray}
 The $N^2$ is a 
kinematical factor reflecting that there are $N^2$ gauge field and 
tensor fermions in QCD(AS/S).  In leading order in $N$, there is a 
factor of two  difference compared to  $U(N) \times U(N)$ QCD(BF), 
 reflecting the difference 
in the total number of degrees of freedom.

For $x< x_c$, the potential is positive definite, and provides a repulsive 
interaction among eigenvalues. Hence, the  eigenvalues 
are uniformly distributed, and the  
 expectation value of the Wilson line vanishes: 
$\langle \tr U \rangle = 0 $. 
The phase 
transition therefore restores both  
the  spatial 
center symmetry and the  charge conjugation symmetry at large radius of 
$S^1$ circle just like the orbifold QCD(BF) theory restores its 
$\G_s \times \I$.   The vacuum energy density 
is of order one  ${\cal E} \sim \O (1)$, and is due  to the 
fluctuations. Therefore, the theory undergoes a  phase transition 
which is associated with an abrupt change in its vacuum energy density from 
$\O(N^2)$ to being $\O(1)$, and triggered by quantum fluctuations rather than 
thermal ones. 

\subsection{Finite temperature phases}
\label{sec:orienti2}
The choice of antiperiodic boundary conditions for fermions on the $S^1$ circle 
corresponds to considering  QCD(AS/S)  at finite temperature.  
The resulting Euclidean functional integral corresponds to the  partition 
function $ Z= \tr e^{- \beta H}$  of a  thermal ensemble on  $S^3$ 
with inverse temperature $\beta$.  

As usual, the change in the boundary condition is reflected as an alternating  
coefficient of the fermionic term  in Eq.\ref{Eq:potorb2}.
\begin{equation}
z_f(x^n) \rightarrow  (-1)^n z_f(x^n) 
\end{equation}
therefore, turning the coefficients in the  effective action  Eq.\ref{Eq:oo}
into Eq.\ref{Eq:universal2}. The thermal effective potential is in terms of the 
temporal Wilson line and determines the temporal center symmetry realization 
$\widetilde \G_t$ along with charge conjugation symmetry $\C$. 

At low temperature (small $x$),   the expectation value of 
the  thermal Wilson line is identically zero, 
$\langle  \frac{1}{N} \tr U \rangle  = 0$. Therefore, 
the temporal center symmetry, 
and   charge conjugation symmetry are unbroken.  
The  free energy density is order one  reflecting the fact that spectral 
density of the color singlets remains $\O(1)$ in the large $N$ limit.
 This is the characteristic of the  confined phase.

As the temperature is  increased,  the mass of the 
$\C$-even mode $\tr \Omega_{+}$ becomes negative leading to a deconfinement
transition. This means, at high 
temperature, the temporal 
center symmetry $\G_t$ is broken, but not the charge conjugation 
symmetry.  In the very high temperature limit 
($x \rightarrow 1$),  the  order parameters are 
\begin{equation}
\langle  \frac{1}{N} \tr \Omega_{+}  \rangle  = \pm 2, \qquad
\langle  \frac{1}{N} \tr \Omega_-  \rangle  = 0, 
\label{Eq:deconQCDAS}
\end{equation}
and the breaking pattern is $ \{1, \widetilde \G_t\}  \times \{1, \C\} 
\rightarrow 
\{1, \C\}$.  In terms of 
eigenvalues  of the Polyakov loop,  this means all the eigenvalues clump 
to either at $v= 0$ or $v= \pi$.  These are the two thermal equilibrium 
states.
The  eigenvalue distribution 
is  given by $\rho(v)= \delta(v)$  or   $\rho(v)= \delta(v - \pi)$
both of which are even under the action of $\C$.   This means both thermal equilibrium state lies in the $\C$ even sector. 
From the partition function, 
The free energy density is particularly simple in the high temperature  
($x\rightarrow 1$) limit: 
\begin{eqnarray}
{\cal F}^{\rm QCD(AS/S)}(x \rightarrow 1)  &&= 
- \frac{\pi^2}{24} T^4  N(N \mp \frac {7}{15})
\label{Eq:freeenergyAS} 
\end{eqnarray}
the expected Stefan-Boltzmann result, $-\frac{\pi^2}{45} T^4  [
N^2 +  \frac {7}{8} N(N\mp 1) ] $

The fact that $\C$ is unbroken at high temperature should not be surprising. 
 The   expectation value of the Polyakov loop (in the 
fundamental representation)
and its charge conjugate 
corresponds to the excess of free 
energy of an external quark whose charge is  $+1 $   and  $-1$ respectively.
Since these charges differs by two,  in the sense of center symmetry of the 
theory, they are in the same conjugacy class as discussed in section 
\ref{sec:gen2}.
This follows from the observation 
that the dynamical  antisymmetric 
representation  quark has charge two under the center  $U(1)$ 
 of $U(N)$,  and can convert  the charge by an additive factor of two 
with no cost. Therefore, 
the dynamical quarks split the charges into two equivalence classes, 
the even ones equivalent to zero and the odd ones equivalent to one. 
Since the excess of 
free energy is a class function taking its values in the center 
symmetry of the theory,  we should have 
$\Delta {\cal F}_{(+1)} =  \Delta {\cal F}_{(-1)}$ implying 
$\langle \tr U \rangle - \langle \tr U^{*} \rangle =0$, and hence 
unbroken charge conjugation symmetry $\C$.

\subsection{Local order parameters}
\label{sec:orienti3}
The symmetries of the nonthermal 
orientifold QCD(AS/S) theory  are, as discussed above, 
 charge conjugation symmetry $\C$ and  spatial center symmetry 
$\widetilde \G_s $.  
The symmetry breaking pattern  discussed in 
section \ref{sec:orienti1} corresponds to 
\begin{equation}
(\Z_2)^2 = \{1, \C, \widetilde \G_s, \C \widetilde \G_s \} \rightarrow  
\Z_{2,D} =  \{1, \C \widetilde \G_s \}
\label{Eq:patternQCDAS}
\end{equation}
leading to two degenerate vacua. These two isolated vacua   are not invariant 
under the individual 
actions of $\C$ and $\widetilde \G_s$, but invariant under   the combined $\C
\widetilde \G_s$ action. 
In analogy with the  orbifold example where unbroken 
$\widetilde \G_s\I$ has nontrivial 
implications for  $\I$ odd local order parameters, unbroken 
$\widetilde \G_s \C $  
has   similar consequences for $\C$ odd local order parameters.

{\bf  Corollary:}  Let  $\O(x)$ be a  local order parameters  
 odd under charge conjugation 
symmetry $\C$. In any phase of the theory in which  $\C \widetilde \G_s$ 
is unbroken,  vacuum expectation value  of such operators vanish. 

Since  $\O(x)$ is  local, it  is singlet  under center symmetry.
Therefore,  we  have 
\begin{equation}
(\C \widetilde \G_s)  \O(x) =  \C  \O(x) =  -\O(x).
\end{equation}
Since  the
vacuum  of  the  theory  is  invariant  under  $\C \widetilde \G_s$,  
we obtain
\begin{equation}
 \langle \O(x)  \rangle =  \langle (\C \widetilde \G_s )\O(x))\rangle 
=  - \langle\O(x) \rangle=0
\end{equation}
as desired. 
Clearly,  the local  operators (which do not wind 
around the $S^1$ circle) are all singlets under the center symmetry. In any 
$\C \widetilde \G_s $ unbroken phase of the theory,  topologically trivial 
but $ \C$ odd order parameters transform nontrivially 
under the  action of  $\C \widetilde \G_s$.  Therefore, 
the vacuum expectation value of all 
such operators must vanishes. 

For thermal QCD(AS/S),   in the  high temperature  
deconfined phase, we found that the  breaking pattern is 
\begin{equation} 
(\Z_2)^2 = \{1, \C, \widetilde \G_t, \C  \widetilde \G_t  \} 
\longrightarrow  \Z_{2} =  \{1, \C \}
\label{Eq:thermalpatternQCDAS}
\end{equation}  
leading to two $\C$ respecting isolated thermal equilibrium states.
In the confined  phase,  the $ (\Z_2)^2 $ symmetry is unbroken. Therefore,  
 no 
 $\C $ odd operator (either local or nonlocal) 
can acquire an expectation value regardless of the phase of the thermal 
theory.

Needless to say, these assertions do not show that $\C$ in the case of
 orientifold QCD(AS/S)  and  $\I$ in the case of orbifold QCD(BF) 
 cannot be spontaneously broken on $\R^4$ via local order parameters. This, 
in principle is possible, but  unlikely.   
Nevertheless,  there is no theorem 
demonstrating that these two symmetries    cannot be 
spontaneously broken  on $\R^4$, unlike the case of parity  \cite{Vafa:1984xg}.

\subsection{On the $N$ dependence  of $\C$ breaking on $\R^3 \times 
({\rm spatial\; } S^1)$ }

For  the $SU(2N+1)$  QCD(AS/S) 
as well as   QCD with fundamental fermions, since the dynamical fermions can 
screen any external charge, 
the center symmetry is absent. Therefore,  $\C$ is  broken as  
 $\{1, \C \}  \rightarrow {1}$.  Unlike the $U(N)$  QCD(AS/S) theory, 
there is no symmetry reason for local order parameters not to acquire a vacuum 
expectation value. 
We expect in this particular class of 
QCD-like  gauge theories and choices of color gauge groups,
both local and nonlocal order parameters to  acquire 
nonzero expectation values in $\C$ broken phase. 
For $SU(3)$ gauge theory with order few  flavors, the two $\C$ 
 breaking vacua  are located at  $U= \exp{( \pm i\frac{2 \pi}{3})}$, and 
the charge conjugation symmetry is spontaneously broken. 
This is unambiguously  demonstrated in recent lattice studies  by 
DeGrand and Hoffmann \cite{DeGrand:2006qb}, who used a low temperature 
asymmetric lattice to probe $\C$  breaking at small volume. These authors 
also demonstrated that there is a $\C$ restoring transition taking place 
around the strong confinement scale of the theory.  
The more  recent 
paper \cite{Lucini:2007as} also examines  local order parameters 
in the  similar setup.  The authors of  \cite{Lucini:2007as} 
use  the   component of the baryonic current along the compact direction, and 
nicely demonstrate that, 
 in the $\C$ broken phase,  there is a persistent current  
correlated with imaginary part of the Wilson line $\tr \Omega_{-}$. 

We want to make few simple observations: Since the fundamental and 
antisymmetric representation of $SU(3)$ coincide,  QCD has few  natural large
 $N$ generalization.  One option is to use  fundamental flavors 
 in  the large $N$, 
another is to antisymmetric, or a mixture of the two. 
The dimensions of these two 
 representations scales differently with $N$, fundamental $\O(N)$ 
and antisymmetric $\O(N^2)$.  Therefore, for a   fixed number of 
flavors the former  is  kinematically suppressed in the large $N$  limit. 
Nevertheless, one can 
take $\frac{n_f}{N}$ for fundamental fermions,
 and keep the total number of fermionic degrees 
of freedom $\O(N^2)$ as well \cite{Veneziano:1976wm}.   
Interestingly, in the presence of the fundamental fermions in $U(N)$ 
gauge theory,  $\C$ do not get broken at small spatial $S^1$.
The unique minimum is  located at $U=-1$  which 
respects $\C$. (not at $U=1$  which  is the case for 
thermal compactification.)
For $U(N)$ gauge theory with
AS/S representation, 
the two minima  are located at $U= \exp{(\pm i \frac{\pi}{2})}$, hence breaks 
$\C$ spontaneously. 
In the case 
of $SU(N)$, the 
minima  of the effective potential is typically  $\O(1/N)$ vicinity of 
 $\pi$ for fundamental fermions and within $\O(1/N)$ vicinity of  
$\pm \frac{\pi}{2}$ for AS/S representation, see figure \ref{fig:wilson2}: 
\begin{equation}
\langle   \Im  \frac{\tr}{N} U
\rangle^{\rm QCD(fun)} = 0 + \O(1/N),  \qquad \langle   \Im  \frac{\tr}{N} U
\rangle^{\rm QCD(AS/S)}  = \pm1 + \O(1/N)  
\end{equation}
In other words, the $\C$  breaking effects in $SU(2N+1)$  QCD is suppressed 
with the use of fundamental  fermions, and is enhanced with AS/S 
representation fermions at large $N$.
  This is 
consistent with the fact that in large $N$  limit, any distinction between 
$U(N)$ and $SU(N)$ should disappear.   

\begin{FIGURE}[ht]
{
  \parbox[c]{\textwidth}
  {
  \vspace*{-30pt}
  \begin{center}
  \raisebox{3pt}{\includegraphics[width=4.0in]{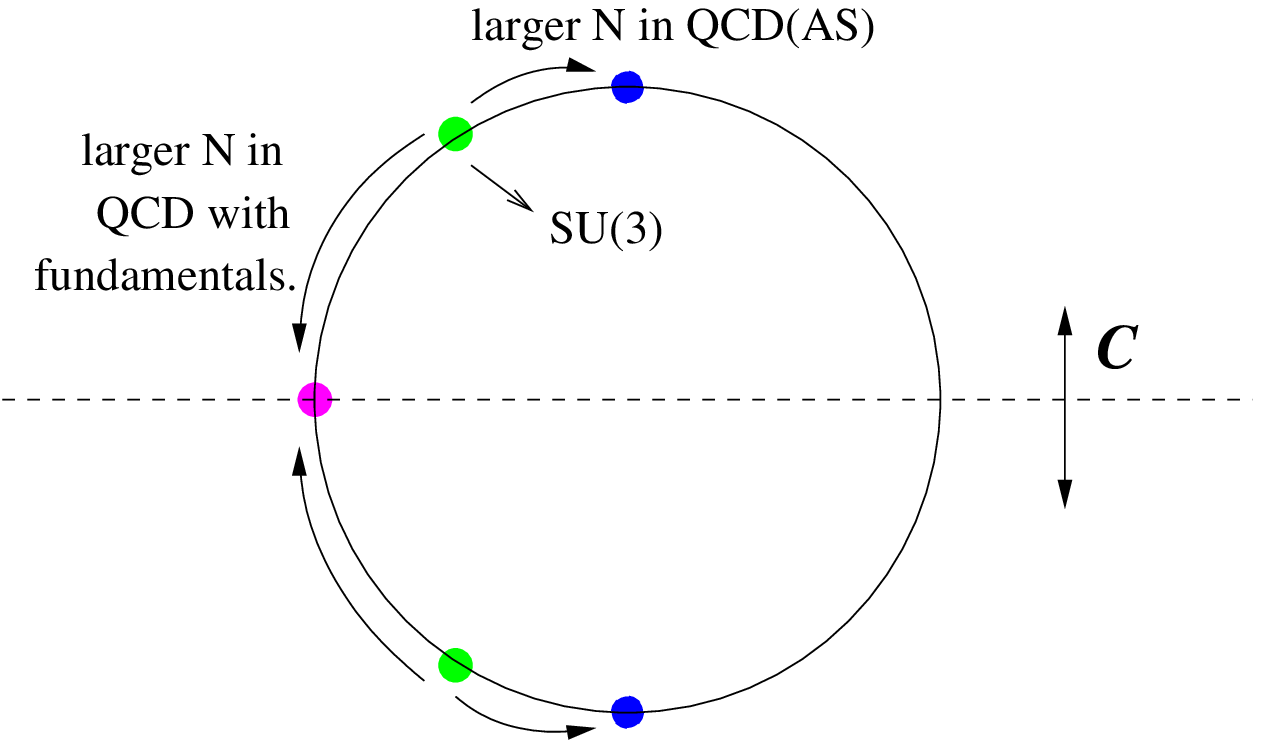}}
  \vspace*{-15pt}
  \caption
    {%
    The $N$ dependence of the $\C$-breaking in two natural generalization of 
    the 
    $SU(3)$ QCD on $\R^3$ times small spatial $S^1$.  
    For $SU(3)$ QCD with few fermions, 
    the two minima are located at  $e^{\pm i \frac{2 \pi}{3}}$. 
    For $SU(2N+1)$ QCD with fundamental fermions, the two minima 
   gradually approach each other with increasing $N$. The $N=\infty$  limit  
    has unique vacuum with unbroken $\C$.  For $SU(2N+1)$ QCD(AS),
   the two minima approaches to antipodal points $\pm e^{i\pi/2}$ and 
    $N=\infty$ theory breaks $\C$ spontaneously, just  like $SU(3)$.     
    }
  \end{center}
  }
\label{fig:wilson2} 
}
\end{FIGURE}

Both the suppression of the $\C$ breaking for fundamental fermions and the 
enhancement for AS/S fermions result in  (at first sight counter-intuitively) 
the suppression  of the vacuum expectation values of the 
  local order parameters.  
Let $ \frac{\tr}{N}  O(x)$  be a local order parameter 
probing $\C$. In QCD with fundamental fermions, this is natural immediate 
since  $\C$ breaking effects diminish with $N$.  
For QCD(AS/S), at sufficiently large $N$, the differences between $SU(2N+1)$ 
and   $U(2N+1)$ is an $\O(1/N)$ effect.  The $U(2N+1)$, unlike $SU(2N+1)$
enjoys a $\G_s=\Z_2$ spatial center symmetry. As shown in the corollary of 
section \ref{sec:orienti3}, the breaking of $\C$ and $\G_s$ are intertwined, 
and the two $\C$ breaking vacua respects $\C\G_s$ preventing any local order parameter from acquiring vacuum expectation value.
 Therefore, for $SU(2N+1)$, 
we must have  
\begin{equation}
\langle 
\frac{\tr}{N}  O(x) \rangle^{\rm QCD(fun/AS/S)}= 0+ \O(1/N)  
\end{equation}

\subsection{A remark on CPT and Vafa-Witten theorems}
A final remark  is  on the details of the symmetry breaking pattern   
Eq.\ref{Eq:patternQCDAS}, in 
 Minkowski space   
$\R^{2,1} \times S^1$ or its Euclidean continuation  $\R^3 \times S^1$.
The latter may be regarded as a zero temperature field theory in which one 
spatial direction is compactified. 
The analysis \cite{Unsal:2006pj} shows that the imaginary part of the Wilson 
loop is not just an order parameter for spatial center symmetry 
$\widetilde \G_s$ and charge 
conjugation  symmetry $\C$, it also monitors parity $\P$ and time reversal 
$\T$.  Therefore, the precise breaking pattern (in the small radius limit)
  for the theory is $(\Z_2)^4 \longrightarrow  (\Z_2)^3$:
\begin{equation} 
 \{1, \C \}\times \{ 1, \P\} \times \{1, \T\} \times \{1,  \widetilde \G_s\} 
\longrightarrow    \{1, \C \P,   \C \T ,   \C  \widetilde \G_s, \P \T, \P  
\widetilde\G_s, 
\T  \widetilde \G_s, \C \P \T  \widetilde \G_s  \}
\label{Eq:fullpatternQCDAS}
\end{equation}  
  where the action 
of even  number of operators leaves the vacuum that it acts invariant, and 
the action of  
odd number of them  switches the two vacua \cite{LGY}. 
In particular, the eight broken symmetries are 
\begin{equation}
  \{ \C , \P,   \T ,   \widetilde \G_s, 
\P  \T \C, \P  \widetilde \G_s \C, 
\T  \widetilde \G_s \P, \C \T  \widetilde \G_s  \} 
\end{equation}
The list includes 
spontaneously broken  parity $\P$, and $\C \P \T$  on 
$\R^{2,1} \times S^1$ at small radius. Does spontaneously broken CPT and 
P  on this (locally) four dimensional theory 
clash with CPT theorem or Vafa-Witten theorem for parity \cite{Vafa:1984xg}?  

The answer is, no. 
CPT assumes the full Lorentz symmetry, which is lost by 
compactification.  The parity argument is based on local order parameters 
on $\R^{3,1}$, in particular, our topologically nontrivial 
 order parameters such as the 
imaginary parts of Wilson lines only emerge upon compactification. 
\footnote{For the $SU(2N+1)$  QCD(AS/S) 
as well as   QCD with fundamental fermions, the breaking pattern 
is different due to the absence of the center symmetry. In particular, we 
expect 
$
 \{1, \C \}\times \{ 1, \P\} \times \{1, \T\}  
\longrightarrow    \{1, \C \P,   \C \T ,  \P \T  \}.
$
Therefore, we expect local \P-odd  order parameters to  acquire vacuum 
expectation value within this phase  
for this specific class of vector-like gauge theories. } 
  Therefore, 
there is no clash  between broken CPT and broken P for QCD-like theories 
formulated on  $\R^{2,1} \times S^1$ and the CPT and P theorems which are 
formulated on $\R^{3,1}$. However,  it would be incorrect  
to make a general statement such as in vector-like  
gauge theories, CPT (or P) cannot be spontaneously 
broken. In fact, for parity, the inapplicability of the 
Vafa-Witten argument, which is originally formulated for gauge theories on 
$\R^4$ \cite{Vafa:1984xg},  
in   finite volumes  is discussed in ref.\cite{Cohen:2001hf}. Even though 
Ref.\cite{Cohen:2001hf} did not demonstrate the existence of a P broken phase, 
it clearly established that there is  no fundamental  principle
 which guarantees  unbroken  P in  finite volume.  
The existence of P broken  
phase in  $U(N)$  QCD(AS/S)  at any $N\geq 3$ \cite{Unsal:2006pj} , 
(or  usual $SU(3)$ QCD with fundamental 
fermions \cite{vanBaal:1988va,vanBaal:2000zc})
justifies 
the argument of   ref.\cite{Cohen:2001hf} by explicit demonstration. 
\footnote{We thank 
Tom DeGrand for pointing the refs.\cite{vanBaal:1988va,vanBaal:2000zc} and
 Misha Shifman for 
bringing the ref.\cite{Cohen:2001hf} to our attention.}
Our discussion shows that, not only the P theorem,  but also CPT theorem is  
inapplicable in finite volume. Both are broken at small spatial 
volume, and should get restored around the the physical 
strong scale of the QCD-like theory.  
The existence of 
this zero  temperature phase transition 
 has been confirmed in recent lattice studies of by DeGrand and Hoffmann 
\cite{DeGrand:2006qb}.

\subsection{Chiral symmetries}
\label{sec:orienti4}
The discussion of the chiral properties of the QCD(AS/S) theory on 
$S^3 \times  S^1$ are verbatim identical 
to the orbifold QCD(BF) theory.  This is true in both 
when  $S^1$ is a thermal circle or just a spatial circle.  
The only difference is in $\O(1/N)$ discrepancy  in the number of vacua.  
In the case of QCD(AS/S), the 
chiral symmetries are 
 $\Z_{2N-4}$ and  $\Z_{2N+4}$, respectively. In the regime 
where  the radii of $S^3 \times S^1$ are large compared to the strong scale, 
these discrete chiral symmetries are spontaneously broken to $\Z_2$, leading to 
$N-2$ and  $N+2$  isolated vacua.  The difference between broken axial 
symmetries $\Z_h$  where   $h= \{ N-2, N+2, N, N \}$ for 
$U(N)$  orientifold QCD(AS/S), $\N=1$  SYM theories and  $U(N) \times U(N)$ 
QCD(BF)  is unimportant in  the large $N$ limit. 

The 
 broken discrete chiral symmetry group  also 
counts the number of ground states. In $U(N)$ SYM theory and  
$U(N)\times U(N)$ QCD(BF), this number is $N$,  and  for QCD(AS/S), it is 
respectively, $N\mp2$.  From the view point of Ref.\cite{Barbon:2005zj}, this 
is a property of nonsupersymmetric QCD(AS/S/BF) which may  
be related to the  index $\tr (-1)^F e^{-\beta H}$ of supersymmetric theory, 
at large $N$. Clearly, there is a  match for the orbifold QCD(BF)   and 
corrections of $\O(1/N)$ for QCD(AS/S).  However, it should be kept in mind 
that  each of  QCD(AS/S/BF) formulated on $\R^3 \times S^1$ (where $S^1$ is 
spatial, implying the theory is at zero temperature) 
undergo  a zero temperature chiral phase transition. This does not happen in 
$\N=1$ SYM. (Ref.\cite{Barbon:2005zj} examined these theories on 
$T^3 \times \R$ where $T^3$ is small, and  
$\C$  in QCD(AS/S) and $\I$ in QCD(BF)  are spontaneously broken. Since the 
necessary symmetry realization conditions do not hold in this phase, 
we do not expect an equivalence. On the other hand, 
for large $T^3$, we expect a 
restoration of $\C$  and $\I$, and consequently the matching of the number of 
the ground states as described above.)    
A prediction of the large $N$ orbifold and orientifold equivalence is that 
the condensates of these theories should coincide in the large volume limit. 
This is demonstrated to be so in lattice simulations \cite{DeGrand:2006uy} for 
QCD(AS/S) and is still undemonstrated for QCD(BF).
For more on  one flavor QCD-like 
theories, see the recent discussion \cite{Creutz:2006ts}.

\section{Multiflavor generalization,  $n_f$  universalities }
\label{sec:multiflavor}
We may generalize the entire previous discussion of vectorlike,
 asymptotically free gauge theories  to the case of multiple fermion flavors. 
Preserving  the asymptotic freedom restricts the number of 
flavors to at most five, $n_f \leq 5$. 
Adding multiple adjoint representation fermions to SYM theory 
will yield a non-supersymmetric $U(N)$ gauge theory with 
$n_f >1$ adjoint Weyl (or Majorana) fermions, which we referred to as 
QCD(adj). The other theories of interest are QCD(AS/S/BF) with $n_f >1$ Dirac 
fermions.  For a fixed  $n_f$,  these theories are related to  one 
another via a web of orbifold and orientifold projections.  If 
$\C$ and $\I$ are unbroken in  QCD(AS/S) and QCD(BF) 
respectively,  we expect complete equivalence of thermodynamics and phase 
diagrams of these theories to QCD(adj) in  the $N=\infty$ limit 
\cite{Kovtun:2004bz}.   

As usual, we may write the thermal partition function 
$Z= \tr e^{-\beta H}$  and the twisted partition function  $\widetilde 
Z= \tr e^{-\beta H} (-1)^F$ for these theories.
Installing appropriate changes in the discussion of effective actions (for 
spatial and temporal Wilson lines) just means restoring the  
fermionic multiplicity:
\begin{equation} 
z_f(x^n) \rightarrow n_f z_f(x^n).
\end{equation}
The phase diagrams of the  vector-like  gauge  theories 
with $n_f>1$ just mimics their 
$n_f =1$ counterparts and    
one qualitatively reaches identical conclusions in the sense of spatial and 
temporal center 
symmetry,  chiral symmetry, 
  charge conjugation symmetry of orientifold QCD(AS/S) 
and interchange symmetry in orbifold  QCD(BF) theories. 
The phase diagrams shown in 
Fig.\ref{fig:phase}  qualitatively   depict  the phases of both 
(nonthermal and thermal) QCD(adj). Analogously,    
Fig.\ref{fig:phase2} describes  the phases of QCD(AS/S/BF).

To make the phase diagrams as quantitatively clear as possible (at least in the 
perturbative regime of phase transitions), let us understand how the  
spatial  and temporal  center symmetry changing phase boundaries move with 
$n_f$. 
In the perturbative regime of the vector-like theories on $S^3 \times S^1$,   
the temperature of the confinement deconfinement transition  (changing the
 temporal center symmetry realization)  and the radius of $R_{S^1}$ at which  
spatial center symmetry realizations change  can be calculated. 
At leading order (in the coupling constant), it is given by 
\begin{eqnarray}
&&\beta_d^{\rm QCD(AS/S/BF/adj)} =R_{S^1, s}^{\rm QCD(AS/S/BF)} = 
\{1.66,\; 1.88,\; 2.04, \;  2.17, \; 2.27 \} \; 
R_{S^3} ,  \cr 
&& R_{S^1, s}^{\rm QCD(adj)} = {\rm none}
\qquad {\rm for}\;\;
 n_f=1, \ldots 5 
\label{Eq:radius}
\end{eqnarray}
respectively.   As $n_f$ increases, the slope of the center symmetry changing  
phase boundary shown in Fig.\ref{fig:phase3} increases.  
Eq.\ref{Eq:radius} is true within 
the domain of validity of the  one loop analysis,
and can only  be altered in small amounts 
by higher order perturbative corrections. 
However,  in the  $\R^3 \times S^1$ limit of these QCD-like theories, 
currently, there are no analytical tools which may be used to determine  the 
radius of (spatial or temporal) $S^1$ at which a phase transition occurs.  
The change in temporal center symmetry realization is associated with 
the  deconfinement confinement transition. 
  The results, based on 
numerical   lattice simulations unambiguously demonstrate \cite{Kogut:1982rt} 
that the transition 
occurs around the strong confinement scale of the theory. (For $SU(3)$, 
numerically, this is reported as $\sim 200 {\rm Mev}$.) 
More recently, the spatial center symmetry realization of the QCD(AS/S)  is 
 examined  on  lattice simulations  and  spatial center symmetry 
changing transition for QCD is observed 
around the strong scale of the theory 
\cite{DeGrand:2006qb, DeGrand:2006uy}. We should also note that, QCD(adj) does 
not undergo a spatial center symmetry changing transition on $\R^3 \times S^1$ 
limit.  At small $S^1$, there is indeed a one-loop effective potential (for  
$2\leq n_f \leq 5$)  
 which, in effect, generates a repulsive interactions among the eigenvalues of 
the Wilson line just like a nonperturbative potential in $\N=1$ SYM does.  
Consequently, all QCD(adj) theories, whether they are supersymmetric ($n_f=1$) 
or not ($n_f \neq 1$), 
shows identical behaviour in the sense of $\G_s$. 

Eq.\ref{Eq:radius}, true within the perturbative regime of these theories on 
$S^3\times S^1$, 
does not  demonstrate that the confinement/deconfinement transition temperature 
should  coincide in the nonperturbative regime on $\R^3 \times S^1$. 
On the other hand, generic  correlators of these theories are
 dependent on the phase. 
As long as the phase transitions in the  QCD(adj) and 
QCD(AS/S/BF) theories are {\it not} driven by the  
$\C$ (or  $\I$)  odd order parameters 
but anything else, the phase diagrams 
of these vectorlike theories must  coincide precisely based on nonperturbative 
large $\Nc$ orbifold/orientifold  equivalence. \footnote{Recall that the nonperturbative large $N$ equivalence is valid provided the ground  (or thermal 
equilibrium) states of both theories lie in their neutral sector. If a $\C$ or 
  $\I$ odd order parameter triggers a phase transition, then the ground 
(or thermal  equilibrium) states of the corresponding theories will move 
to the  non-neutral sector. In the thermal phase diagram of QCD(AS/S/BF/Adj), 
all three phases   lie in the neutral sector, and transition are driven 
by neutral sector operators. In contrast, 
in the nonthermal  QCD(AS/S/BF/), there exist a phase in which $\C$ and $\I$ 
are  broken, and the vacua are in the non-neutral (twisted) sector.} 
In particular, this implies the strong coupling phase transition on 
$\R^3 \times S^1$ must occur at the same temperature in QCD(AS/S/BF/adj).  
This is a prediction of large $\Nc$ orbifold equivalence  which may be tested 
on the lattice. 

A watered-down version of why  such universal behaviour is plausible 
may  be surmised in perturbation theory.   
All of our vectorlike theories with a given number of flavors $n_f$ possess 
identical   
renormalization group  beta functions (for the 't Hooft couplings) 
in the  $N= \infty$  limit.  Let 
$\alpha(\mu) = \lambda(\mu) / {4 \pi} $, then 
\begin{equation}
\frac {d \; \alpha }{d\log \mu} = - \frac{b_0}{2 \pi} \alpha^2 -    
\frac{b_1}{ 4 \pi^2} \alpha^3 + \O(\frac{1}{N}, \alpha^4),  \qquad {\rm where} \;\;
b_0= \frac{11}{3} - \frac{2}{3}n_f, \qquad   
b_1= \frac{17}{3} - \frac{8}{3}n_f, \qquad
\label{Eq:pert}
\end{equation}
The  typical corrections in renormalization group flow of couplings is 
$\O(1/N)$ for QCD(AS/S) and $\O(1/N^2)$ for 
QCD(BF). In the large $N$ limit, these  subleading terms 
 can be safely neglected. (See  \cite{Armoni:2004uu}, for example.)
Therefore, within a specific perturbative 
scheme for the QCD(AS/S/BF/adj)  theories with a given number of flavor $n_f$,  
the scale at which the theory becomes strong is 
independent of what the underlying theory is. This is an
 obvious manifestation of the perturbative matching 
between the orbifold and orientifold partners. 
The result also reflects that the scale at which these theories
 become strongly coupled, say larger than 
$\alpha \geq 1/4$, only depends on  $n_f$  and not up on 
the  color quantum number $\cal R$ of the fermions.  (So long as fermions 
are in one of the four double index representations AS/S/BF/adj.)

Let us assume for 
simplicity that    the theories at some large scale, say $M$ (the lattice 
cut-off),  are assigned 
small couplings, $\alpha_i(M) = \alpha (M)$ independent of the 
number of flavors 
$n_f$ in the theory. Then there is an exponential hierarchy between the scales 
at which strong coupling takes over. Let $Q^{*}_{n_f}$ denote the energy scale 
 where the coupling constant becomes large, 
and the perturbative analysis is no longer reliable. 
To lowest order in perturbation theory, we find 
\begin{equation}
\frac{Q^{*}_{n_f}}{Q^{*}_{n_f=1}}=  
\left[\frac{Q^{*}_{n_f=1}}{M}\right]^{\frac{2 n_f -2 }{11-2n_f}}
\end{equation}
Since $ \frac{ Q^{*}_{n_f=1}}{M}  \ll 1$, the scale at 
which $n_f > 1$ 
flavor QCD 
theory becomes strongly coupled is parametrically large  in length 
(small in energies)
 with respect to   QCD $n_f=1$. Therefore,   physical quantities 
such as hadron sizes,  deconfinement radius (inverse temperature)  
in QCD with $n_f>1$  
compared to  QCD with $n_f=1$ are   larger. This also implies a larger 
region of validity of perturbation 
theory with the increasing number of flavors. The 
cartoon of the $n_f$ dependence of the phase boundaries for $n_f =1$ and 
$n_f=5$ is illustrated in Fig.\ref{fig:phase3}.  

The phase diagram Fig.\ref{fig:phase3}  also demonstrates the 
chiral symmetry realizations. For thermal QCD(AS/S/BF/Adj), the chiral symmetry 
is unbroken at  high curvature (small $S^3$)  or at high temperature 
(small thermal $S^1$).  
This is also true for the nonthermal compactification of  
QCD(AS/S/BF) where $S^1$ is spatial. The QCD(adj) endowed with 
periodic boundary conditions for fermions  
is simply different.  On large $S^3$, 
(to zeroth order, the theory on  $\R^3 \times S^1$), 
 we expect broken  chiral symmetry 
regardless of the the size of the spatial $S^1$, just like 
$\None$ SYM. 

In the sense of presence or absence of 
chirally symmetric and asymmetric phases, 
Fig.\ref{fig:phase3}  provides a fair 
caricature.  However, 
unlike the $n_f=1$ case in which only discrete chiral symmetries are present, 
the    $n_f >1 $ also brings in continuous chiral symmetries. The spontaneous 
breaking of continuous chiral symmetries are associated with the Goldstone 
bosons. Below, we will discuss the discrete and continuous 
chiral symmetry breaking patterns, the associated domain walls and 
Goldstone bosons.  

\begin{FIGURE}[ht]
{
  \parbox[c]{\textwidth}
  {
  \vspace*{-30pt}
  \begin{center}
  \psfrag{inf}{$\infty$}
  \raisebox{3pt}{\includegraphics[width=3.0in]{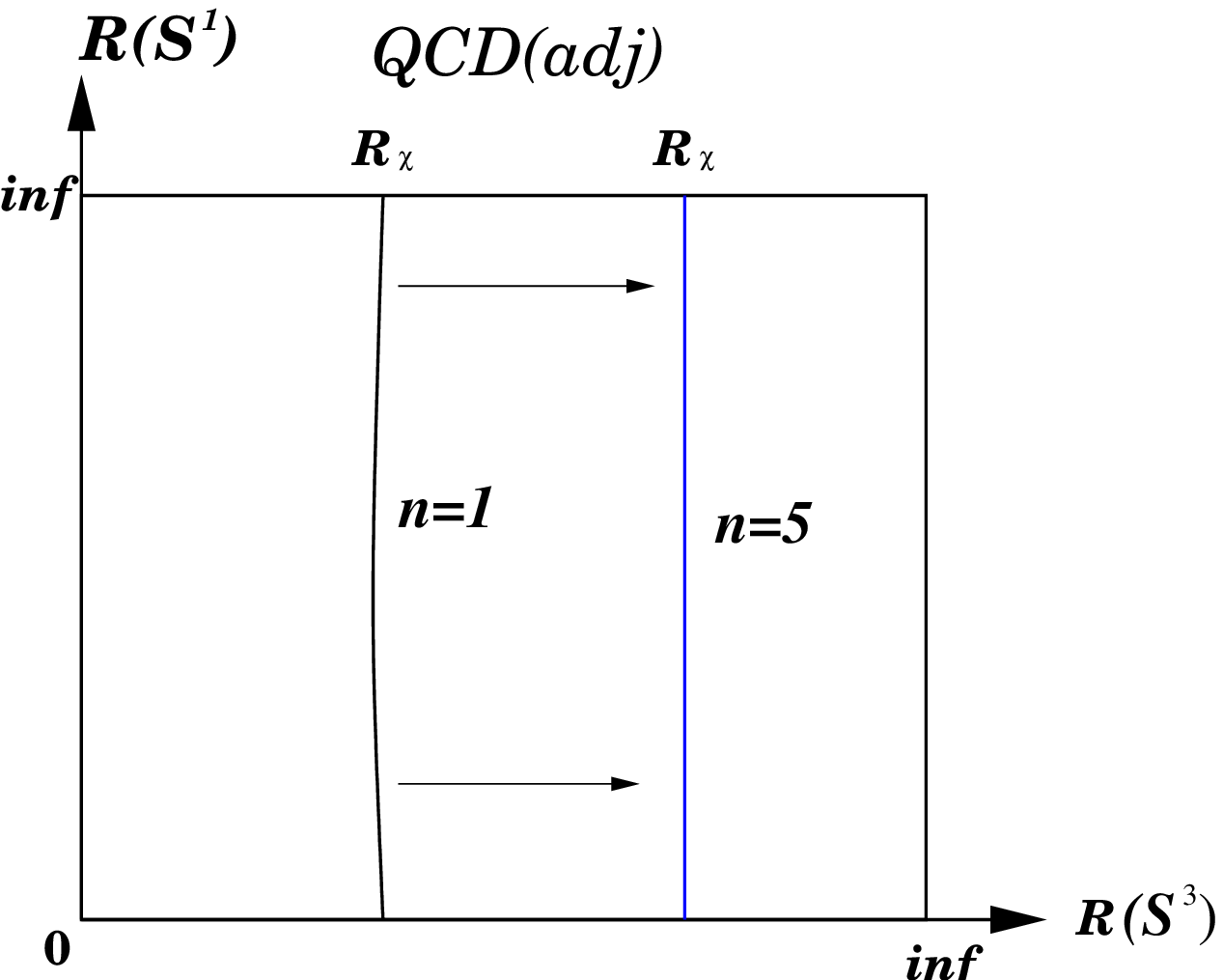}}
  \includegraphics[width=3.0in]{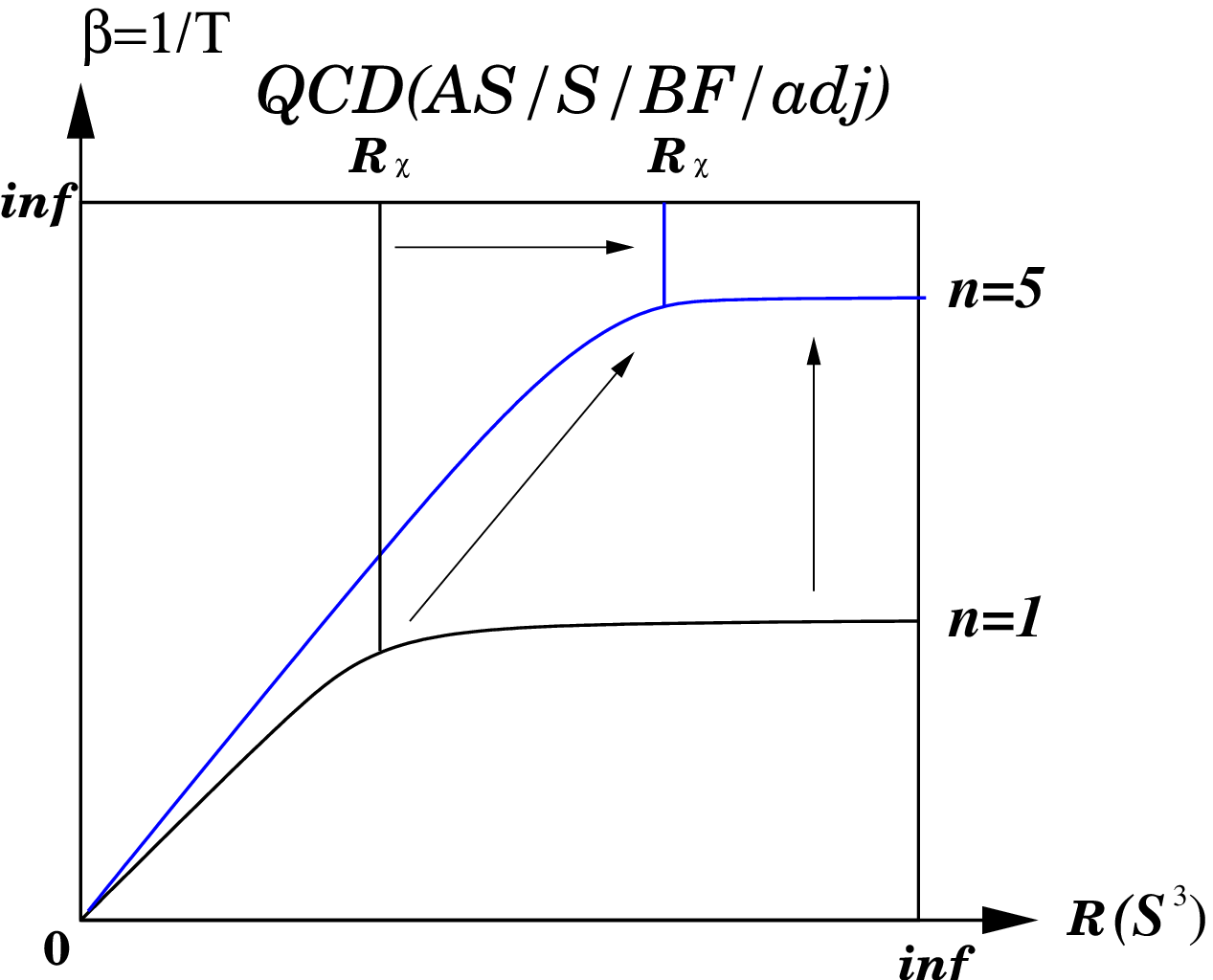}
  \vspace*{-15pt}
  \caption
    {%
    Multiflavor generalization of the phase diagram given in 
    Fig.\ref{fig:phase} 
    and Fig.\ref{fig:phase2}. The left figure  depicts how 
    the   addition of multiple adjoint  fermion 
     flavors alters QCD(adj) with periodic boundary conditions by gradually 
     pushing the 
    chiral transition radius to larger values. 
    The right figure may be thought of as corresponding 
     to  multiflavor QCD(AS/S/BF/adj)  with thermal boundary conditions or 
    QCD(AS/S/BF) with periodic boundary conditions.   
    As seen in the figure, the net effect of adding flavors is to increase 
    the slope of the phase transition line (associated with the appropriate 
    center symmetry realization) in the perturbative regime and to 
   push the nonperturbative  
   strong confinement length  to  higher values (or to lower energy scales).  
    If the theory were genuinely conformal,  then there would not be such a 
    scale 
    and the confinement deconfinement phase boundary 
     would not bend down around any 
    particular scale. The phase boundaries  for $n_f=2,3, 4$ should be between 
    $n_f=1$ and $n_f= 5$ in increasing order. (The $n_f \geq 6$ theories are
     not 
    asymptotically free and is not examined in this paper.)
    }
  \end{center}
  }
\label{fig:phase3} 
}
\end{FIGURE}

\subsection{Chiral  condensates  in  QCD(adj)}
The    QCD(adj) theory with $n_f$  flavor Majorana fermions  
possess both continuous and discrete global symmetry given by 
\begin{equation} 
\bigl(SU(n_f) \times \mathbb Z_{2n_f N} \bigr) \; /  \; 
\mathbb Z_{n_f} 
\end{equation}
The adjoint fermions, under $SU(n_f) \times  \Z_{2n_f N}$  transform as 
(suppressing the flavor indices)
$ \psi \rightarrow  V \psi, \;\; \psi \rightarrow  e^{i2\pi/(2N_cn_f)} \psi$ 
where $V \in SU(n_f)$. Since 
the $\Z_{n_f}$ subgroup of the 
 discrete chiral rotation can be undone by an element of the center of 
$SU(n_f)$, it should be modded out to prevent double counting. 

The  chiral  symmetry  is expected to break down to 
$SO(n_f)$ (times $\Z_2=(-1)^F$ if $n_f$ is odd. For even $n_f$, $(-1)^F$ is 
an element of  $SO(n_f)$.)  
by  the formation of the  fermion bilinear condensate 
$\langle \tr \psi_i\psi_j\rangle $, a symmetric tensor of $SU(n_f)$. 
This gives rise to a vacuum manifold with $\Nc$ disjoint components (islands),
each of which is the coset space $SU(n_f)/SO(n_f)$.  
The breaking pattern of the chiral symmetry is 
\begin{equation}
 (SU(n_f) \times \Z_{2\Nc n_f})/ \Z_{n_f} \rightarrow SO(n_f) (\times \Z_2)
\end{equation}
This is an interesting pattern which shows both discrete and continuous 
chiral symmetry breaking. The continuous symmetry breaking gives rise to 
$\frac{n_f(n_f+1)}{2} -1 $ Goldstone bosons. The discrete symmetry breaking
is responsible for the $N$ isolated vacuum manifolds 
and the domain walls interpolating 
among them.  
The determinant of the condensate, 
which is insensitive to flavor rotations, 
 distinguishes the vacuum manifolds which cannot be continuously connected 
to each other via flavor rotation.  
\begin{equation}
\arg \det \langle \tr \psi_i\psi_j\rangle   = \frac{2 \pi }{N} k \in \Z_N, 
\qquad  k= 0, \ldots, N-1 
\end{equation}
Notice that for the $n_f=1$ case, the  $\Nc$ 
isolated coset spaces 
 shrink to $\Nc$ isolated  points, and 
we recover the  well-know result for $\N=1$  
SYM  theory  where there are no Goldstones,  just $N$ isolated vacua 
with   domain walls interpolating among them.   It would be interesting to 
study the domain walls in QCD(adj) with $2 \leq n_f \leq 5$ in more detail. 

At sufficiently large $S^3$, where the base space 
may be  approximated by $\R^3 \times S^1$, the chiral condensate must  be 
independent of $S^1$ radius. This large $\Nc$ volume independence  
is discussed in detail  in \cite{Kovtun:2007py}, and 
the necessary and sufficient condition for its validity is unbroken 
center symmetry. This condition is satisfied in spatial compactifications 
of QCD(adj) at any $S^1$. In   thermal compactifications, the unbroken center 
symmetry also implies a confined phase, therefore,  the chiral 
condensate has to be independent of the temperature in the confined phase
\cite{Kovtun:2007py}.   

However, we do not expect the chiral condensate to be independent of $S^3$ 
radius.  The technical reason for this, 
 we cannot formulate the volume change 
of sphere $S^d, d\geq 2 $ as an orbifold projection. This is unlike the case 
of the d-torus $T^d$, including $S^1$\cite{Kovtun:2007py}.  
  In fact, if $S^3$ is 
sufficiently small, we expect a chirally symmetric phase. 
The independence of the chiral symmetry realization from the $S^1$ radius 
(shown in the left panel of Fig:\ref{fig:phase3}) in 
QCD(adj) (including SYM) is particularly interesting. 
Keeping $S^1$ small,  and   
varying $S^3$ from small radius all the way to $\R^3$,  we observe that 
a chiral transition should takes place in a regime 
where the theory is perturbative at the scale of  $S^1$. 
It may therefore be possible to investigate this chiral transition 
by using perturbative techniques, combined with the effective field theory 
considerations of ref.\cite{Pisarski:1983ms}.  This is left for 
the future work. 

\begin{FIGURE}[ht]
{
  \parbox[c]{\textwidth}
  {
  \vspace*{-30pt}
  \begin{center}
  \raisebox{3pt}{\includegraphics[width=2.0in]{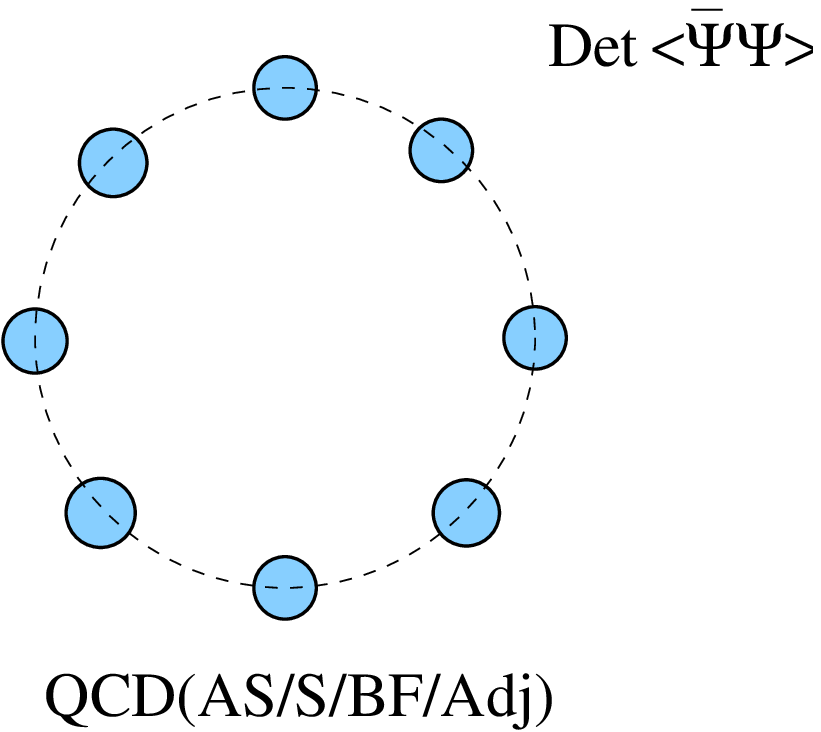}}
  \vspace*{-15pt}
  \caption
    {%
The vacuum structure of the $U(N)$  QCD-like gauge theories 
for  $1 \leq n_f \leq 5$.
The islands  represents  coset spaces associated with continuous 
symmetry breaking. There are $h= N-2, N+2, N, N$ such islands  
(corresponding to QCD(AS/S/BF/adj) respectively)  distinguished 
by the phase of the determinant of the condensate, which is a $\Z_h$ valued 
object.  Since these theories show both discrete and continuous 
chiral symmetry breaking, there exist both Goldstone bosons (for $n_f >1$ ) 
and domain walls $n_f \geq 1$. Notice that   the coset spaces shrink to  
points in the case $n_f =1$, and the Goldstone bosons disappears. 
    }
  \end{center}
  }
\label{fig:islands} 
}
\end{FIGURE}

\subsection{Chiral condensates in  QCD(AS/S/BF)} 
Let us consider the chiral symmetry breaking pattern of multi-flavor 
QCD(AS/S/BF) theories. These are all vectorlike gauge theories 
with $n_f$ two index complex representation fermions.  Even though the color 
gauge structures of these theories are different, their chiral properties 
are  almost identical.   Here, almost means up to $\O(1/N)$.   
At the classical level,    both 
 QCD(AS/S) and   QCD(BF) 
 possess a $U(n_f)_L \times  U(n_f)_R$ global symmetry. 
At the quantum level, because of the chiral anomaly, the  symmetry reduce to  
$SU(n_f )_L \times SU(n_f )_R \times U(1)_B \times 
\mathbb \Z_{2 h n_f }$ where 
\begin{equation}
h=N-2, N+2, N  \;\; {\rm for \;\; QCD(AS/S/BF)}
\end{equation} 
respectively. The difference in the large $N$ limit 
is a negligible $\O(1/N)$ in the discrete chiral symmetry. 
  
By factoring out the  doubly-counted symmetries, we obtain 
\begin{equation}
\frac{ SU(n_f)_L \times SU(n_f)_R \times U(1)_B \times
\Z_{2hn_f} }{ \left( \Z_{n_f} \times \Z_{n_f} \times \Z_2 \right)} . 
\end{equation}
The $\Z_2$  is  common in  $U(1)_B$ and $\Z_{2hn_f}$. 
One  $\Z_{n_f}$ is common to  the center of the axial  $A=L-R$
 and 
$\Z_{n_f}$ subgroup of axial $\Z_{2hn_f}$. 
The other is common to the center of the vector $V=L+R$ 
 and $\Z_{n_f}$ subgroup  of vectorial $U(1)_B$. 
It is expected, when $\min(R_{S^3}, R_{S^1}) \gg \Lambda^{-1}$ , 
this symmetry should be spontaneously broken by the formation of a 
fermion bilinear condensate down to 
\begin{equation}
\frac{SU(n_f)_V \times U(1)_B 
\times \Z_2} {  ( \Z_{n_f} \times \Z_2 )}.  
\end{equation}
This leads to  a vacuum manifold with $h$
components each of which is the coset space
$[SU(\Nf)_L \times SU(\Nf)_R]/ SU(\Nf)_V$. The $h$ condensates are 
distinguished 
by the phase of the determinant of the chiral order parameter, 
\begin{equation}
\arg \det \langle \tr \bar \Psi^i \Psi^j \rangle = 
\frac{2 \pi }{h} k  \in \Z_h, \qquad  k=0, 1, \ldots h-1 
\end{equation}
The  continuous chiral symmetry breaking leads to $n_f^2 -1$ 
Goldstone bosons. The  discrete chiral symmetry breaking is responsible for the 
existence of the domain walls interpolating between different components of the 
vacuum manifolds.  Notice that in the $n_f=1$  case,  
there are no Goldstone bosons 
and the breaking pattern correctly reproduces the result derived for one-flavor 
theories.

In the context of large $N$ orbifold and orientifold equivalences, 
there has been a 
certain amount of  confusion  about the mismatch of the number of 
Goldstone bosons  between QCD(adj) and QCD(AS/S/BF). 
These points are clarified  in  
\cite{Kovtun:2005kh} and \cite{Unsal:2006pj} where it is shown that 
the equivalence only applies  to symmetry invariant channels: In the case of 
the orientifold, this corresponds to $\C$-invariant subsector of QCD(AS/S) 
 and  in the case of orbifolds,  to the 
$\I$ interchange symmetry invariant channel 
of the orbifold QCD(BF) theory.  
Therefore, we have to grade the Goldstone bosons 
according to the $\C$ symmetry of orientifold  
and  $\I$ interchange symmetry of 
orbifold theory.   
The currents which may generate the corresponding Goldstone bosons are  
given by
$J^{\mu5}_a = \bar \Psi^i\gamma^{\mu} \gamma^{5} (t_a)_{ij} \Psi^j$ where 
$t_a$ are generators of the $SU(n_f)$ algebra.  The even 
(odd) graded Goldstone can be generated by 
the currents $J^{\mu5}_a$ for which  $t_a$ is a symmetric (antisymmetric) 
matrix.   Let $\Z_2$ represent either $\C$ or $\I$. The $\Z_2$ eigensystem 
of currents is given by  
\begin{equation}
 \Z_2: J^{\mu5}_a =\left \{ \begin{array}{ll}
&  - J^{\mu5}_a, \;\;   {\rm for } \;\; t_a \in SO(n_f) \cr  
&  + J^{\mu5}_a,  \;\;   {\rm for } \;\;t_a \in SU(n_f)/SO(n_f) 
\end{array} 
\right.
\end{equation}
Therefore, the odd Goldstones reside in the subalgebra $SO(n_f)$ 
and the even ones in the complement $SU(n_f)/SO(n_f)$. 
In QCD(AS/S/BF),  the currents corresponding to an
$\half n_f(n_f+1) -1$ of the Goldstone bosons 
 are $\Z_2$ even  while those corresponding to     $\half n_f(n_f-1) $  
of them are 
$\Z_2$ odd.  
The total number of $\Z_2$ even Goldstones in QCD(AS/S/BF) 
correctly reproduces 
the total number  Goldstone bosons in QCD(adj). 

Since the chiral dynamics are  dictated by the symmetries of the 
underlying theory, and the orbifold QCD(BF) and QCD(AS/S) theories possess 
identical chiral symmetry, their chiral Lagrangian  and dynamics 
should  coincide.  (This assumes unbroken $\C$ and $\I$ on $\R^4$, both of 
which are unproven.)
Recall that in the previous  sections, we have demonstrated  that 
 the  effective actions describing the spatial and temporal 
center symmetry realizations of QCD(AS/S) 
and QCD(BF) maps to  identical problem in terms of $\C$ and $\I$ 
eigenstates, respectively.   
Therefore, for a given 
$n_f$,  the infrared physics of the  QCD(AS/S) and QCD(BF) seems to be 
remarkably close 
on $S^3 \times S^1$ at arbitrary radius in the  large $N$ limit. We should 
state that  our point of view on the orbifold QCD(BF), and orbifold 
equivalences by and  large,   is opposite to 
the one presented so far in literature 
\cite{Armoni:2004uu,Armoni:2005wt}. In particular, we are highly optimistic 
that orbifold equivalences  are  useful source 
in understanding the dynamics of large $N$ QCD-like gauge theories. 

\section{Structure of  Large N limits, supersymmetry,
  and nonperturbative large  N  equivalence.
 }
\label{sec:why}

The study of the dynamics of vectorlike gauge theories reveals that certain 
aspects of these theories  become indistinguishable in the $N=\infty$ limit.
  In particular, for a given number of flavors $n_f$, the bosonic 
($(-1)^F$ even) subsector 
of QCD(adj) becomes indistinguishable from the $\C$-even 
subsector of QCD(AS/S) and the $\I$ even subsector of QCD(BF) in cases 
where the respective $\Z_2$ symmetries are unbroken.   This is a consequence 
of the nonperturbative orbifold and orientifold equivalence proven rigorously 
by using lattice regularization  in \cite{Kovtun:2003hr, Kovtun:2004bz}.

The nonperturbative large $N$ equivalence is valid provided the ground  
(or thermal  equilibrium) states of orbifold and orientifold 
partners  lie in their respective neutral sectors. 
In the thermal phase diagram of QCD(AS/S/BF/Adj), 
in all three phases, the ground (or thermal equilibrium)  states   are 
in the neutral sector, and all the order parameters of these transitions  
are  neutral sector operators. (see Fig.\ref{fig:phase} and \ref{fig:phase2} right panels).  Therefore, the phase diagrams must be identical 
in the $N=\infty$ limit.   

If  the  $(-1)^F$,  $\I$,   $\C$ 
odd order parameter probes  a phase transition, respectively in SYM, QCD(BF), 
and QCD(AS/S),  this implies  the ground 
(or thermal  equilibrium) states of the corresponding theories will move in 
or out of the neutral sector.
For SYM, this does not happen, since $(-1)^F$ is  unbroken. 
 In contrast, the  nonthermal  
QCD(AS/S/BF/) has a phase in which the ground states are in nonneutral sector.
 (see Fig.\ref{fig:phase} and \ref{fig:phase2} left panels).  
Hence, in this phase, there is no equivalence between SYM and the others. 
In the other phases, there is an equivalence. 
 
Let us make this more crisp with a  simple example 
in the $\R^3 \times S^1$ limit.  The 
matching of the free energies  
in the high temperature phase (small temporal $S^1$) 
\begin{equation} 
 \lim_{N \rightarrow \infty} (N^2)^{-1}  {\cal F}^{\rm SYM} =
 \lim_{N \rightarrow \infty} (2N^2)^{-1}  {\cal F}^{\rm QCD(BF)} 
=  \lim_{N \rightarrow \infty} 
(N^2)^{-1}  {\cal F}^{\rm QCD(AS/S)} = -\frac{\pi^2}{24} T^4  
\end{equation}
 is a consequence   of the unbroken $(-1)^F$, 
$\C$ and   $\I$ symmetry, respectively, and just follows from the fact that the thermal equilibrium states in these theories are in their neutral sector. 
 On the other hand,  the mismatch of ground state 
energies  in case where $S^1$ is a small spatial circle 
\begin{equation}
 0=\lim_{N \rightarrow \infty} (N^2)^{-1}  {\cal E}^{\rm SYM} \neq 
    \lim_{N \rightarrow \infty} (2N^2)^{-1}  {\cal E}^{\rm QCD(BF)} =  
\lim_{N \rightarrow \infty} 
(N^2)^{-1}  {\cal E}^{\rm QCD(AS/S)}=  -\frac{\pi^2}{24 (R_{S^1})^4}   
\end{equation}
  results due to broken 
$\I$ and $\C$. In this regime of 
QCD(AS/S/BF), the ground states do not lie in their respective 
neutral sectors.  
 This  nicely illustrates the symmetry realization 
conditions which are both necessary and sufficient for large $\Nc$ orbifold and 
orientifold equivalence. 

Certain aspects of the large $\Nc$  equivalence are surprising, 
and may teach us important lessons on    
 the 
structure of large $N$ (and hopefully smaller $N$) gauge theories.  One aspect 
we did not sufficiently emphasise in the analysis of phases, yet  is of 
 importance, is vanishing of certain correlators and expectation values 
in the $N=\infty$ 
limit, not because of symmetry, but because of the dynamics of the  
$N=\infty$  theory. We will first allude to a  few of such examples, and try to 
understand the underlying reasons afterwords. 

\subsection{Center symmetries}
As explained in footnote \ref{footnote:centerAS},  in QCD(AS/S) the center 
symmetry is just $\Z_2$,  unlike $\None$ 
SYM and QCD(BF) where the center symmetry is $U(1)$. In QCD(AS/S), the 
unbroken 
$\Z_2$ center symmetry does guarantee that  the Wilson lines with odd 
winding number will have zero vacuum expectation value, however, it
does not enforce this on the  Wilson lines with even winding number. 
On the other hand, for QCD(BF) or $\None$ SYM,  any Wilson line (with an 
arbitrary  winding number) will have zero expectation value just due 
to unbroken $U(1)$  center symmetry. This just means, in QCD(AS/S), the even 
winding number Wilson lines must be suppressed in the large $N$ limit, 
$\langle \frac{1}{N} \tr U^{2k}\rangle = \O(1/N) $ rather than being $\O(1)$. 
Therefore,  in the confining phase of  QCD(AS/S) 
 $\langle \frac{1}{N} \tr U^{2k}\rangle = 0  $ at $N=\infty$ 
without a  symmetry reason. More explicitly, in the confined phase, 
the large $N$  orbifold/orientifold  
equivalence  implies  
 \begin{eqnarray}
&&\langle \frac{\tr}{N} U^k \rangle^{\rm SYM} = 0,  \;\; 
 \langle \frac{1}{2N} (\tr U_1^k  \pm   \tr U_2^k) \rangle^{\rm QCD(BF)} = 0 \cr
&&  \langle \frac{\tr}{N}  U^{2k+1} \rangle^{\rm QCD(AS/S)} = 0, \;\;
\langle \frac{\tr}{N}  U^{2k} \rangle^{\rm QCD(AS/S)} = \O(\frac{1}{N}), 
 \end{eqnarray}
where $k=1, \ldots \infty$. Lattice simulations can easily check this 
assertion.
Identical considerations are also true for the equivalences relating the 
unitary,  orthogonal and symplectic gauge theories. In 
the last two, since their center symmetries are finite groups, symmetry 
 only implies  the expectation value of the 
Wilson lines with odd winding number has to vanish, and the loops 
with even winding number are in principle $\O(1)$. However, large $N$ 
equivalence implies the expectation value of the even-winding loops in 
$SO/Sp$ 
theories, in confined phases, should be $\O(1/N)$, and hence vanish in 
the $N=\infty$ limit.   

\subsection{Gluon condensate and vacuum energy}

Another example of this kind is the gluon condensate $\frac{1}{N} \tr F^2$. 
For  $\None$ SYM on $\R^{3,1}$, the vacuum expectation value of 
$\frac{1}{N} \tr F_{\mu \nu}^2$ is zero because of the unbroken supersymmetry.  
The easiest way to see this is to recall the trace anomaly 
as an operator identity, 
\begin{equation}
T^{\mu}_{\; \mu} = \frac{3N }{16 \pi^2} \left\{ 
\begin{array} {ll}
\tr F_{\mu \nu}    F^{\mu \nu} & \qquad \N=1 {\rm SYM } \\
(1+ \O(1/N)) \tr F_{\mu \nu}    F^{\mu \nu} & \qquad  {\rm QCD(AS/S)} \\
(1+ \O(1/N^2)) (\tr   F_{1}     F_{1} + 
\tr   F_{2}   F_{2}) & \qquad   {\rm QCD(BF)}  
\end{array} 
\right. 
\label{Eq:tranomaly}
\end{equation}
In  $\N=1$ SYM,  the energy momentum tensor  can be written as 
$2 \sigma_{\mu, \alpha {\dot \alpha} } T^{\mu \nu} = \{\bar Q_{\dot \alpha}, 
J^{\nu}_{\alpha} \}$ where  $J^{\nu}_{\alpha}$ is supercurrent and  
$Q_{ \alpha}= \int J^{0}_{\alpha}$ is the supercharge. 
  Since 
the supercharge $Q$ annihilates the vacuum, the vacuum expectation value of 
the energy momentum tensor, and hence gluon condensate $\langle \tr F^2 \rangle $ 
is identically zero 
for   $\None$ SYM.    On the other hand,  QCD(AS/S/BF) are 
nonsupersymmetric and there is no symmetry reason for   
$\langle \frac{1}{N}\tr F^2 \rangle$ to vanish, therefore, it should  not.
For the $n_f=1 $ QCD-like theories  on  $\R^{3,1}$ (or  $\R^{4}$), 
we expect 
 \begin{equation}
\langle \frac{\tr}{N}  F_{\mu \nu}^2 \rangle^{\rm SYM} = 0,  \;\; 
 \langle \frac{\tr}{N}  F_{\mu \nu}^2 \rangle^{\rm QCD(AS/S)} = \O(\frac{1}{N}), 
\;\;
 \langle \frac{1}{2N} (\tr F_1^2  +   \tr F_2^2) \rangle^{\rm QCD(BF)} = 
\O(\frac{1}{N^2}), 
\label{Eq:condensate}
 \end{equation}
  The implication of the  nonperturbative 
large $N$  orbifold and orientifold equivalence is that the  gluon condensate 
is suppressed in the $N=\infty$ limit for QCD(AS/S/BF).   In these 
 $N= \infty$ one flavor QCD-like 
theories, the condensate should vanish identically without any symmetry 
reasons. 

In a Lorentz invariant theory, 
the vacuum energy density is 
$${\cal E}= \langle T^{00} \rangle = \frac{1}{4} \langle T^{\mu}_{\mu} 
\rangle.  $$ Therefore, using the  trace anomaly  Eq.\ref{Eq:tranomaly} 
and  using the $N$ dependence of condensates Eq.\ref{Eq:condensate}, we reach 
$
{\cal E}^{\rm SYM}=0, \; {\cal E}^{\rm QCD(BF)}=  \O(1),  \;
{\cal E}^{\rm QCD(AS/S)}= \O(N).
$
This demonstrates two things: First, the leading $\O(N^2)$  
ground state energy is zero for both SYM, and QCD(AS/S/BF), rather than 
its natural $\O(N^2)$ scale. 
 Second, the 
 vacuum energy density of  QCD(BF) is  $\O(N)$ better than the QCD(AS/S) 
theory. [The observation that vacuum energy should be $\O(N)$ in QCD(AS/S) 
is made in \cite{Armoni:2003jk,Armoni:2004uu}.]
 Despite the divergent 
 vacuum energy of QCD(AS/S), the large $N$ orientifold equivalence is 
valid (assuming $\C$ is unbroken in $\R^4$), since the large $N$ equivalence 
only applies to  
the leading $\O(N^2)$ contribution to vacuum energy which happens to be zero: 
\begin{equation}
\lim_{N \rightarrow \infty} \left[ (N^2)^{-1}  {\cal E}^{\rm QCD(AS/S)}= 
\frac{\O(N)}{N^2}\right] =
\lim_{N \rightarrow \infty} \left[ 
(2N^2)^{-1}  {\cal E}^{\rm QCD(BF)}= \frac{\O(1)}{N^2} \right] =0  
\end{equation}
Physically, a more interesting  quantity to look at is the quark mass and 
$\theta$  
angle dependence of the vacuum energy density in the presence of a small mass 
perturbation \cite{Armoni:2004uu,Kovtun:2005kh}. In this case,  the vacuum 
degeneracy between $\{N, N, N-2, N+2\}$ vacua of the 
 SYM, QCD(BF), QCD(AS/S), do get lifted by an amount proportional to mass times 
the chiral condensate of the respective theory. 
The vacuum energy density as a function 
of $\theta$ has indeed its natural $\O(N^2)$ scaling, and 
  is identical for SYM and QCD(BF), and  deviates only at subleading 
$\O(1/N)$ level from  QCD(AS/S). Explicitly, 
\begin{equation} 
\lim_{N \rightarrow \infty} \left[ 
\frac{{\cal E}^{\rm SYM}}{N^2}= \frac{{\cal E}^{\rm QCD(BF)}}{2N^2}=
  \frac{{\cal E}^{\rm QCD(AS/S)}}{N^2}\right](\theta)=
 \min_k \frac{m}{\lambda} \Lambda^3 
\cos[\frac{2 \pi}{N}k + \frac{\theta}{N} ]
\end{equation}
 The $N$ degenerate vacua splits 
into $N$ branches,  each of which is $ 2\pi N$ periodic.    Consequently, 
the vacuum energy is just $2 \pi$  periodic, i.e., 
${\cal E}(\theta)= {\cal E}(\theta+ 2 \pi)$ as expected.  The level crossings
 takes place at $\theta=\pi$ where two states becomes degenerate.

The main point is that 
the vanishing of the vacuum energy in the massless limit of SYM 
 is due to supersymmetry. On the other hand, there is no symmetry reason 
for the $\O(N^2)$ contribution to vacuum energy to vanish for the 
nonsupersymmetric QCD(AS/S/BF) theories. When supersymmetry is softly broken 
by a small mass term, then the vacuum energy regains its natural $\O(N^2)$ 
scaling for SYM, and coincides with the natural $\O(N^2)$ 
vacuum energy densities of the 
 nonsupersymmetric QCD(AS/S/BF).

 \subsection{Spectral degeneracies }
The nonperturbative particle spectrum of any QCD-like theory, including 
$\N=1$  SYM is beyond the reach of our current analytical capabilities. 
Unfortunately, supersymmetry has not been greatly helpful to embark on  
this  hardest front of gauge theory either.  
Nonetheless, unbroken supersymmetry  implies 
spectral degeneracies among the bosonic  and fermionic 
excitations.  Only the bosonic $(-1)^F$ even particles lie in  the neutral 
sector of SYM.
In $N= \infty$ QCD(AS/S/BF), all the gauge invariant 
operators (with finitely many field insertions) 
that one can construct are necessarily bosonic. 
Since the large $N$ equivalence is between bosonic subsector of the 
$\None$  SYM and $\Z_2$ ($\C$ or $\I$)  invariant bosonic subsectors of 
QCD(AS/S/BF),  this means that the bosonic spectrum of  QCD(AS/S/BF) and SYM
must match. Let ${\cal B}$ denote the bosonic Hilbert space. 
Then the implication of 
the large $N$ equivalence on the nonperturbative 
 particle spectra is 
\begin{equation}
{\rm spec}[{\cal B}]^{\rm SYM}= {\rm spec}[{\cal B^+}]^{\rm QCD(AS/S)}=
{\rm spec}[{\cal B^+}]^{\rm QCD(BF)}
\label{Eq:specB1}
\end{equation}
where ${\cal B^+}$ labels $\C$-even physical states  in QCD(AS/S) 
and  $\I$-even states  in QCD(BF). 
This brings us to our next point.  The spectral matching of bosons 
(and fermions)  in SYM is due to supersymmetry, however, there is no symmetry 
reason  for bosonic pairs to coincide in QCD(AS/S/BF) theory. \footnote{ 
In 
multiflavor case, the generalization of the nonperturbative spectrum relation 
among our vectorlike gauge theories is 
$ 
{\rm spec}[{\cal B}]^{\rm QCD(adj)}= \ 
{\rm spec}[{\cal B^+}]^{\rm QCD(AS/S/BF)} $
In particular, this implies the number of the massless (or light) 
nonperturbative states corresponding to the Goldstone bosons 
in ${\rm spec}[{\cal B}]^{\rm QCD(adj)}$ 
should coincide with ${\rm spec}[{\cal B^+}]^{\rm QCD(AS/S/BF)}$. This is 
indeed  demonstrated  to be the case as discussed
in   section \ref{sec:multiflavor}, reflecting the importance of the even 
subsector (neutral) of bosonic Hilbert space. 
In other words, the $\C$ and $\I$ odd 
Goldstone bosons and  mesons living in 
${\rm spec}[{\cal B^-}]^{\rm QCD(AS/S/BF)}$ has no image in the Hilbert  space 
of the QCD(adj).} 
One can continue this list further and applications for supersymmetric 
theories are given in \cite{Kovtun:2005kh}.

To expand on this idea a bit further, consider 
 a supersymmetric Ward identity  for $\N=1$  SYM theory.  Let $W$  be the 
field strength  superfield of one chirality, $\bar D W =0$. A particular 
supersymmetric Ward  identity is 
 $
\langle{\rm tr} D^2 W^2 (x) \; {\rm tr} D^2 W^2(0)\rangle =  0 \, 
$
implying $\langle ( \tr F^2(x) + i \tr F \widetilde F(x))  
(\tr F^2(0) + i \tr F \widetilde F(0)]\rangle_{\rm conn}=0$. 
 Due to unbroken parity in vectorlike gauge theories 
\cite{Vafa:1984xg}, the cross terms in the connected correlator 
vanish and we obtain 
\begin{equation}
\langle  \tr F^2(x) \tr F^2(0) \rangle_{\rm conn} - 
\langle \tr F \widetilde F(x) \tr F \widetilde F(0) \rangle_{\rm conn}=0.
\end{equation}
Notice that this connected correlator does not vanish due to fermion-boson 
degeneracy, but due to same spin, (spin zero) boson-boson  degeneracy of the 
spectrum.   
The unbroken supersymmetry implies that massive scalar particles in the 
smallest supermultiplets of $\N=1$ SYM should be degenerate with spin zero 
pseudoscalar, as well as with spin $\half$ fermions.    
The spin-zero, positive (negative) parity 
color singlet glueballs  exhaust the first (second) correlator. 
 Since this is true at arbitrary separation,  it implies 
\begin{equation}
\sum _{\rm even\,\, parity} \frac{|\alpha_n^{+}|^2} { k^2 + m_{n,+}^2} = 
\sum _{\rm odd\,\, parity} \frac{|\alpha_n^{-}|^2}{k^2+ m_{n,-}^2}
\label{Eq:glueballs}
\end{equation}
at arbitrary momenta $k$. Here, ${\alpha_n^{\pm}}$ is the amplitude to 
create a positive  (negative) parity glueball state at level
 $n$ and $m_{n, \pm}^2$ is the mass  of parity even/odd glueball. 
Notice that the relation Eq.\ref{Eq:glueballs} is not a sum rule, 
since it is true at arbitrary momentum.  
Therefore, the implication of the equivalence is 
$\frac{m_{n,+}^2}{m_{n,-}^2}=\left\{ 1, 1+\O(1/N),  1+\O(1/N^2) \right\}$ 
respectively for SYM, QCD(AS/S) and QCD(BF). The nonsupersymmetric QCD(AS/S/BF) 
theories must exhibit parity doubled scalars and pseudoscalar provided 
the respective $\Z_2$ symmetries are unbroken \cite{Kovtun:2005kh,
Armoni:2003fb}, in the 
$N= \infty$ limit.  
The spectral 
degeneracies  in QCD(AS/S/BF) is not protected by symmetries, but just 
due to the equivalence. 

\subsection{On the structure of large $N$ limit } 

One natural question is,  why is it that in  the $N= \infty$ limit, 
distinct quantum theories 
with very  different fundamental symmetries 
(the orbifold and orientifold partners of SYM, for example),  behave
as if they have  some higher symmetry, even though in 
reality    they  do  not?    Does it make sense to ask whether  there are 
accidental symmetries emerging in the $N= \infty$ limit?

The proof of large $N$ equivalence in fact has the key ingredient.
 In brief, this  follows 
from the  fact that the large $N$   is a classical limit 
in the sense that  the root mean square quantum fluctuations of 
reasonable  operators are suppressed \cite{Yaffe:1981vf}. (The reasonable 
 operators are the ones with a smooth large $N$ limit.)  
 At $N= \infty$, appropriately identified 
operators in  neutral sectors of QCD(AS/S/BF/adj) 
satisfy {\it identical} Schwinger-Dyson or loop equations 
\cite{Kovtun:2003hr} if the symmetries defining their neutral sectors are 
unbroken, leading  to the 
equivalence of the theories.
(or  their classical Hamiltonians and coherence algebras coincide as 
described in \cite{Kovtun:2004bz}.) 
At any finite $N$, there are corrections  to the 
loop equations which are different for each individual theory and such 
corrections do not match. 
A subset of these  subleading 
corrections may be regarded as a consequence of the quantum or thermal 
fluctuations in the large $N$ limit. This is most easily seen by recalling  
the factorization (or cluster decomposition) property 
that   is used in obtaining the $N= \infty$ loop equations. 
  The factorization means,  if $O_i(x)$ is some reasonable  gauge invariant 
operator,  then  
\begin{equation} 
\langle O_1(x) O_2(x) \rangle -\langle O_1(x) \rangle 
\langle O_2(x) \rangle  = \O(1/N^p) 
\end{equation} 
where $p=1,2$ depending on the theory.
  Therefore,  
 $\langle O(x)^2 \rangle -\langle O(x) \rangle^2 = \O(1/N^p)  $ 
implying the suppression of the  quantum fluctuations 
in the 
$N \rightarrow \infty$ limit. In this sense, the quantum theories in the 
$N=\infty$ limit behave as classical  \cite{Yaffe:1981vf}.  
\footnote{The 
clustering  property breaks down if the theory is in the 
mixed phase (or there are 
more than one vacuum)  in the theory. All of our $n_f=1$  QCD-like theories, 
SYM, QCD(AS/S/BF) has  $N$ isolated  vacua (in leading order in large $N$) 
characterised by their chiral condensate in $\R^4$.  
In such cases, one can add a small mass term for fermions and that will lift 
the degeneracy completely leading to unique vacuum for arbitrary $\theta$ 
angle (except for $\theta= \pi$).   The large $N$ cluster decomposition will 
be valid at any value of the mass. Analogous operation on 
$\R^3 \times (\rm thermal \; S^1)$ will pick a single 
thermal equilibrium state. Both operations,  in essence, 
restricts the Hilbert space of the corresponding theories over a 
particular ground (or thermal equilibrium) state, hence consequently the 
cluster decomposition  property is  restored. The addition of these small 
perturbations  (which explicitly breaks  chiral and center symmetry) 
also changes the  phase transitions of the corresponding $N= \infty$ SYM, 
QCD(AS/S/BF) into rapid crossovers.}
The difference in the 
loop equations of  QCD(AS/S) and QCD(BF) and QCD(adj) essentially arises in 
these subleading terms.  
Such fluctuations  are typically of $\O(1/N)$ for QCD(AS/S) and $\O(1/N^2)$ 
for QCD(BF/adj).   
At $N= \infty$,   there is no difference in 
the neutral sector dynamics of   QCD(AS/S/BF/adj)   as demonstrated by 
using  loop equations or the coherent state approach 
 \cite{Kovtun:2003hr, Kovtun:2004bz}, and dynamics of the theories coincide. 

In that regard, the  large $N$ orbifold/orientifold 
partners with very different fundamental symmetries, 
enjoys  in their neutral sector,  the protection of the 
highest symmetry  available in the equivalence  chain. 
For example, the Wilson lines 
  of QCD(AS) clearly behaves as if they are protected by a   $U(1)$ center 
symmetry (of its partners, SYM or QCD(BF)) 
rather than $\Z_2$ (its own truthful symmetry). There are spectral 
degeneracies in the 
bosonic Hilbert spaces  as if  it is protected by a supersymmetry,
 i.e.,  ${\rm spec}[{\cal B^+}]^{\rm QCD(AS/S/BF)} 
={\rm spec}[{\cal B}]^{\rm SYM}$ . 
If a phase transition is driven by a neutral operator, it will concurrently 
take place in all the theories in the equivalence chain. In other words, 
there is a well-defined mapping among the  connected correlators of all neutral 
sector operators.  
If a progress leads to the solution of one of these $N=\infty $ theories, 
say $\N=1$ SYM,  it implies solution for the neutral sectors of all. This is 
the essence of the nonperturbative orbifold and orientifold equivalence. 

 The  large $N$ theories also possess  
    non-neutral subsectors, as well. The loop equations of operators 
in these  subsectors are unrelated to one another, and 
there is no map between them.  For example, QCD(AS/S/BF) are non-supersymmetric 
and do not have fermionic excitations at all. The large $N$ equivalence do not 
apply to nonneutral symmetry channels. 
The presence of the nonneutral 
sector, and its dynamics do not alter the dynamics of the neutral sector so 
long as  the symmetries defining the neutral sectors are unbroken.  If the 
symmetries are unbroken, then the ground (and thermal equilibrium) state of the 
corresponding theories are in the neutral sector. Otherwise, they are outside 
the neutral sector, and there is no equivalence. 

\section{Prospects}

 The phase diagrams of QCD-like gauge theories with fermions in two index 
representations  QCD(adj/AS/S/BF) shown in Figs.\ref{fig:phase}, 
\ref{fig:phase2}, \ref{fig:phase3}  and understanding 
their interrelation via the nonperturbative 
orbifold and orientifold equivalences 
 are the main  results  of this 
work. As stated in the introduction,  the nonperturbative 
large $N$ orbifold and orientifold 
equivalences are only valid if certain symmetry realizations are satisfied. 
In our examples, this translates into unbroken 
 charge conjugation symmetry $\C$ in 
$U(N)$ QCD(AS/S) and unbroken $\Z_2=\I$ interchange   symmetry in 
$U(N) \times U(N)$ QCD(BF) \cite{Kovtun:2004bz,Unsal:2006pj}.  
The phase  diagrams 
 are in part derived within the regime of applicability of 
perturbative analysis thanks to the work of 
Refs.\cite{Sundborg:1999ue,Aharony:2003sx} on $S^3 \times S^1$,  
and in part borrowed from lattice gauge 
theory in the strongly coupled regime  in cases where data exists 
\cite{Narayanan:2005en,DeGrand:2006qb,Kogut:1982rt}.  
All of our findings support the smooth volume dependence 
conjecture for (spatial and temporal) center symmetry realizations: 
{\it In asymptotically free, confining vectorlike large $N$ gauge theories, 
if a weak coupling 
phase transition changing the (spatial or temporal) center symmetry 
realization exists at small volume on $S^3 \times S^1$, it will evolve into 
a full-blown  nonperturbative (strong coupling) phase transition in 
$\R^3 \times S^1$.   Inversely, if we do not find any such transition on small 
$S^3\times S^1$, there will not be a center symmetry changing transition 
on $\R^3 \times S^1$ either. }
Consequently, our phase diagrams  merely reflect the  simplest 
possibilities consistent 
with our current knowledge of vectorlike gauge theories and in that regard, 
should be viewed as  conjectural rather than fully demonstrated.  

In particular, we were unable to demonstrate that $\C$ in the orientifold 
QCD(AS/S), and  $\I$ in the case of orbifold QCD(BF) cannot be spontaneously 
broken in the  large $S^1$ limit on 
$\R^3 \times S^1$.  However, we were able to 
show that if $S^1$ is a thermal (or temporal)  circle (endowed with 
antiperiodic boundary conditions for fermions), neither $\I$ nor $\C$ 
are broken at small radius. On the other hand, if $S^1$ is a spatial 
circle (where periodic boundary conditions are used for fermions), 
then both $\I$ and $\C$    are broken on small radius. Therefore, this symmetry 
breaking is sensitive to the choice of the boundary conditions for fermions 
along the $S^1$ circle and is clearly  a finite size effect, albeit physical 
as the transition is expected to occur around the strong scale 
$\Lambda^{-1}$.  In such cases, we 
chose the simplest consistent possibility at large radius, i.e, unbroken 
$\I$ and $\C$. \footnote{
Currently, there is no proof
demonstrating  $\I$ in orbifold QCD(BF) and  $\C$ in QCD(AS/S) cannot be spontaneously 
broken on $\R^4$ or $\R^3\times$ (large $S^1$). The essential obstacle for the 
proof of such argument is the existence of ``real local'' order parameters 
probing $\C$ and $\I$, 
unlike  the case of parity \cite{Vafa:1984xg} 
where all local order parameters are purely 
``imaginary'' in Euclidean formulation in $\R^4$.  
We  find the spontaneous 
breaking of $\C$ and $\I$ on large radius  strongly unlikely and everything 
that we know about these theories is consistent with unbroken $\C$ and $\I$.
[This last viewpoint is also emphasised for $\C$ without providing a proof in 
\cite{Armoni:2007rf,Armoni:2007vb}. My intuition is that, the two problems are 
intimately connected. If it is possible to prove one rigorously, the other 
will  follow.] We should also note that 
 in theories involving fundamental scalars and formulated on $\R^4$, 
it is fairly easy to construct theories which spontaneously break 
both $\C$ and $\I$ via local order   parameters.}
 We were also able to show that the effective potentials 
for QCD(AS/S) and QCD(BF) on $S^3 \times S^1$ map to identical problems 
Eq.\ref{Eq:orb2} and Eq.\ref{Eq:oo} in the 
respective $\C$ and $\I$ eigenstates of the Wilson lines. This demonstrates 
that  within the regime of applicability of perturbation theory, all $\C$ and 
$\I$ symmetry  realizations of orientifold QCD(AS/S) and orbifold QCD(BF) 
are identical, hopefully altering unjustified prejudice
 in the literature about orbifold QCD.  

We have also seen some remarkable new phenomena. Some of these are  
artifacts of the  topology of the sphere such as 
low temperature confining phases without chiral symmetry breaking in 
QCD(adj/AS/S/BF)  in the 
 small $S^3$ limit. It is likely that  this phase is not 
visible in lattice simulations 
 traditionally  formulated on $T^4$. On the other 
hand, we have also seen zero temperature chirally symmetric phases of QCD(AS/S) 
on $\R^3 \times S^1$ (where one of the $\R$ is the decompactified temperature 
circle)  which also breaks  symmetries such as CPT and P.  The existence of 
this phase for $N=3, n_f=4$  QCD has been demonstrated in recent lattice 
simulations in ref.\cite{DeGrand:2006qb, Lucini:2007as}. 
It should be emphasised that 
this  phase transition is {\it not} due to thermal fluctuations, since on 
$\R^3 \times\; ({\rm spatial}\;  S^1)$  the temperature is zero, 
and rather quantum mechanical fluctuations trigger an instability. 
In general,  the center symmetry changing 
phase transitions probed by working with the twisted partition 
function $\widetilde Z= \tr [(-1)^F e^{-\beta H}]$ can be thought of as the 
transitions induced by quantum fluctuations,  and the  
transitions monitored by  working with the thermal partition 
function  $Z= 
\tr e^{-\beta H}$   are typically induced  by thermal fluctuations. 

To our knowledge,  there is no analysis of phase diagrams of large $N$, 
asymptotically free, confining  QCD-like  gauge  theories formulated on 
$S^3 \times S^1$ as a function of volume in the string theory 
literature. This is true for  $\N=1$ SYM, as well.   
Realizing that the phase diagrams of 
$\N=1$ SYM  shown in Fig.\ref{fig:phase} (both thermal and nonthermal) 
are as rich  as  
any other QCD-like  theory,  it  may be worth pursuing this further.
The most interesting aspect  arises in the 
$\R^3 \times ({\rm spatial}) \; S^1$ limit. 
In particular,  we observe that 
for QCD-like theories with complex representation 
Dirac fermions, a change in spatial center symmetry and chiral symmetry 
realizations should occur at some critical spatial radius (not temperature) 
\cite{DeGrand:2006qb,Unsal:2006pj}. 
On the other hand, 
for real representation fermions endowed with periodic boundary 
conditions, a spatial center symmetry changing transition does  not occur. 
(For adjoint representation fermions and gauge groups $SU/SO/Sp$, 
this is demonstrated in \cite{Kovtun:2007py} and used to demonstrate that  in 
such QCD-like theories the dynamics is independent of the $S^1$ volume.)
 It would be  interesting to  see how such a phenomenon might arise 
in the 
stringy setups of Refs.\cite{Aharony:2006da,Parnachev:2006dn,Sakai:2004cn}.

Finally, the phase diagram of conformal 
$\N=4$ SYM on $S^3 \times S^1$
 is currently known \cite{Aharony:2003sx,Sundborg:1999ue} 
at least at weak coupling and presumably 
at strong coupling due to AdS/CFT.  Examining the phases of the 
simplest orbifold and orientifold partners of  $\N=4$ SYM where 
gauge-gravity duality is probably better understood 
(in comparison to  the asymptotically free, confining gauge theories)
may help us to find new 
gravitational phases, as well as  phase transitions among 
different  geometries. Research in this direction is ongoing. 

\acknowledgments
I am indebted to Larry Yaffe, who taught me large $N$ limits, for many 
enlightening discussions in the course of this work. 

I  benefited from conversations with 
  Maarten Golterman,  Andy Cohen,  Pavel Kovtun, Ofer Aharony, 
Michael Peskin, Du{\v s}an  Simi{\'c}, Misha Shifman, Adi Armoni, 
Matt Headrick, Erich Poppitz, and Tim Hollowood. 
I  thank Tom DeGrand and  Roland Hoffmann for sharing their simulation 
results  on $\C$ breaking and restoration prior to the publication. 
I would like to acknowledge 
the Aspen Center for Physics where portions of this paper were initiated.
This work was supported by the
U.S.\ Department of Energy Grants DE-AC02-76SF00515

\appendix

\section{Independence of dynamics from the boundary 
 conditions on  $\R^3 \times (\rm large \;  S^1)$}
\label{sec:appendix}
In the  paper, we made an assertion that the $N= \infty$ dynamics of the 
QCD-like theories must be independent of the boundary condition of fermions on 
$\R^3 \times S^1$  so long as $S^1$ is sufficiently large.  This is an 
immediate corollary to the volume independence of large $N$ gauge theories 
\cite{Kovtun:2007py}. Below, we give a sketch of the proof. 
In order to make our arguments well defined, we use a lattice regularization.  
The underling assumptions of the proof are  
{\it i)} a smooth large $N$ limit, {\it ii)}  unbroken center symmetry. 

Let us consider a  $U(2N)$ QCD(adj) to be specific on $\R^3 \times S^1$ where 
the lattice along $S^1$ has size $2L$. 
Assume the fermions are endowed with periodic 
boundary condition,  $\psi(2L) = \psi(0)$ along the $S^1$ circle.  Then, 
one can construct a volume reducing  $\Z_2$ projection  which is equivalent 
to imposing the constraints:
$\psi(x+L) = -\psi(x) $ for fermions and $U (x+L) = U(x)$ 
for link fields. The outcome is $U(2N)$ QCD(adj) on $S^1$ with size $L$. 
The validity  of this large $N$  equivalence relies on translation symmetry 
and fermion number symmetry in large volume theory (which are unbroken) 
and center symmetry in the 
daughter, small volume theory. 
One can also ``undo'' the volume reduction, by a  blow-up projection, 
which takes $U(2N)$  gauge 
theory on size $L$ lattice to a $U(N)$ gauge theory on size $2L$ lattice while 
preserving the antiperiodic boundary conditions for fermions. (For details, see
\cite{Kovtun:2007py}.) Now, small volume theory is regarded as a parent and large volume theory as the daughter. The validity of the equivalence again relies on 
identical symmetry realizations. 

The first $\Z_2$ projection lost half of the degrees of freedoms,  so did 
the second $\Z_2$. 
\begin{equation}
 [U(2N), 2L] 
\mathop{\longrightarrow}\limits^{\Z_2} 
 [U(2N), L] \mathop{\longrightarrow}\limits^{\Z_2} [U(N), 2L] 
\end{equation}
Assuming the 't Hooft large $N$ limit is smooth,  the net effect of the 
two projection is to change the periodic boundary conditions of fermions to 
antiperiodic ones, while reducing the number of color by a factor of four. 

The condition of the unbroken center symmetry in the small volume theory 
is equivalent to the requirement that the theory be in a low temperature 
confining phase.  Therefore, so long as the temperature 
is lower than the deconfinement temperature, 
the connected correlators and expectation values 
 are independent of the boundary conditions along the 
$S^1$ circle. Combining this boundary value independence with volume 
independence,   we observe that 
in the large $N$ limit,  the spectrum of particles, chiral 
condensates, gluon condensate  in the QCD-like theories 
(and in particular, in $\N=1$ SYM)  should be temperature, and boundary value 
 independent.

\bibliographystyle{JHEP} 

\bibliography{orientifold1}

\end{document}